%% file: ms5.tex
\documentclass[12pt,preprint]{aastex}
\newcommand{\bdv}[1]{\mbox{\boldmath$#1$}}

\def\au{{\rm au}} 
 
\def\kms{{\rm km}\,{\rm s}^{-1}}
\def\masyr{{\rm mas}\,{\rm yr}^{-1}}
\def\kpc{{\rm kpc}}
\def\mas{{\rm mas}}

\def\muas{\mu{\rm as}}

\def\max{{\rm max}}

\def\rel{{\rm rel}}

\def\eff{{\rm eff}}
\def\rot{{\rm rot}}

\def\fwhm{{\rm FWHM}}
\def\e{{\rm E}}
\def\bxi{{\bdv\xi}}
\def\bpi{{\bdv\pi}}
\def\bmu{{\bdv\mu}}

\def\bgamma{{\bdv\gamma}}

\def\btheta{{\bdv\theta}}

\begin{document}
\title{Systematic KMTNet Planetary Anomaly Search. X.  Complete Sample of 2017 Prime-Field Planets}

\input author.tex

\begin{abstract}

  We complete the analysis of planetary candidates found by the KMT
  AnomalyFinder for the 2017 prime fields that cover $\sim 13\,{\rm
    deg}^2$.  We report 3 unambiguous planets: OGLE-2017-BLG-0640,
  OGLE-2017-BLG-1275, and OGLE-2017-BLG-1237.  The first two of these
  were not previously identified, while the last was not previously
  published due to technical complications induced by a nearby
  variable.  We further report that a fourth anomalous event,
  the previously recognized
  OGLE-2017-BLG-1777, is very likely to be planetary, although its
  light curve requires unusually complex modeling because the lens and
  source both have orbiting companions.
  One of the 3 unambiguous planets, OGLE-2017-BLG-1275 is the
  first AnomalyFinder discovery that has a {\it Spitzer} microlens
  parallax measurement, $\pi_\e\simeq 0.045\pm0.015$, implying that
  this planetary system almost certainly lies in the Galactic bulge.
  In the order listed, the four planetary events have planet-host mass
  ratios $q$, and normalized projected separations $s$, of $(\log
  q,s)$ = $(-2.31,0.61)$, $(-2.06,0.63/1.09)$, $(-2.10,1.04)$, and
  $(-2.86,0.72)$.  Combined with previously published events, the 2017
  AnomalyFinder prime fields contain 11 unambiguous planets with
  well-measured $q$ and one very likely candidate, of which 3 are
  AnomalyFinder discoveries.  In addition to these 12,
  there are three other unambiguous planets with
  large uncertainties in $q$.
  
\end{abstract}


\section{{Introduction}
\label{sec:intro}}

We present the analysis of all planetary events that were
identified by the KMTNet AnomalyFinder algorithm \citep{ob191053,af2}
%I, IV,
and occurred during the 2017 season within the 6 prime KMTNet fields, which
contiguously cover $\sim 13\,{\rm deg}^2$ of the richest microlensing
region of the Galactic bulge, with cadences $\Gamma=2$--$4\,{\rm hr}^{-1}$.
This work follows the publications of
complete samples of the 2018 prime \citep{ob180383,kb190253,2018prime},
%II, III, V,
and sub-prime \citep{2018subprime} AnomalyFinder events,
% VI, 
the 2019 prime \citep{ob191053,kb190253,af2} and sub-prime
\citep{2019subprime} events, the 2016 prime \citep{2016prime} events,
% VIII, IX
as well as a complete sample of all events from 2016-2019 with
planet-host mass ratios $q<10^{-4}$ \citep{logqlt-4}.  There are 21 sub-prime
% VII
fields that cover $\sim 84\,{\rm deg}^2$ of the Galactic bulge with
cadences $\Gamma=0.2$--$1\,{\rm hr}^{-1}$. The above references are
(ignoring duplicates) Papers I, IV, II, III, V, VI, VIII, IX, and VII, in the
AnomalyFinder series.  The locations
and cadences of the KMTNet fields are shown in Figure~12 of \citet{eventfinder}.
Our immediate goal, which we expect to achieve within a year, is to publish
all AnomalyFinder planets from 2016-2019.  Over the longer term, we plan
to apply AnomalyFinder to all subsequent KMT seasons, beginning 2021.

%2019 subprime has 1895 events
For the 2017 prime fields, the AnomalyFinder %\citep{af2}
identified a total of 124 anomalous events (from an underlying sample of 998
events), which it classified as
``planet'' (16), ``planet/binary'' (10), ``binary/planet'' (14),
``binary'' (77), and ``finite source'' (7).
Among the 77 in the ``binary'' classification, 33 were judged by eye to be
unambiguously non-planetary in nature.  Among the 16 in the ``planet''
classification, 8 were previously published, one of which had been newly
discovered by AnomalyFinder, while one planetary event is
in preparation but has a very large uncertainty in $q$.
Among the 10 in the ``planet/binary''
classification, none were previously known planets,
and among the 14 in the ``binary/planet'' classification 
one was a previously published planet.  Among the 77
classified as ``binary'', one was previously published, and two others
had been recognized as probably planetary
(OGLE-2017-BLG-1237 and OGLE-2017-BLG-1777).
None of the 7 ``finite source'' events were previously published planets.
In sum, the AnomalyFinder recovered 9 previously published planets that
had been discovered by eye,
as well as two other planets that had been recognized but not published.
In addition, one of its new discoveries (KMT-2017-BLG-0428) has already
been published \citep{logqlt-4}, while another  (OGLE-2017-BLG-0448)
is in preparation but with a very large uncertainty in $q$.
We find that among the remaining
candidates two are unambiguously planetary (OGLE-2017-BLG-0640
and OGLE-2017-BLG-1275), while two others have planetary solutions but
cannot be unambiguously interpreted (OGLE-2017-BLG-0543 and OGLE-2017-BLG-1694).
Hence, we here report analyses of a total of six anomalous events,
three that are unambiguously planetary, one that is very likely to
be planetary, and two that have competitive planetary and non-planetary
solutions.
% previously recognized: OB171237, OB171777
% new: OB171275, OB170640

Finally, we note that while there are no known planets from among 2017
prime-field KMT events that were not identified as candidates by the
AnomalyFinder, there is one known planet,
OGLE-2017-BLG-0604 \citep{kb162397}, that does not enter the AnomalyFinder
sample because the underlying event was not found by the EventFinder system
\citep{eventfinder} and so was not subjected to the AnomalyFinder search.
Judging by the strength of the planetary signal
(Figure~6 of \citealt{kb162397}), it almost certainly would have been found.
However, this omission will have no effect on statistical analyses based on
this paper (and other AnomalyFinder papers) because both the detections and
planet sensitivity calculations are restricted to events cataloged by
EventFinder and AlertFinder \citep{alertfinder}.

\section{{Observations}
\label{sec:obs}}

The description of the observations is nearly identical to that in 
\citet{2018prime} and \citet{2018subprime}.
The KMTNet data are taken from three identical 1.6m telescopes,
each equipped with cameras of 4 deg$^2$ \citep{kmtnet} 
and located in Australia (KMTA),
Chile (KMTC), and South Africa (KMTS).  When available, our general policy is
to include Optical Gravitational Lensing Experiment (OGLE) and 
Microlensing Observations in Astrophysics (MOA) data in the analysis.
However, none of the 6 events analyzed here were alerted by MOA.
OGLE data were taken using their 1.3m telescope
with $1.4\,{\rm deg}^2$ field of view at Las Campanas Observatory in Chile.
For the light-curve analysis, we use only the $I$-band data.

As in those papers, Table~\ref{tab:names}
gives basic
observational information about each event.  Column~1 gives the
event names in the order of discovery (if discovered by multiple teams),
which enables cross identification.  The nominal cadences are given in column 2,
and column 3 shows the first discovery date.  The remaining four columns
show the event coordinates in the equatorial and galactic systems.
Events with OGLE names were originally discovered by the OGLE Early
Warning System \citep{ews1,ews2}.  In 2017, the KMT AlertFinder
system \citep{alertfinder} was not yet operational.  Hence all KMT events
were discovered post-season by the EventFinder system \citep{eventfinder}.

Two of the events (OGLE-2017-BLG-1275 and OGLE-2017-BLG-0640)
were observed by the UKIRT Microlensing Survey in the $H$ and $K$ bands
\citep{ukirt17} using the UKIRT 3.6m telescope in Hawaii.  We use these
data for color determination in both cases (see Section~\ref{sec:cmd}).
For the case of OGLE-2017-BLG-1275, we also incorporate them into the
light-curve modeling.  These data are available through their public
archive\footnote{https://exoplanetarchive.ipac.caltech.edu/docs/UKIRTMission.html}.

OGLE-2017-BLG-1275 and OGLE-2017-BLG-1237 were both observed by {\it Spitzer}
as part of a large-scale microlensing program \citep{yee15}.  The {\it Spitzer}
observations of OGLE-2017-BLG-1275 are described in
Section~\ref{sec:spitzer-ob171275}, while
those of OGLE-2017-BLG-1237 do not show a discernible signal and so are not
used in this paper.

To the best of our knowledge, there were no ground-based follow-up observations
of any of these events.  

The KMT and OGLE data were reduced using difference image analysis 
\citep{tomaney96,alard98},
as implemented by each group, i.e., pySIS \citep{albrow09} and
DIA \citep{wozniak2000}, respectively.  To prepare for publication,
KMT data were rereduced using a tender loving care (TLC) version of pySIS
(Yang et al., in preparation), while OGLE data were rereduced using their
standard pipeline but centroided on the lens rather than the baseline object.

\section{{Light Curve Analysis}
\label{sec:anal}}

\subsection{{Preamble}
\label{sec:anal-preamble}}

With one exception that is explicitly noted below, we reproduce here
Section~3.1 of \citet{2018subprime}, which describes
the common features of the light-curve analysis.  We do so
(rather than simply referencing that paper) to provide easy access to
the formulae and variable names used throughout this paper.
The reader who is interested in more details should consult
Section~3.1 of \citet{2018prime}.  Readers who are already familiar
with these previous works can skip this section, after first reviewing
the paragraph containing Equation~(\ref{eqn:probmu2}), below.

All of the events can be initially approximated by 1L1S models, which
are specified by three \citet{pac86}
parameters, $(t_0,u_0,t_\e)$, i.e., the time of lens-source closest
approach, the impact parameter in units of $\theta_\e$ and the Einstein
timescale,
\begin{equation}
t_\e = {\theta_\e\over\mu_\rel}; \qquad
\theta_\e = \sqrt{\kappa M\pi_\rel}; \qquad
\kappa\equiv {4\,G\over c^2\,\au} \simeq 8.14\,{\mas\over M_\odot},
\label{eqn:tedef}
\end{equation}
where $M$ is the lens mass, $\pi_\rel$ and $\bmu_\rel$ are the
lens-source relative parallax and proper-motion, respectively,
and $\mu_\rel \equiv |\bmu_\rel|$.  The notation ``$n$L$m$S'' means $n$ lenses
and $m$ sources.   In addition, to these 3 non-linear parameters, there are
2 flux parameters, $(f_S,f_B)$, that are required for each observatory,
representing the source flux and the blended flux.

We then search for ``static'' 2L1S solutions, which generally require 4
additional parameters $(s,q,\alpha,\rho)$, i.e., the planet-host separation
in units of $\theta_\e$, the planet-host mass ratio, the angle of the
source trajectory relative to the binary axis, and the angular source
size normalized to $\theta_\e$, i.e., $\rho=\theta_*/\theta_\e$.

We first conduct a grid search with $(s,q)$ held fixed at a grid of values
and the remaining 5 parameters allowed to vary in a Monte Carlo Markov chain
(MCMC). After we identify one or more local minima, we refine these by
allowing all 7 parameters to vary.

We often make use of the   heuristic analysis introduced by \citet{kb190253}
and modified by \citet{kb211391} based on further investigation in 
\citet{2018prime}.
If a brief anomaly at $t_{\rm anom}$
is treated as due to the source crossing the planet-host axis,
then one can estimate two relevant parameters
\begin{equation}
s^\dagger_\pm = {\sqrt{4 + u_{\rm anom}^2}\pm u_{\rm anom}\over 2}; \quad
\tan\alpha ={u_0\over\tau_{\rm anom}},
\label{eqn:heuristic}
\end{equation}
where $u_{\rm anom}^2= \tau_{\rm anom}^2 + u_0^2$ and
$\tau_{\rm anom} = (t_{\rm anom}-t_0)/t_\e$.
Usually,
$s^\dagger_+>1$ corresponds to anomalous bumps and
$s^\dagger_-<1$ corresponds to anomalous dips.   
This formalism predicts that if there are two degenerate solutions, $s_{\pm}$,
then they both have the same $\alpha$ and that there exists a $\Delta\ln s$
such that
\begin{equation}
s_\pm = s^\dagger_{\rm pred}\exp(\pm \Delta \ln s),
\label{eqn:heuristic2}
\end{equation}
where $\alpha$ and $s^\dagger$ are given by Equation~(\ref{eqn:heuristic}).
To test this prediction in individual cases, we can compare the purely
empirical quantity $s^\dagger\equiv \sqrt{s_+ s_-}$ with prediction
from Equation~(\ref{eqn:heuristic}), which we always label with a subscript,
i.e., either $s^\dagger_+$ or $s^\dagger_-$.  This formalism can also be used
to find ``missing solutions'' that have been missed in the grid search,
as was done, e.g., for the case of KMT-2021-BLG-1391 \citep{kb211391}.

For cases in which the anomaly is a dip, the mass ratio $q$ can be estimated,
\begin{equation}
q = \biggl({\Delta t_{\rm dip}\over 4\, t_\e}\biggr)^2
{s^\dagger\over |u_0|}|\sin^3\alpha|,
\label{eqn:qeval}
\end{equation}
where $\Delta t_{\rm dip}$ is the full duration of the dip.  
In some cases, we investigate whether the microlens parallax vector,
\begin{equation}
\bpi_\e\equiv {\pi_\rel\over \theta_\e}\,{\bmu_\rel\over\mu_\rel}
\label{eqn:piedef}
\end{equation}
can be constrained by the data.  When both $\pi_\e$ and $\theta_\e$ are
measured, they can be combined to yield,
\begin{equation}
M = {\theta_\e\over\kappa\pi_\e}; \qquad
D_L = {\au\over \theta_\e\pi_\e + \pi_S},
\label{eqn:mpirel}
\end{equation}
where $D_L$ is the distance to the lens and
$\pi_S$ is the parallax of the source.

To model the parallax effects due to Earth's orbital motion, we add
two parameters $(\pi_{\e,N},\pi_{\e,E})$, which are the components of
$\bpi_\e$ in equatorial coordinates.  We also add (at least initially)
two parameters $\bgamma =[(ds/dt)/s,d\alpha/dt]$, where
$s\bgamma$ are the first derivatives of projected lens orbital position 
at $t_0$, i.e., parallel and perpendicular to the projected separation of the
planet at that time, respectively.
In order to eliminate unphysical solutions, we impose a  constraint
on the ratio of the transverse kinetic to potential energy,
\begin{equation}
\beta \equiv \bigg|{\rm KE\over PE}\bigg|_\perp
= {\kappa M_\odot {\rm yr}^2\over 8\pi^2}\,{\pi_\e\over\theta_\e}\gamma^2
\biggr({s\over \pi_\e + \pi_S/\theta_\e}\biggr)^3 < 0.8 .
\label{eqn:betadef}
\end{equation}
It often happens that $\bgamma$ is neither significantly constrained
nor significantly correlated with $\bpi_\e$.  In these cases, we suppress
these two degrees of freedom.

Particularly if there are no sharp caustic-crossing features in the light curve,
2L1S events can be mimicked by 1L2S events.  Where relevant, we test for
such solutions by adding at least 3 parameters 
$(t_{0,2},u_{0,2},q_F)$ to the 1L1S models.
These are the time of closest approach and impact parameter of the
second source and the ratio of the second to the first source flux
in the $I$-band. 
If either lens-source approach can be interpreted
as exhibiting finite source effects, then we must add one or two further
parameters, i.e., $\rho_1$ and/or $\rho_2$.  And, if the two sources
are projected closely enough on the sky, one must also consider
source orbital motion.

In a few cases, we make kinematic arguments that solutions are unlikely
because their inferred proper motions $\mu_\rel$ are too small.   If planetary
events (or, more generally, anomalous events with planet-like signatures)
traced the overall population of microlensing events, then the fraction with
proper motions less than a given $\mu_\rel\ll \sigma_\mu$ would be, 
\begin{equation}
p(\leq\mu_\rel) ={(\mu_\rel/\sigma_\mu)^3\over 6\sqrt{\pi}}
\rightarrow 4\times 10^{-3}\biggl({\mu_\rel\over1\,\masyr}\biggr)^3
\qquad {\rm (old)},
\label{eqn:probmu}
\end{equation}
where (following \citealt{gould21})
the bulge proper motions are approximated as an isotropic Gaussian
with dispersion $\sigma_\mu = 2.9\,\masyr$.

However, subsequent to the work of \citet{2018prime} and \citet{2018subprime},
\citet{masada} showed that the proper-motion distribution of observed
planetary microlensing events scales $\propto d\mu_\rel\,(\mu/\sigma_\mu)^\nu$
where $\sigma_\mu=3.06\pm 0.29\,\masyr$ and $\nu=1.02\pm0.29$.  Hence,
in place of Equation~(\ref{eqn:probmu}), we adopt
\begin{equation}
p(\leq\mu_\rel) ={(\mu_\rel/2\sigma_\mu)^{\nu+1}\over [(\nu+1)/2]!}
\rightarrow {\mu_\rel^2\over 4\sigma_\mu^2}  
\rightarrow 2.8\times 10^{-2}\biggl({\mu_\rel\over1\,\masyr}\biggr)^2,
\label{eqn:probmu2}
\end{equation}
where we have evaluated at $\sigma_\mu=3.0\,\masyr$ and $\nu=1$.
For example, $p(\leq 0.5\,\masyr) = 0.7\%$ and
$p(\leq 0.1\,\masyr) = 0.03\%$.

{\subsection{OGLE-2017-BLG-1275} % kb170314 
\label{sec:anal-ob171275}}

\begin{figure}
\plotone{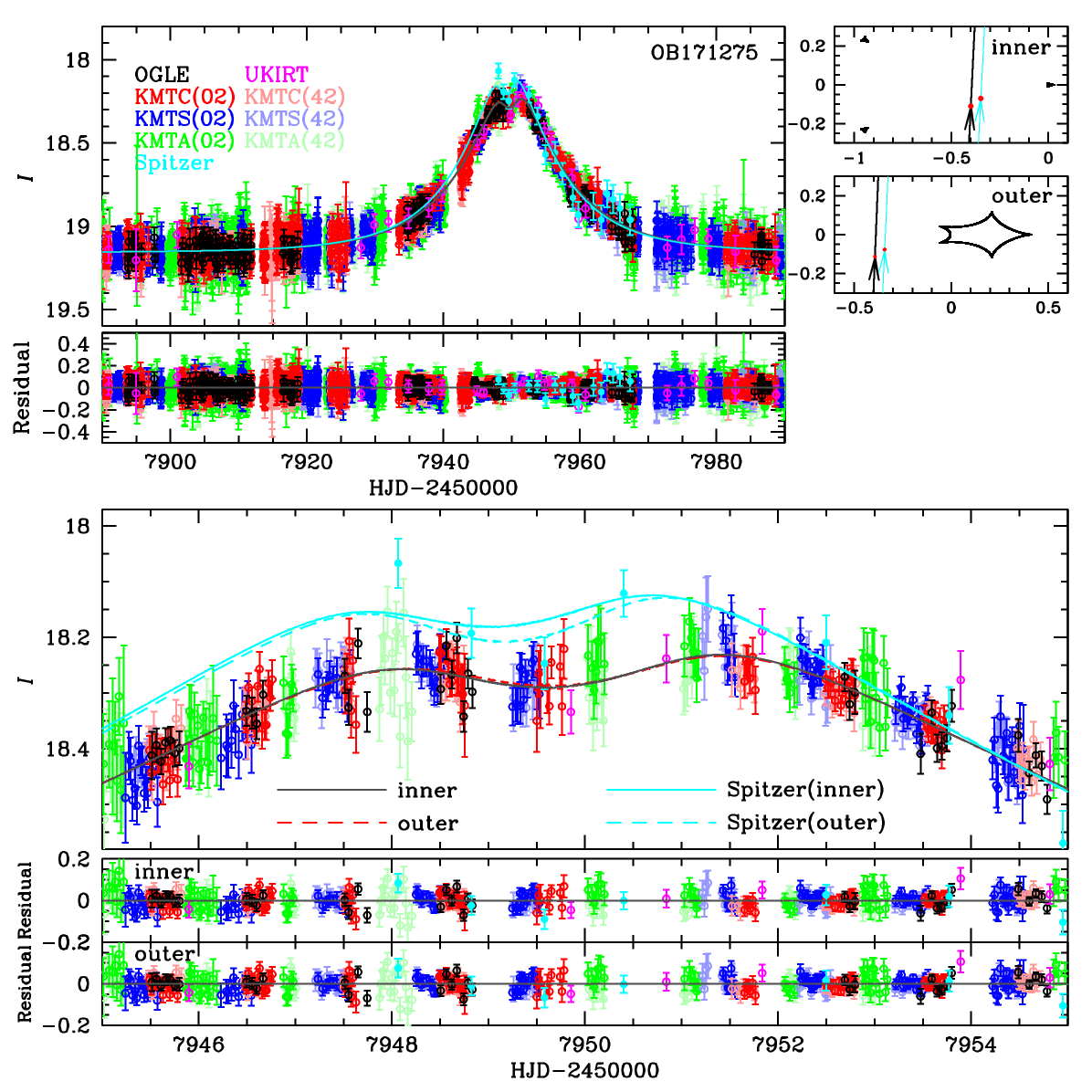}
\caption{Data (color-coded by observatory) together with the predictions
and residuals for 2 of the 3 models of OGLE-2017-BLG-1275 specified in 
Table~\ref{tab:ob171275parms}.
i.e., an ``inner''/''outer'' pair that explain the ``dip'' near the peak by
a minor-image pertubation.   We do not show the third  (``wide'') solution,
which explains this dip by a major-image pertubation, to avoid clutter and
because it is formally excluded by $\Delta\chi^2=35$.  The {\it Spitzer} data
(cyan), which were taken from solar orbit, show a similar dip near the peak
followed by a similar decline, which together imply that the source trajectory 
as seen by {\it Spitzer} was similar to one from the ground (right insets), and
hence that the microlens parallax $\pi_\e$ is small.}
\label{fig:1275lc}
\end{figure}

Figure~\ref{fig:1275lc} shows a relatively low-amplitude 1L1S microlensing
event punctuated by a dip just $\Delta t_{\rm anom}\simeq -0.3\,$day
before peak, with a full width, $\Delta t_{\rm dip}\simeq 3.0\,$day.
A \citet{pac86} fit (with the dip excised) yields $t_\e=13.3\,$day and
$u_0=0.40$.  Using the formulae from Section~\ref{sec:anal-preamble},
we obtain $\tau_{\rm anom}=-0.02$, $u_{\rm anom}=0.40$, $\alpha=273^\circ$,
$s^\dagger_-=0.82$, and $q=6.5\times 10^{-3}$.

The grid search yields three minima, two of which (when refined)
form a classic ``inner/outer'' degenerate pair, whose parameters
are in excellent agreement with the heuristic predictions for $\alpha$ and
$s^\dagger=\sqrt{s_{\rm inner}s_{\rm outer}}$, and in qualitative agreement 
for $q$.  See Table~\ref{tab:ob171275parms}.
Note that, as is often the case \citep{ob190960}, the source passes
outside the planetary wing of a resonant caustic in the ``outer'' model
and inside the planetary caustics in the ``inner'' model.

Somewhat surprisingly, it also yields a third solution characterized
by a caustic-crossing (``wide'') major-image perturbation.  In this solution,
the ``dip'' is produced by the declining magnification just outside
of the caustic walls, while the violent caustic crossing itself is
finessed by a $0.4\,$ day gap in the data.  In addition to being
somewhat implausible, this fit is substantially worse,
with $\Delta\chi^2=35$, so we do not further consider it.

While dip-type anomalies are rarely well fitted by 1L2S models, this
can happen.  Therefore, as a matter of due diligence, we check such models
but find that they are excluded by $\Delta\chi^2=95$.

{\subsubsection{OGLE-2017-BLG-1275: A {\it Spitzer} Planet} % kb170314 
\label{sec:spitzer-ob171275}}

OGLE-2017-BLG-1275 was one of about 1000 microlensing events that were
observed by {\it Spitzer} during a six-year campaign to obtain
microlens parallax measurements for planetary events, using the
method proposed by \citet{refsdal66} of observing simultaneously
from Earth and a satellite in solar orbit.  In the
AnomalyFinder series of papers, we usually leave the analysis
of the {\it Spitzer} data to future papers, in large part because
the {\it Spitzer} analysis is usually ongoing independently.
However, OGLE-2017-BLG-1275 is the first planet with viable
{\it Spitzer} data that was newly discovered by AnomalyFinder
rather than being recognized while the event was ongoing or during
a post-season by-eye review\footnote{The first AnomalyFinder planet,
OGLE-2019-BLG-1053, showed a {\it Spitzer} flux change of 1.8 units.
Because this is only a few times larger than the systematic errors,
\citet{ob191053} left the investigation of whether it had a measurable
parallax to future work.}. We therefore
break with this tradition and analyze these data here.  

By special arrangement, OGLE made a practice of issuing
its Monday alerts (i.e., the day for {\it Spitzer} uploads) immediately
after the end of observations in Chile (UT 10:30), which enabled the
{\it Spitzer} team to evaluate new candidates by the {\it Spitzer} operations
deadline, i.e., UT 15:00.  Despite the fact that OGLE-2017-BLG-1275
was a relatively short and (optically) faint event, OGLE was able to
issue its alert 5 days before peak at HJD$^\prime=7944.94$, just a few
hours before the deadline for submitting targets for {\it Spitzer}
observations.  At the time, it was
not yet bright enough to be observable by Spitzer, but it was
recognized that it had the potential to be a high-magnification
event. As such, it was selected as a ``secret'' {\it Spitzer} target. Later,
the Spitzer team recognized that it had brightened enough to meet the
criterion for Spitzer detection, and so announced the event as a
``subjective, immediate'' target at UT 20:03 on 11 July, corresponding
to HJD$^\prime = 7946.34$. Hence, by an unusual combination of good luck and
aggressive monitoring, Spitzer observations began just 3.25 days after
the data that enabled the alert were taken.

According to the protocols of \citet{yee15}, planets can only be included
in the {\it Spitzer} statistical sample if the decision to observe the event
was not influenced by the presence of the planet.  In one sense, this
condition is obviously satisfied, as the planet remained unrecognized for
5 years.  Nevertheless, to be more precise, we see that the {\it Spitzer}
team announced  this event more than a day before the
onset of the ``dip''.  Hence, this criterion is fully satisfied.

Often the {\it Spitzer} team organizes follow-up observations to enhance
the prospects for detecting planets, but they did not do so in this case.
It also usually observed the target in the $H$ band using the SMARTS ANDICAM
camera in order to measure
the source color.  However, in this case, it noted that this field
was being observed 1--2 times per day by the UKIRT Microlensing Survey
and so saw no need to duplicate these observations.
We include the $H$-band observations in the modeling and show them in
Figure~\ref{fig:1275lc}.  They contribute modestly
to excluding the ``wide'' model, but their main contribution is to the
measurement of the source color.  See Section~\ref{sec:cmd-ob171275}.

The {\it Spitzer} data comprise 17 epochs spanning 21.3 days and beginning
on 7948.07, i.e., 1.9 days {\it before} $t_{0,\oplus}$.  They 
overall fall by about 5 flux units during the first half of this period and
are roughly flat in the second half.  Within the first period, they
are roughly flat (possibly with some structure) during the first 2 days.
Thus, prior to any detailed analysis, they appear to be broadly consistent
with the ground-based light curve, i.e., peaking at about the same time, and
affected by the dip-type anomaly that is seen from the ground.
Hence, they appear to imply a very small $\bpi_\e$.

We proceed cautiously in several steps because {\it Spitzer} microlensing
data are known to exhibit systematics at roughly the 0.5 flux-unit level.
While well below the overall flux variation, these systematics could
nevertheless affect the analysis at some level.

{\subsubsection{Test of Color Constraints} % kb170314 
\label{sec:color-constraint}}

A key point is that, in contrast to most {\it Spitzer} microlensing
planets, the case of OGLE-2017-BLG-1275 permits an independent check
from the so-called color-color flux constraint.  To briefly review,
when \citet{refsdal66} first advocated microlens parallax observations
from solar orbit, he implicitly assumed that the peak and (at least one)
wing of the event would be observed from space.  Then, the times of peak
($t_{0,\oplus},t_{0,\rm sat}$) and normalized impact parameters
($u_{0,\oplus},u_{0,\rm sat}$) could be directly extracted from both light curves.
The microlens parallax (in modern notation) could then be ``read off''
\begin{equation}
  \bpi_\e = {\au\over D_\perp}(\Delta\tau,\Delta\beta);
  \qquad \Delta\tau \equiv {t_{0,\rm sat} - t_{0,\oplus}\over t_\e},
  \qquad\Delta\beta \equiv u_{0,\rm sat} - u_{0,\oplus},
  \label{eqn:dtaudbeta}
\end{equation}
where ${\bf D}_\perp$ is the Earth-Satellite vector offset projected on the
sky and the two components are, respectively, parallel and perpendicular to
this vector.
As \citet{refsdal66} already recognized, but is better illustrated
in Figure~1 from \citet{gould94}, there is a four-fold degeneracy because
$u_0$ is a signed quantity, but only its amplitude can normally be extracted
from a single-observatory light curve.  However, while some {\it Spitzer}
microlensing light curves do essentially meet the conditions implicitly
assumed by \citet{refsdal66} (see \citealt{ob140939} for a spectacular
example that looks eerily similar to Figure~1 from \citealt{gould94}),
the great majority do not.  The main reason for this is that operational
constraints imposed a 3--10 day delay from the time that the event was
recognized until {\it Spitzer} observations could begin.  A secondary
reason is that for disk (although not bulge) lenses, typically $\pi_{\e,E}>0$,
while {\it Spitzer} was in an Earth-trailing orbit, i.e., west of Earth,
and so these events typically peaked earlier as seen from {\it Spitzer}.

It is well known that for rising events, one cannot accurately predict
either $t_0$ or (especially) $u_0$ at times $t$ such that
$(t_0-t)\gg t_\eff\equiv u_0 t_\e$.  By the same token, it is equally impossible
to recover $t_{0,\rm sat}$ and $u_{0,\rm sat}$ for satellite data that begin
well after peak.  However, these difficulties can be greatly ameliorated
if the source flux of the space observatory ($3.6\,\mu$m, i.e., $L$ band,
in the present case) can be strongly constrained by $VIL$ or $IHL$ color-color
relations.  The $\bpi_\e$ contours then take the form of circular arcs
\citep{osculating} which are more or less extended according to the size of
$(t_{\rm start}-t_0)/t_\eff$, where $t_{\rm start}$ marks the commencement of the
space observations. See \citet{kojima1c} for an extreme example.

In the present case, one could take the orientation that color-color
relations are superfluous because the peak is covered.  But this also
means that color-color relations can provide an external check on systematics:
if the unconstrained fits yield {\it Spitzer} fluxes that are in strong conflict
with the constraint, this would be strong evidence that these fits are
dominated by systematics, whereas the contrary result would be evidence that
they are not.

{\subsubsection{{\it Spitzer} Analysis} % kb170314 
\label{sec:spitzer-anal-ob171275}}

We begin by undertaking a {\it Spitzer}-``only'' analysis, in which
the 7 standard 2L1S parameters are held fixed at the values given
in Table~\ref{tab:ob171275parms}, and the 17 {\it Spitzer} data points
are fit to four parameters, $(\pi_{\e,N},\pi_{\e,E},f_{S,\rm sat},f_{B,\rm sat})$.
We consider the ``inner'' and ``outer'' solutions separately.
We do not employ a flux constraint.  In this initial test, we consider
only the $u_{0,\oplus}>0$ solutions, in the expectation (confirmed below)
that the $u_{0,\oplus}<0$ solutions will be essentially symmetric in $\pi_{\e,N}$.
To ensure that we completely cover the relevant parameter space, we conduct this
modeling by a dense grid in $(\pi_{\e,N},\pi_{\e,E})$, fully covering $\pi_\e<1$.

The main result of this test is that for each of the
inner/outer pair of solutions,
and even without a flux constraint, there is a single, well-localized minimum.
These are at $(\pi_{\e,N},\pi_{\e,E})\simeq (-0.08,+0.02)$. and $(-0.06,+0.03)$,
for the outer and inner solutions, respectively, with the first favored
by $\Delta\chi^2\sim 7$.  See Table~\ref{tab:1275spitzer}\footnote{
Note that, as is usually the case in the analysis of {\it Spitzer} microlensing
planets, the {\it Spitzer} magnitudes are uncalibrated.  Specifically, we use
$L_{\it Spitzer}=18 - 2.5\log10(f_{\it Spitzer})$, where $f_{\it Spitzer}$ is
the instrumental flux.}.
That is, there is no $\pm u_{0,\rm sat}$ degeneracy, such
as was predicted by \citet{refsdal66} for 1L1S events.  The physical reason for
this is that the single minima have $u_{0,\rm sat} = +0.025$ and $+0.03$,
respectively, i.e., both passing on the same minor-image side of the caustic
as is seen from Earth.  In the alternate solutions (for 1L1S), these
would have $u_{0,\rm sat} = -0.025$ and $-0.03$, i.e., passing on the
major-image side of the caustic.  Once the planet is included this would
have a completely different structure,
with a bump instead of a dip.  Indeed, for the outer-solution case, this
impact parameter would have the source passing directly over the caustic.
For many {\it Spitzer} planets, this 1L1S degeneracy survives because the
source does not pass close to any caustics as seen from {\it Spitzer}.
However, as in the case of the first {\it Spitzer} planet, OGLE-2014-BLG-0124
\citep{ob140124}, this degeneracy is broken here.

We refine this test by seeding an MCMC at each of the minima derived from the
grid.  This approach allows us to evaluate the $(I-L)$ color and its
uncertainty (in the absence of a constraint),
$(I-L)_{\rm outer} = 2.05 \pm 0.16$ and $(I-L)_{\rm inner} = 2.14 \pm 0.20$.
These can be compared with the constraint, which is derived in
Section~\ref{sec:cmd-ob171275}, $(I-L)_{\rm constraint} = 2.43 \pm 0.08$.
These are therefore in $2.1\,\sigma$ and $1.3\,\sigma$ tension, respectively.
Of course, only one of these solutions can be correct, so one might
infer that the overall tension is at $1.3\,\sigma$, which would be hardly
notable.  However, we should keep in mind that it is the outer solution that is
overall preferred by $\Delta\chi^2\sim 7$ (for the {\it Spitzer}-``only'' data)
that is in greater tension.
We conclude that the constraint-free measurement and the constraint are
in qualitative agreement, but we await investigation of the full fits to
make a final assessment.
% L_outer = 17.295 +- 0.164
% L_inner = 17.168 +- 0.205
% I_s,ogle = 19.36, I_kmtc02 = 19.36 - 2.5*log(0.3006/0.2882) = 19.32,19.31
% (I-L) = 2.05 +- 0.16 outer
% (I-L) = 2.14 +- 0.20 inner

Next we conduct full fits by seeding an MCMC, fitting
both the ground and {\it Spitzer} data, and then allowing 11 parameters 
$(t_0,u_0, t_\e,\rho,s,q,\alpha,\pi_{\e,N},\pi_{\e,E},f_{S,\rm sat},f_{B,\rm sat})$
to vary,  %:\hfil\break\noindent,
while determining the ground-observatory flux parameters by a standard linear
fit.  We seed eight different models, i.e., four for each of the outer and
inner solutions.  In each case, the first seed is the model just described,
which we label $(++)$ because $u_{0,\oplus}>0$ and $u_{0,\rm sat}>0$.  We seed the
$(--)$ model by reversing the signs of $(u_0,\alpha,\pi_{\e,N})$.
Although, based on the $\bpi_\e$ grid search, we do not expect to find
$(+-)$ and $(-+)$ solutions, as a matter of due diligence we seed them
according to the prescription of \citet{refsdal66}.  However, we find
that, indeed these do not converge.

For the four solutions that do converge, i.e.,
(outer \& inner)$\times$($(++)$ \& $(--)$), we repeat the fit with and
without the flux constraint.  The results, shown in Tables~\ref{tab:ob171275pp}
and \ref{tab:ob171275mm},
confirm the preliminary assessment based on the {\it Spitzer}-``only''
analysis.  For the inner $(++)$ solution, imposing the constraint increases
$\chi^2$ by $\Delta\chi^2=4.3$ for 1 dof, while for the outer solution,
$\Delta\chi^2=7.2$.  The numbers are similar for the $(--)$ solution.
While these results are broadly consistent with
statistical fluctuations, they could reflect low level systematics.
However, even if so, these are no more severe than is  typical for ground-based
data in microlensing events.  Hence, we accept the constrained fits from
Tables~\ref{tab:ob171275pp} and \ref{tab:ob171275mm},
at face value.  Note also that the 7 standard parameters
are hardly affected by the addition of {\it Spitzer} data relative
to what was derived in Table~\ref{tab:ob171275parms} based on the
ground-only fits.

\begin{figure}
\plotone{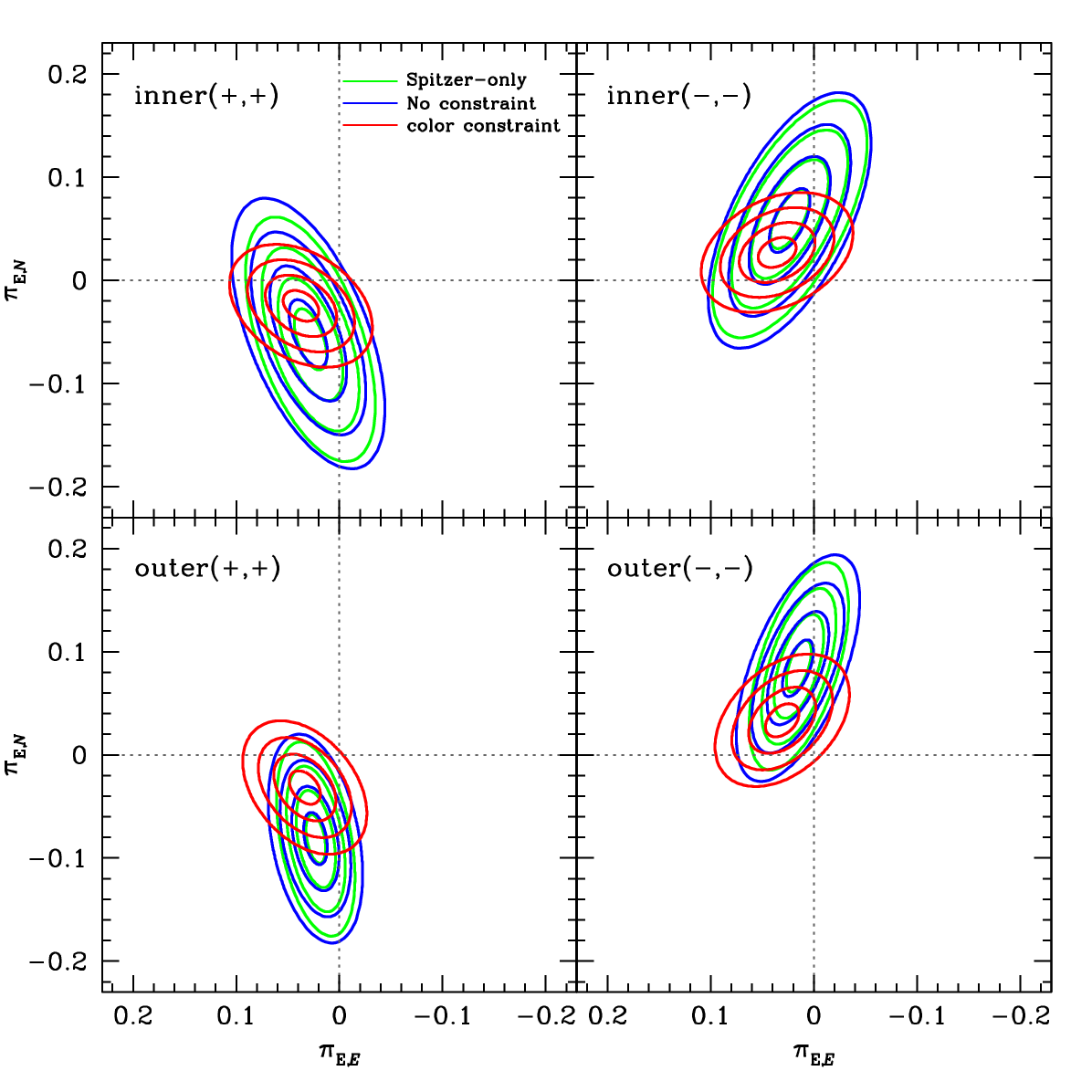}
\caption{Contours of $\Delta\chi^2=(1,2,3,4)$ for three different fits
  (as indicated by the colors in the legend), for each of four different
  geometries (labeled in the four panels).  The contours are derived from
  the means and covariances matrices of the MCMCs from the 12 different
  models.  In all four cases, the {\it Spitzer}-''only'' (green) and
  unconstrained ground+{\it Spitzer} (blue) fits are in close agreement.
  After imposing the color constraint (red), the minimum of the ``inner''
  solution is displaced by $\sim 1\,\sigma$, while the minimum of the ``outer''
  solution is displaced by $\sim 2\,\sigma$.  Note that the $(+,+)$ and
  $(-,-)$ solutions are approximately related by reflections in $\pi_{\e,N}$.}
\label{fig:ob1275-pie}
\end{figure}

In Figure~\ref{fig:ob1275-pie}, we show the $\bpi_\e$ contours for the
three different fits ({\it Spitzer}-``only'', ground+{\it Spitzer}
without color constraint, and ground+{\it Spitzer} with color
constraint)\footnote{We also tested for the annual parallax effect in
the ground-based data alone. However, due to the short duration of the
event, $\pi_{\e,N}$ was effectively unconstrained and the uncertainties in
$\pi_{\e,E}$ were large ($\pm 0.35$). Thus, the ground-only constraints are fully
consistent with the Spitzer parallax.}   for each of the four
geometries.  Figure~\ref{fig:ob1275-pie} shows that the color
constraint has three effects.  First, it reduces the error bars,
primarily in the north direction.  It is expected from
Equation~(\ref{eqn:dtaudbeta}) that the main impact will be on the
north direction because the Earth-satellite separation is mainly
east-west, so that $\pi_{\e,E}$ is mainly constrained by the location
of $t_{0,\rm sat}$, which is directly fit from the light curve and
does not depend on the source flux.  Second, the best fit parallax
amplitude, $\pi_\e$, is driven to lower values, by factors of about
1.4 and 1.9 for the ``inner'' and ``outer'' solutions, respectively.
For the ``inner'' solutions, this change is within the error bar while
for the ``outer'' solution, it is in mild tension.  Third, the
parallax amplitude, $\pi_\e$, is brought into closer agreement among
the four solutions.  

The very small parallax amplitudes, $\pi_\e\sim 0.045\pm 0.015$ 
imply that the host lies in or very near the Galactic bulge.  That is, the
projected velocities are $\tilde v\equiv \au/\pi_\e t_\e\sim 2900\,\kms$,
which is typical of bulge lenses.  By comparison, for disk lenses moving
according to an idealized flat rotation curve with $v_\rot \simeq 220\,\kms$
(and for typical bulge sources),
$\pi_\rel =\au\,\mu_\rel/\tilde v\rightarrow\pi_S(v_\rot/\tilde v)\sim 0.01\,\mas$,
i.e., indicating a lens location where the bulge strongly dominates over the
disk.  We present a full Bayesian analysis in Section~\ref{sec:phys-ob171275}.

\begin{figure}
\plotone{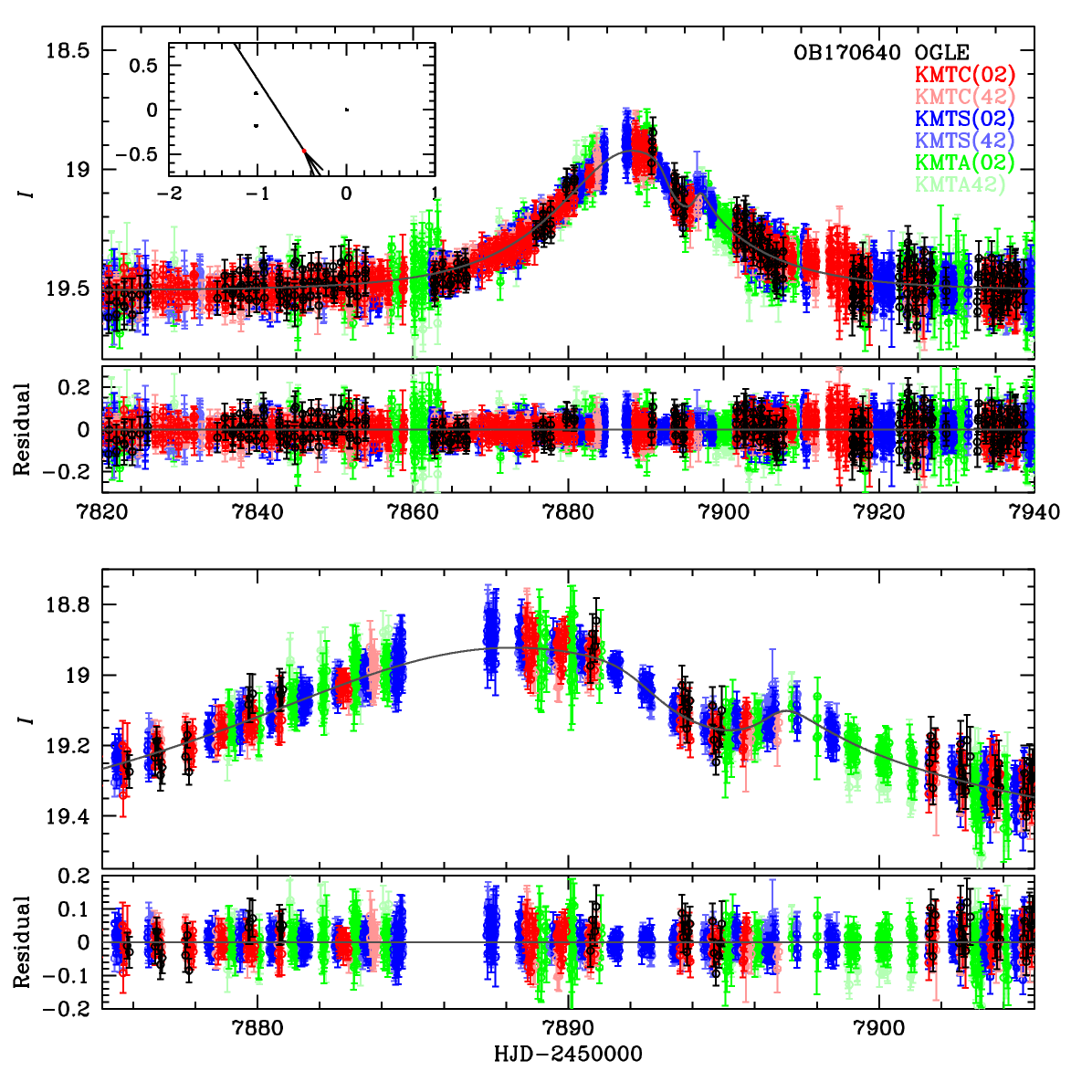}
\caption{Data (color-coded by observatory) together with the model prediction
  for OGLE-2017-BLG-0640.  The dip in the light curve at HJD$^\prime\sim 7895$
  is caused by the source passing over a channel of depressed magnification
  that thread the two planetary caustics on the opposite side of the host
  relative to the planet, as illustrated in the inset.
}
\label{fig:0640lc}
\end{figure}

{\subsection{OGLE-2017-BLG-0640}% kb171726
  \label{sec:anal-ob170640}}
Figure~\ref{fig:0640lc} shows a relatively short, relatively
low-amplitude 1L1S event, peaking at $t_0\simeq 7888.0$ and punctuated
by a dip at $t_{\rm anom}\simeq 7894.5$ with full width
$\Delta t_{\rm dip}\simeq 5.5\,$days.  Although the baseline
$I_{\rm base}\sim 19.5$ appears faint, the KMT tabulated extinction is unusually
high, $A_I=4.69$, implying that the source is likely to be a giant, in which
case the blending is likely to be small.  Under this assumption a 1L1S
fit to the event with the anomaly excised yields $u_0=0.64$ and $t_\e=15\,$days.
Then, $\tau_{\rm anom}= 0.43$ and $u_{\rm anom}= 0.77$, implying
(from Equations~(\ref{eqn:heuristic}) and (\ref{eqn:qeval})) that
$\alpha=236^\circ$, $s^\dagger_-= 0.69$, and $q=5.2\times 10^{-3}$.

The grid search returns only one competitive local minimum,
whose refined parameters
are given in Table~\ref{tab:ob170640parms} and which has an ``inner'' geometry,
as illustrated in Figure~\ref{fig:0640lc}.  The heuristic predictions for
$\alpha$ and $q$ are confirmed, while the fit value $s_{\rm inner}$ suggests
that there may be another solution at
$s_{\rm outer} =(s^\dagger_-)^2/s_{\rm inner} = 0.78$.  In fact, the grid search has a 
minimum with this topology, but it was too strongly disfavored to
warrant refinement.  As a matter of due diligence, we seed an MCMC with
this prediction for $s_{\rm outer}$ but find that it is excluded by
$\Delta\chi^2=227$.  The fundamental reason for this is that the inner/outer
degeneracy is much less severe when (as in the present case), $\alpha$
is far from $\pm 90^\circ$ \citep{zhang22}.  In particular, for
OGLE-2017-BLG-0640, the ``inner'' trajectory passes close to the second
caustic, which generates a weak bump at the end of the ``dip'',
whereas the ``outer'' trajectory would generate such a weak bump at the
beginning of the ``dip''.

{\subsection{OGLE-2017-BLG-1237}% kb170422 
\label{sec:anal-ob171237}}

\begin{figure}
\plotone{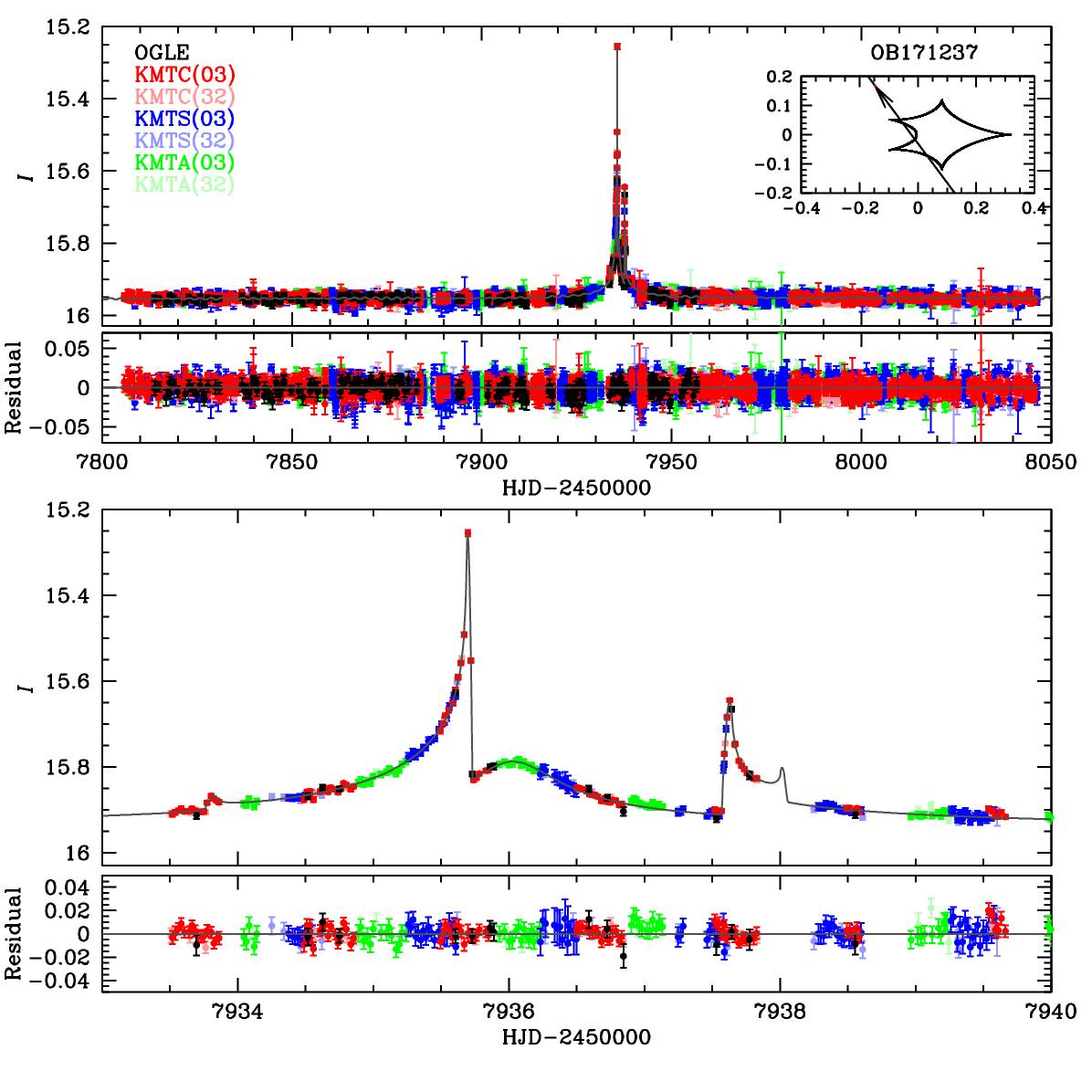}
\caption{Data (color-coded by observatory) together with the model prediction
  for OGLE-2017-BLG-1237.  There is a low-amplitude bump at HJD$^\prime\sim 7936$
  flanked by two U-shaped troughs.  Together, these imply that the source
  has passed the central cusp of a resonant caustic and has intersected
  sections of this caustic before and after the passage.  See inset.
}
\label{fig:1237lc}
\end{figure}

OGLE-2017-BLG-1237 shows a clear caustic exit peaking at HJD$^\prime =7935.70$,
followed by a caustic entrance peaking at HJD$^\prime =7937.63$, with a
low-amplitude bump between these at HJD$^\prime= 7936.0$.  Because of the
time ordering of these two clear caustic features, there must be an
additional entrance before the first (which was, in fact, observed
at HJD$^\prime =7933.8$) and an additional exit after the second (which
occurred during a gap in the data).  The bump is then almost certainly
due to an approach to the cusp associated with the more massive component
of the binary lens.  That is, the geometry of the caustic system can
be inferred by eye.  A systematic grid search confirms that there is only
one minimum, whose refinement is illustrated in Figure~\ref{fig:1237lc}
and whose parameters are given in Table~\ref{tab:171237parms}.

In general, it would be unusual for such an obvious and well-constrained
planetary event to escape publication for 6 years following its occurrence.
However, in this case, there is a nearby bright star, $I=15.88$, that lies
$1.4^{\prime\prime}$ from the lens, which is a double-mode pulsator.
The two closely spaced modes have periods $P=2.9\,$days, with a beat
period of 41 days and some additional long-term structure as well.
This star lies on the foreground main sequence
(see Section~\ref{sec:cmd-ob171237}), and it has a Gaia parallax
$\pi=0.60\pm 0.14\,\mas$, seemingly confirming that it is a nearby disk
star.  While the full amplitude of its variations is only
$\Delta I\sim0.06\,$mag, relative to its own baseline flux, this amplitude
is about 15 times larger than the unmagnified flux of the source star.
Thus, the variable created significant difficulties for the light-curve
analysis.  By chance, the caustic features of the light curve occur near the
minimum of the 41-day beat period, so contamination by the variable does
not have a major impact on their interpretation.  However, the effect,
particularly the fractional effect, on the wings of the light curve is much
larger, which directly impacts the measurement of $t_\e$, and thus
would indirectly impact $\rho$, $q$, and other parameters.

To remove this impact, we fit first carried out photometry of the
nearby variable over two years, restricted to KMTC and KMTS, which are
both high cadence and high quality.  We then fit the resulting light curve
(excluding the portion during the microlensing event)
to the sum of 9 sine waves, each characterized by three parameters
(amplitude, period, and phase).  When fitting the microlensing light curve,
we added a term consisting of a free parameter multiplied by the
mathematical representation of the variable light curve.  We expect
that the amount of contamination from the variable may be a function of
seeing, and therefore we tried to find more complicated models that would
include both seeing variations and the variable-star function.  However,
these efforts did not lead to significant improvement.

{\subsection{OGLE-2017-BLG-1777}% kb170282 
\label{sec:anal-ob171777}}

OGLE-2017-BLG-1777 was early recognized as a potentially 2L1S planetary event
by C.\ Han based on his combined analysis of OGLE and KMT data.
However, he judged that the light curve could in principle also have been
generated by a 1L2S binary source event.   Because of this complication
(and the difficulty of ultimately resolving it), detailed analysis was deferred.

Such a detailed analysis was initiated by us as part of the
AnomalyFinder series of papers.
The orientation of these papers is to thoroughly analyze {\it all}
events that are identified by the KMT AnomalyFinder \citep{ob191053}
and that have competitive planetary solutions, regardless of whether
the events are unambiguously planetary.  For example, for 2018,
\citet{2018prime} and \citet{2018subprime} analyzed (or cataloged from
earlier AnomalyFinder papers) a total 17 clear planets, but also
6 other events for which the interpretation of the anomaly was ambiguous.
% TW 6; 7; earlier 3+1; 3 gould ambigous 3 jung ambigous 

However, even in this context, OGLE-2017-BLG-1777 is unusually complex.
Its light curve is best explained
by the lens and source both having companions, with the former being
a Jovian mass-ratio planet and the latter being a very low mass (VLM) star or
brown dwarf (BD).  The event has a number of peculiar features,
such as an exceptionally small Einstein radius, $\theta_\e$, and an
exceptionally small lens-source relative proper motion, $\mu_\rel$.
As these two parameters derive from completely independent physical
characteristics, but could in principle be generated by an incorrect
estimate of the normalized source radius, $\rho$,
their confluence invites caution.
If the $\theta_\e$ measurement is correct, then the lens host is (like the
source companion) a VLM star or BD.  Much, but not all, of the evidence for
the planetary companion comes from a single discrepant data point, which was
the last one taken by OGLE during 2017.  In fact, there are additional
issues related to the long-term behavior of the OGLE data that required
careful investigation.  Hence, it is not a simple matter to properly address
all of these issues within the context of a paper that systematically
analyzes many planetary (and possibly planetary) events.

Nevertheless, because the goal of the AnomalyFinder papers is to present
comprehensive analyses of all planetary and possibly planetary events, we
must address all of these issues.  We do so by putting the main thread of the
light-curve analysis in this section, while deferring complex technical
and semi-technical points to appendices.

The first point is that the light curve is affected in two different ways
by a neighbor that lies $\sim 800\,\mas$ to the west of the source and is
brighter than it by $\sim 4.3\,$mag.  We analyze these effects and correct
the light curve for them, as we discuss in Appendix~\ref{sec:lc-corr}.

\begin{figure}
\plotone{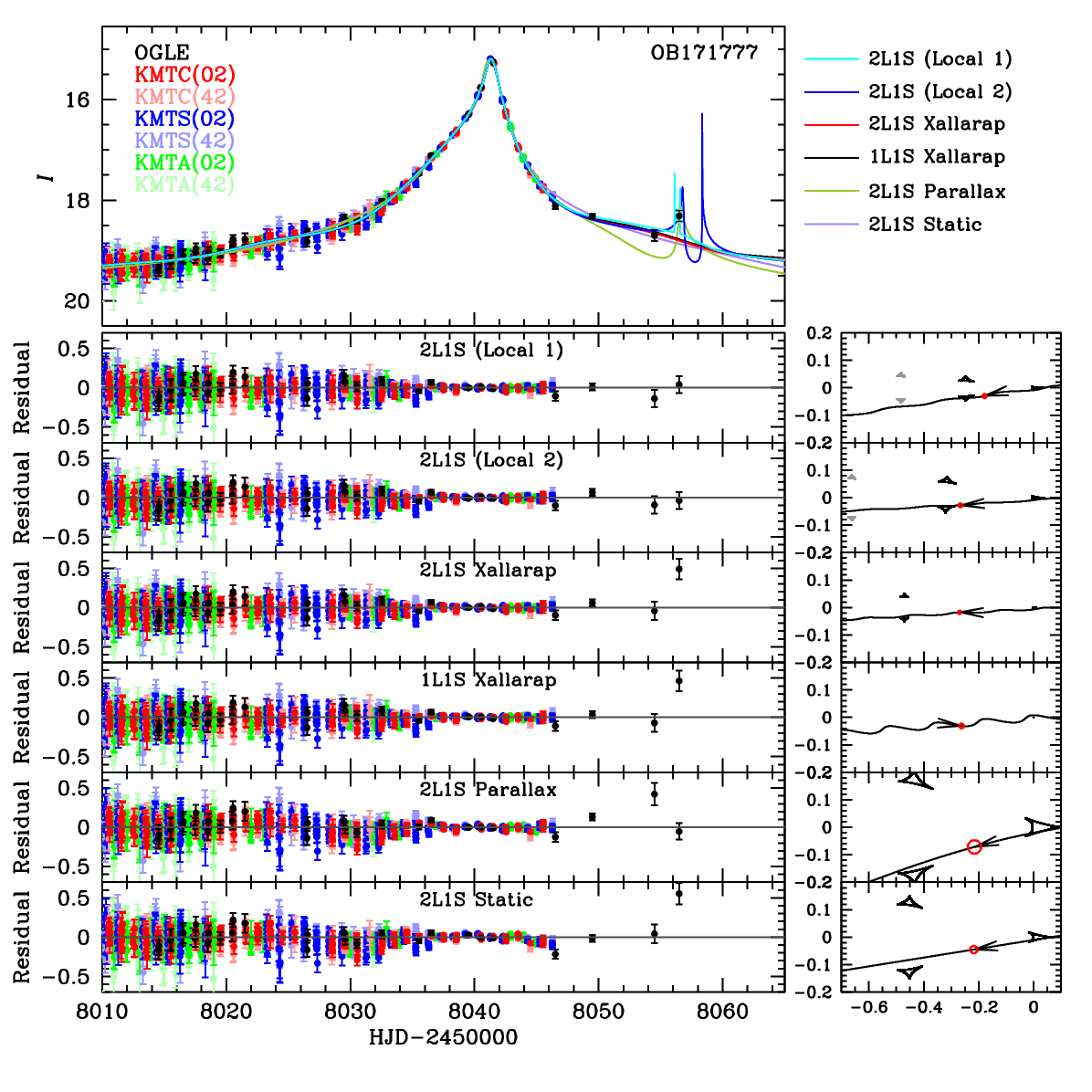}
\caption{Data (color-coded by observatory) together with the model predictions
  for OGLE-2017-BLG-1777. Simple 2L1S models (two bottom panels) leave a
  $P\sim 13\,$day wave of residuals.  Adding xallarap but without lens orbital
  motion (third panel) removes these but cannot explain the last data point.
  However, 2L1S models that include xallarap, parallax, and lens orbital
  motion (top two panels) give a significantly better fit.
}
\label{fig:1777lc}
\end{figure}

Next, contrary to the usual practice, we begin by introducing all the
parameters that are required to describe our final model, rather than
recapitulating the history of their gradual introduction as simpler models
failed, one-by-one.  This will allow us to comprehensively present the
relationship between the final model and these simpler models.
The final models as well as several intermediate models are illustrated
in Figure~\ref{fig:1777lc}, while their parameters are given in
Tables~\ref{tab:ob171777parms1} and \ref{tab:ob171777parms2}.

The final model is described by 18 parameters.  Seven of these are the
standard parameters that are always required to analyze 2L1S events,
including all of the other events in this paper,
$(t_0,u_0,t_\e,s,q,\alpha,\rho)$.  Four of the parameters are the microlens
parallax vector $\bpi_\e$ and the linearized transverse lens-orbital motion,
$\bgamma$, as described in Section~\ref{sec:anal-preamble}.  Finally, seven
parameters are required to describe the Kepler orbit of the source around
the center of mass of itself and its dark companion, i.e.,
``xallarap effect''.

From a microlensing standpoint, xallarap is the inverse
of parallax: light-curve distortions are induced by the orbital motion of
the source rather than the observer.  Hence, in principle, an annual
parallax signal can always
be imitated by xallarap, provided that the source orbital motion mimics that of
Earth.  Because of this, standard xallarap parametrization is set so that
the measured xallarap parameters will mimic the corresponding parallax
parameters for the case that the apparent xallarap is caused by parallax.
Thus, while the microlens xallarap parameters are closely related to standard
Kepler parameters, they are expressed somewhat differently.

The amplitude and orientation of the source motion (scaled by $\theta_\e$)
is $\bxi\equiv(\xi_N,\xi_E)$.  As with the parallax, $\bpi_\e$, the direction
is expressed with respect to the lens-source relative motion at $t_0$.
The phase and inclination of the orbit are expressed as $(\alpha_S,\delta_S)$,
so that in the case that the apparent xallarap is due to parallax, these
will be exactly equal to the ecliptic coordinates of the event.  Note in
particular that edge-on orbits have $\delta_S=0^\circ$, whereas in standard
Kepler parametrization, the inclination would be $i=90^\circ$.  The period $P$
and eccentricity $e$ are exactly the same as in the standard Kepler
parametrization.  Finally the ``phase'' of the periastron is measured
relative to $t_0$.  See Appendix~\ref{sec:bd-xallarap} for more details
about xallarap.

We briefly summarize how we were led to such a complex model.
We began with a standard 2L1S model, which provided a reasonable first
approximation to the data, but (1) failed to match the final OGLE point,
and (2) left an oscillatory residual with a period of order 13 days near
the peak.  See the magenta curve and the bottom panel of residuals in
Figure~\ref{fig:1777lc}.
As single-point discrepancies are quite common in microlensing,
and this data point was taken at high airmass and in twilight, we initially
suppressed the final OGLE point at HJD$^\prime=$8056.4965.
We then added xallarap to the 2L1S fit and found
substantial improvement (red curve and third panel of Figure~\ref{fig:1777lc}).
We then investigated whether a second lens was
really required to explain the light curve, or whether the entire light-curve
distortion could be explained just by 1L1S plus xallarap
(black curve and fourth panel).  We found
$\Delta\chi^2 = \chi^2({\rm 1L1S+xallarap})-\chi^2({\rm 2L1S+xallarap})=21$
for 3 dof, which would imply a marginal detection of a planet.  However,
we noticed that the 2L1S+xallarap model predicted a caustic crossing within
a few days of the final OGLE data point (which we had previously suppressed).
Hence, we carefully examined this point at the image level.  Although the
observing conditions were not optimal, there proved to be no indicators
that artifacts of non-astronomical origin were corrupting the measurement.
We then re-included this point and added lens orbital-motion and parallax
to the fit, thereby finding two different models, in each of which the
final OGLE data point was explained by the source passing through the
planetary caustic.

The two 18-parameter 2L1S models have an improvement over the 1L1S+xallarap
model of $\Delta\chi^2=36$ for 7 dof.  Formally (for Gaussian statistics),
this would imply a false-alarm probability of $p\sim 10^{-5}$.  However,
there are many reasons for caution.  First, microlensing data are known
to be non-Gaussian, although it is rare that $\Delta\chi^2=36$ differences
between competing models would be questioned.  Second, much of this
$\Delta\chi^2$ is due to a single data point that was taken under somewhat
difficult conditions.  While we have closely vetted this point at the image
level, the possibility that we have missed something about it cannot be
absolutely ignored.  Third, there are several features of the solution,
which we summarize in the next two paragraphs, that are very unusual.
While unusual things do happen, it is worrisome that they occur in an
event for which the planetary interpretation relies on somewhat uncertain
evidence.  While the AnomalyFinder papers do not aim to make final decisions
about including any particular event in subsequent statistical analyses,
they do attempt to provide comprehensive accounts of all the information
that is required to make such decisions.  We therefore alert the reader
to the nature of these issues in the next two paragraphs, while providing
systematic accounts in the appendices.

The first issue is that both $\theta_\e$ and $\mu_\rel$ are unusually small.
We will show in Section~\ref{sec:cmd-ob171777} that $\theta_*\simeq 0.5\,\muas$,
which implies $\theta_\e \simeq 70\,\muas$ and $\mu_\rel \simeq 0.4\,\masyr$.
These are both rare for planetary microlensing events.
The first depends only on the lens mass and the distances to the lens and
source, while the second depends only on the proper motions of the lens and
source.  Because these are physically independent, but both depend on the
same $\rho$ measurement, they could be a warning sign of a major error
in the model.  We address this issue in Appendix~\ref{sec:rho}.

The second issue is that, as we show in Appendix~\ref{sec:bd-xallarap},
BD and VLM-star companions to the source can only be detected via xallarap
if both $\theta_\e$ and $\mu_\rel$ are unusually small.  Therefore, given
that we report
the detection of such a xallarap companion, it is not a further surprise
that $\theta_\e$ and $\mu_\rel$ are small.  On the other hand, it is the
case that the presence of such objects in close-in orbits is rare.
In particular, the probability of transit for companions in $P=12.5\,$day
orbits about Sun-like stars is $\sim 4.5\%$.  Such transits would give rise to
strong signals in the great majority of stars observed by {\it Kepler} and would
appear as Jovian-planet sized companions, which would certainly be investigated.
Hence, their low observed frequency shows they are intrinsically rare.

On the other hand, the fact that these two major questions about the event
are both entangled with the prediction of the models that the source has
a BD or VLM-star companion (i.e., $M_{\rm comp}\sim 0.075\,M_\odot$,
see Appendix~\ref{sec:bd-xallarap}) implies that confirmation of this companion
would greatly increase confidence in the solution.  We note that
the predicted semi-amplitude of radial velocity (RV) variations of the
source star is $v\sin i\sim 6.5\,\kms$ (see Appendix~\ref{sec:bd-xallarap}),
which should be measurable on
large telescopes despite the faintness of the source, $I_S\sim 20.3$.

% 181 1153 -> 457*exp(-18) = 

\begin{figure}
\plotone{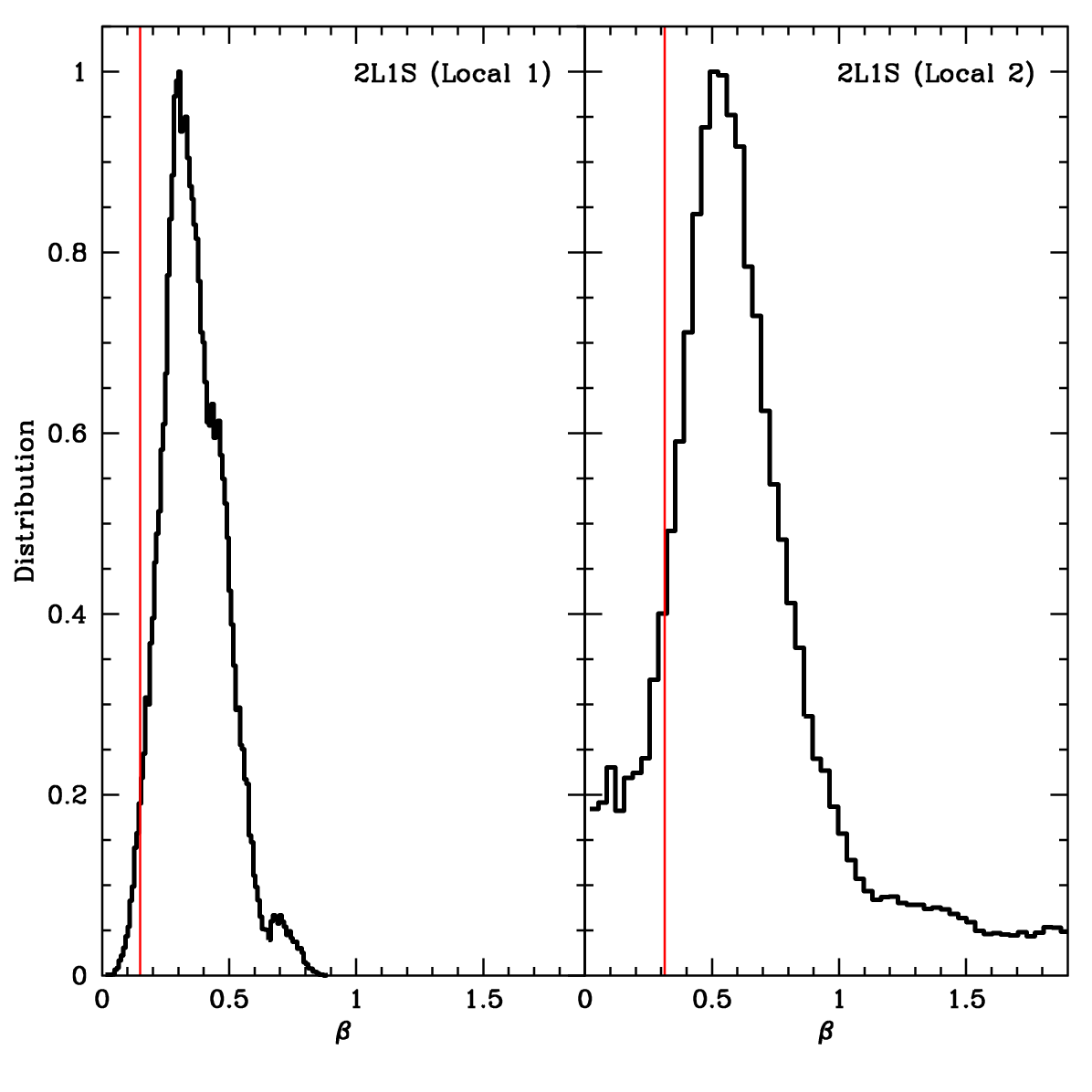}
\caption{Distributions of $\beta\equiv |{\rm KE/PE}|_\perp$, i.e., the
  ratio of transverse kinetic to potential energy
  (Equation~(\ref{eqn:betadef})) for the two solutions of OGLE-2017-BLG-1777.
  The orbital-motion determination is strongly influenced by the need to match
  a single, otherwise-discrepant, point.  See Figure~\ref{fig:1777lc}.  If
  this point were an artifact, one would expect the inferred $\beta$ to either
  be unphysically high ($\beta>1$) or improbably low ($\beta\la 0.1$).
  Instead, the distributions are peaked near values that are physically
  expected, thus increasing the plausibility of these solutions.
}
\label{fig:1777beta}
\end{figure}

Finally, there is one further test that we can make regarding the
reality and/or plausibility of this unusually complex lens-source system.
One can see from the top two caustic diagrams of Figure~\ref{fig:1777lc}
(i.e., orbital-motion models) that the planetary caustics moved a substantial
distance (0.2 or 0.4 Einstein radii) between $t_0$ (gray triangles) and
the time of the final OGLE point (black contours), a motion that is determined
directly by the data (in particular, the time of the final ``discrepant''
OGLE data point), but which is interpreted as being caused by physical
motion of the planet within the context of these models.  If the point
were spurious, which is the main concern about the reality of these models,
then most likely the inferred planetary motion would be either too large
to be consistent with Kepler's Laws or would be so small that it would
require an improbable projection on the plane of the sky to be consistent
with Kepler's Laws.  This test can be made in terms of the ratio
of the transverse kinetic-to-potential energy parameters, $\beta$, which is
defined in Equation~(\ref{eqn:betadef}).  For face-on circular orbits,
$\beta=0.5$, and for a typical range of random projections,
$0.1\la\beta\la 0.5$.  Figure~\ref{fig:1777beta} shows the distributions of
$\beta$ based on the MCMCs for the two solutions.  In both cases, we see that
the peaks of these distributions are both relatively compact and
overlap the expected range.  Logically, this test would only require
that this be true of at least one of these solutions, but in any case,
it is true of both.  The fact that the inferred orbital motion is
consistent with expectations from Kepler's Laws adds to the credibility
of these planetary solutions, but does not prove that they are correct.

{\subsection{OGLE-2017-BLG-0543}% kb170140 
  \label{sec:anal-ob170543}}

\begin{figure}
\plotone{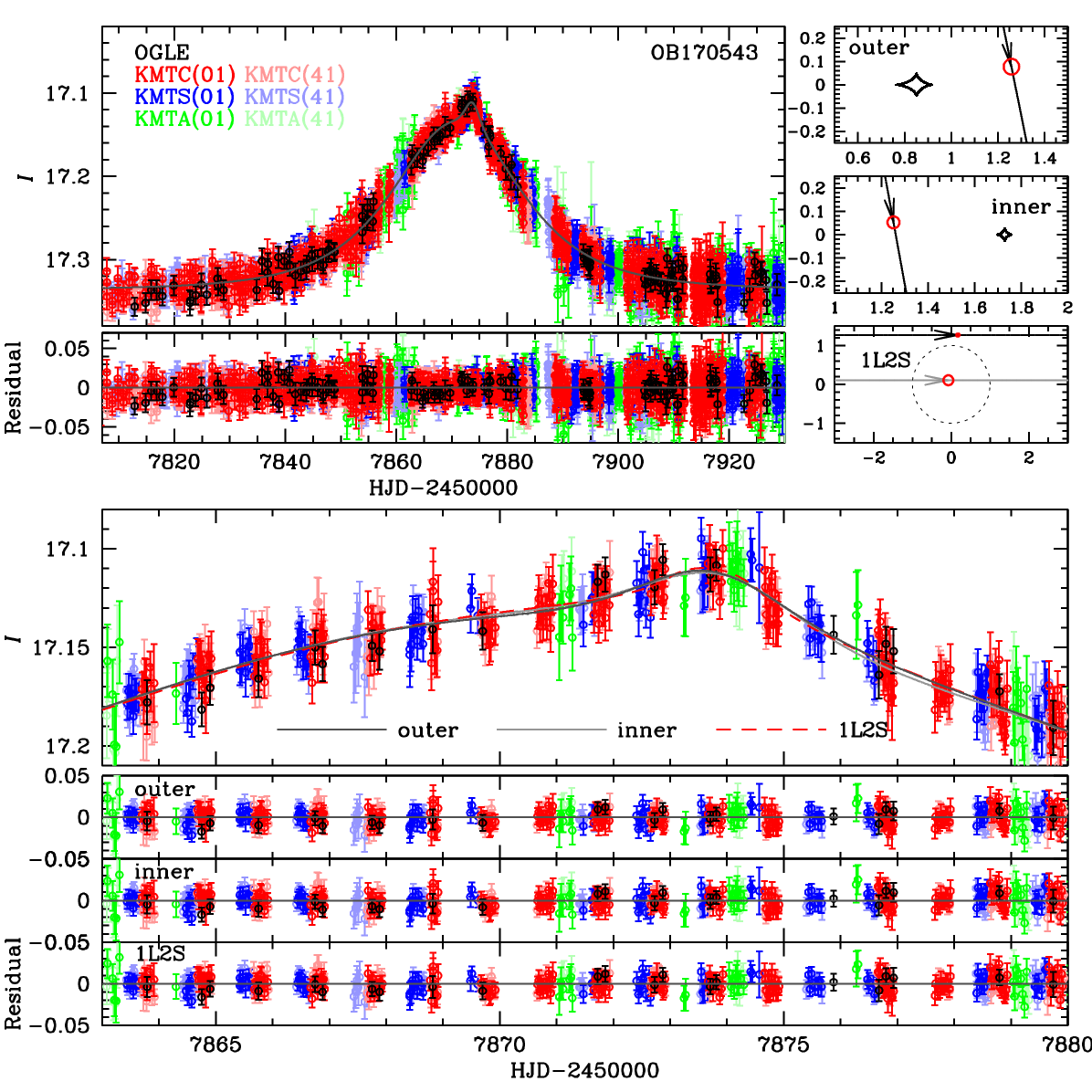}
\caption{Data (color-coded by observatory) together with the model predictions
  for OGLE-2017-BLG-0543. The bump at HJD$^\prime\sim 7873.7$ can be fit
  either by a weak cusp approach due to a planet (upper two inset panels)
  or a companion of the source passing the primary lens.  Because these,
  respectively, 2L1S and 1L2S models cannot be distinguished from the data,
  the nature of the lens system remains undetermined.
}
\label{fig:0543lc}
\end{figure}

Figure~\ref{fig:0543lc} shows a relatively short, low-amplitude 1L1S
event that is punctuated by a short bump at $t_{\rm anom}=7873.5$, i.e.,
$\Delta t_{\rm anom} =+2.5\,$days after the peak at $t_0 = 7871$.  Taking
account of the KMT tabulated extinction, $A_I=2.24$, the baseline object
has $I_{0,\rm base}\sim 15.1$, corresponding to a giant.  Assuming, as is
likely, it is unblended or only weakly blended, $u_0=1.25$ and $t_\e=12\,$days.
Then $\tau_{\rm anom}=0.21$, $u_{\rm anom}=1.27$, $\alpha= 80.5^\circ$ and
$s_+^\dagger=1.82$.

The grid search returns two local minima.  We find (as is common
for such low-amplitude events) that the division of the baseline flux
into source and blend fluxes is poorly constrained, and so we set $f_B=0$,
which (as mentioned above) is plausible.  The resulting parameters are
shown in Table~\ref{tab:ob170543parms}.  The heuristic predictions for
$\alpha$ are correct to within $\sim 2^\circ$.  The prediction
$s^\dagger_+=1.82$ is in excellent agreement with
$s^\dagger\equiv \sqrt{s_{\rm outer} s_{\rm inner}}=1.82$.

As for any smooth-bump anomaly, we must check for alternative 1L2S solutions.
After applying the same $f_B=0$ constraint, we indeed find such a solution,
which is illustrated in Figure~~\ref{fig:0543lc} and summarized in
Table~\ref{tab:ob170543parms}.  
This solution is disfavored by only $\Delta\chi^2=2.9$, which is well below
the level needed to securely claim the detection of a planet.  In some
cases there can be additional arguments that could be made against the 1L2S
solution.  For example, it might predict an implausibly low proper motion,
$\mu_\rel=\theta_*/t_*$.  In Section~\ref{sec:cmd-ob170543}, we will show
that $\theta_*\simeq 5\,\muas$.  From Equation~(\ref{eqn:probmu2}), we would
have to be able to constrain $\mu_\rel \la 1\,\masyr$ to contribute significantly
to such an argument, which would in turn require a restriction $t_* > 44\,$hr.
However, this value is nearly within the $1\,\sigma$ range for both $t_*$ and
$t_{*,2}$.  Hence, no such argument can be made.

Another possible argument could be made if the 1L2S model predicted a
substantially different color than the one measured from the light curve.
However, first, the second source is predicted to lie $-2.5\log q_F = 6.0$
mag below the primary.  Because the primary is a giant in or just below
the clump, such a secondary would have a similar color to the primary, so
even a precise color measurement could not distinguish between the 2L1S
and 1L2S predictions.  Second, the $V$-band signal from the secondary
(assuming that the 1L2S model is correct) is too weak to measure its color.

Therefore, while it is possible that the anomaly in OGLE-2017-BLG-0543 is
due to a planet, the 1L2S model also provides a plausible solution, and
it certainly cannot be confidently excluded.  Hence, this event should
not be cataloged as planetary.

{\subsection{OGLE-2017-BLG-1694}% kb172126 
\label{sec:anal-ob171694}}

\begin{figure}
\plotone{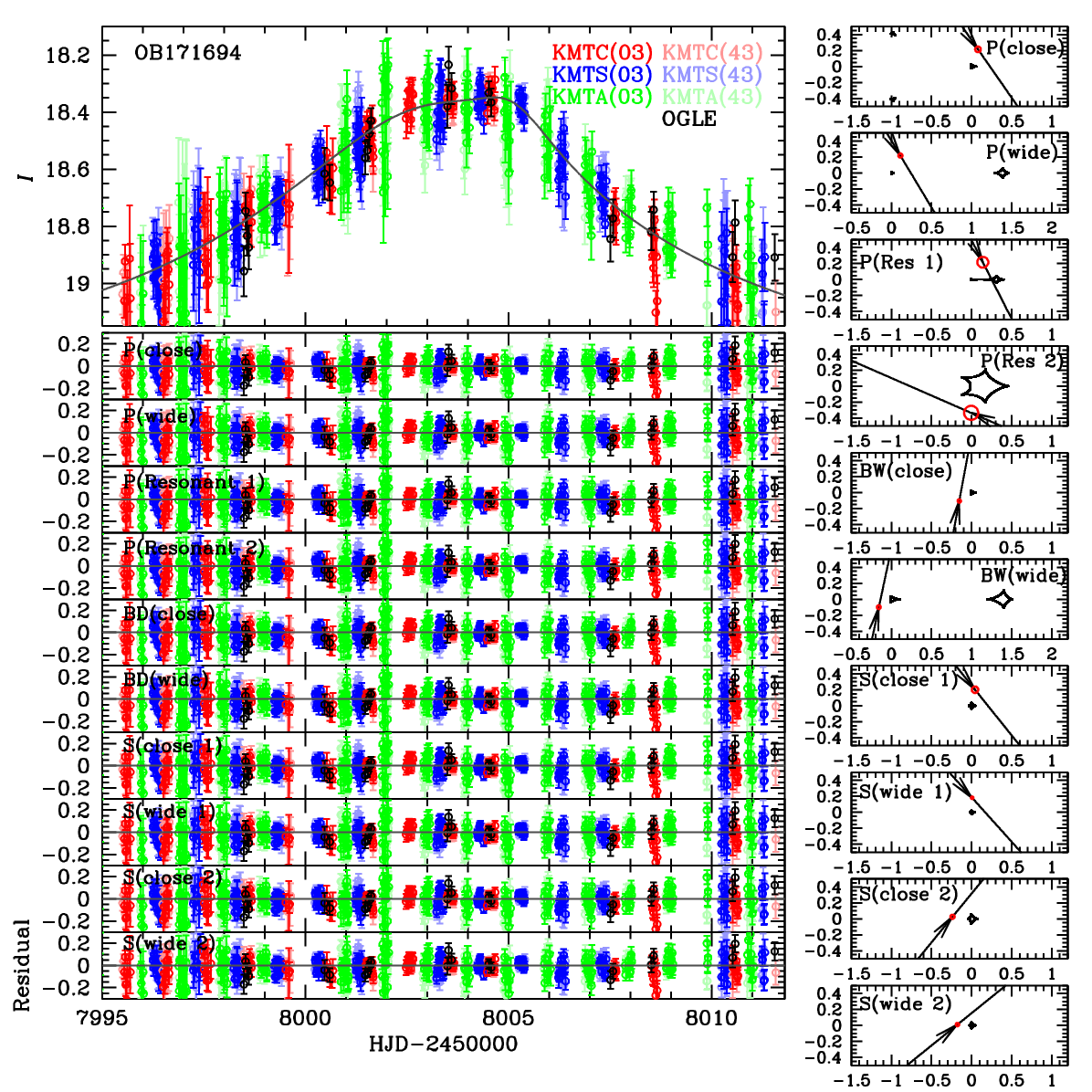}
\caption{Data (color-coded by observatory) together with model predictions
  for OGLE-2017-BLG-1694. Although the light-curve distortion is unambiguously
  detected, it can be fit by numerous, widely differing models.  Hence, no
  definite conclusions can be drawn about the lens system.
}
\label{fig:1694lc}
\end{figure}

Figure~\ref{fig:1694lc} shows an approximately 1L1S event with a
weak anomaly near
peak, plausibly a post-peak bump.  The underlying 1L1S curve is itself
somewhat ambiguous because it is consistent with a range of timescales,
$t_\e$, or (because $t_\eff = u_0 t_\e$ and $f_S t_\e$ are approximate
invariants), equivalently, a range of $f_S$ or $u_0$.  Because both
the 1L1S event and the anomaly are ambiguous, we dispense with a heuristic
analysis and proceed directly to the grid search.  This yields nine different
solutions that are within $\Delta\chi^2<10$ of the minimum, plus some
additional ones that are somewhat worse.  Two of these, with
$(s,q)=(0.637,0.023)$ and $(1.926,0.019)$, form a close-wide pair of a
major-image perturbation and are planetary in nature. A third is also
planetary, but with a resonant caustic geometry and $(s,q)=(1.16,0.0016)$.
See Table~\ref{tab:ob171694parmsplan}.  
For this reason, the
event is included in the present paper as a ``possible planet''.
However, another two, with $(s,q)=(0.492,0.051)$ and $(1.892,0.062)$, form a
close-wide pair of a minor-image perturbation and are in the brown-dwarf
regime.  See Table \ref{tab:ob171694parmsbd}.
And the remaining four are clearly in the stellar regime.
See Table~\ref{tab:ob171694parmsstar}.  Finally, in
Table~\ref{tab:ob171694parmsplan}, we show one other planetary model that
is just beyond our $\Delta\chi^2<10$ threshold.
The residuals of all these models are
shown in Figure~\ref{fig:1694lc}.  As there is no way
to distinguish among them, the event cannot be cataloged as planetary.

\section{{Source Properties}
\label{sec:cmd}}

As in Section~\ref{sec:anal-preamble}, above, we begin by reproducing
(with minor modifications) the
preamble to Section~4 of \citet{2018subprime}.  Again, this is done
for the convenience of the reader.  Readers who are familiar with
\citet{2018subprime} may skip this preamble.

If $\rho$ can be measured from the light curve, then 
one can use standard techniques \citep{ob03262} to determine the
angular source radius, $\theta_*$ and so infer
$\theta_\e$ and $\mu_\rel$:
\begin{equation}
\theta_\e = {\theta_*\over \rho}; \qquad
\mu_\rel = {\theta_\e\over t_\e}.
\label{eqn:mu-thetae}
\end{equation}
However, in contrast to the majority of published by-eye discoveries
(but similarly to most of new AnomalyFinder discoveries reported
in \citealt{ob191053,af2,kb190253,2018prime}), most of the planetary
events reported in this paper have only upper limits on $\rho$,
and these limits are mostly not very constraining.  
As discussed by \citet{2018prime}, in these cases,
$\theta_*$ determinations are not likely to be of much use, either now
or in the future.  Nevertheless, the source color and magnitude measurement
that are required inputs for these determinations may be of use in the
interpretation of future high-resolution observations, either by space
telescopes or adaptive optics (AO) on large ground-based telescopes
\citep{masada}.
Hence, like \citet{2018prime}, we calculate $\theta_*$ in all cases.

Our general approach is
to obtain pyDIA \citep{pydia} reductions of KMT data at 
one (or possibly several) observatory/field combinations.  These
yield the microlensing light curve and field-star photometry on the
same system.  We then determine the source color by regression of the
$V$-band light curve on the $I$-band light curve, and the source magnitudes
in $I$ by regression on the best-fit model.  Similarly to
\citet{2018prime}, we calibrate these color-magnitude diagrams (CMDs)
using published
field star photometry from OGLE-III \citep{oiiicat} or OGLE-II 
\citep{oiicat1,oiicat2,oiicat3} photometry whenever these are available.
However, while 5 of the 6 events analyzed in this paper have OGLE-III photometry,
for two of these (OGLE-2017-BLG-1275 and OGLE-2017-BLG-0640),
the field is heavily extincted.  For the first, this leads us to rely on a
combination of $VIH$ data, as described in Section~\ref{sec:cmd-ob171275}.  
For the second, we carry out the analysis using an $I/H$ CMD, as
described in Section~\ref{sec:cmd-ob171237}.  
For the sixth event, OGLE-2017-BLG-0543,
we work directly in the KMTC pyDIA magnitude system.  Because the
$\theta_*$ measurements depend only on photometry relative to the clump,
they are unaffected by calibration.  In the current context, calibration
is only needed to interpret limits on lens light, which is not an
issue for this event because it is not reliably detected as a planet.

We then follow the standard method of \citet{ob03262}.  We adopt the
intrinsic color of the clump $(V-I)_{0,\rm cl}= 1.06$ from \citet{bensby13}
and its intrinsic magnitude from Table~1 of \citet{nataf13}.
We obtain 
$[(V-I),I]_{0,\rm S} = [(V-I),I]_{\rm S} + [(V-I),I]_{0,\rm cl} - [(V-I),I]_{\rm cl}$.
We convert from $V/I$ to $V/K$ using the $VIK$ color-color relations of
\citet{bb88} and then derive $\theta_*$ using the
relations of \citet{kervella04a,kervella04b} for giant and dwarf sources,
respectively.  After propagating errors, we
add 5\% in quadrature to account for errors induced by the overall method.
These calculations are shown in Table~\ref{tab:cmd}.
Where there are multiple
solutions, only the one with the lowest $\chi^2$ is shown.  However,
the values of $\theta_*$ can be inferred for the other solutions by noting
the corresponding values of $I_S$ in the event-parameter tables and using
$\theta_*\propto 10^{-I_S/5}$.  In any case, these are usually the same within
the quoted error bars.

Where relevant, we report the astrometric 
offset of the source from the baseline object.

Comments on individual events follow.

{\subsection{OGLE-2017-BLG-1275}% kb170314 
  \label{sec:cmd-ob171275}}

For OGLE-2017-BLG-1275, we start by analyzing the source color primarily
using $I/H$ data (as opposed to $V/I$ data).  We do so for two reasons.
First, the source is very reddened, which leads to substantially smaller error
bars on individual $H$-band observations compared to those in the $V$ band.
Second, contrary to most the most AnomalyFinder planets, this event has
{\it Spitzer} data, for which we require a color-color relation.  These are
overall both easier to determine and more reliable in $IHL$ than $VIL$
because the former is based on a shorter extrapolation.

\begin{figure}
\plotone{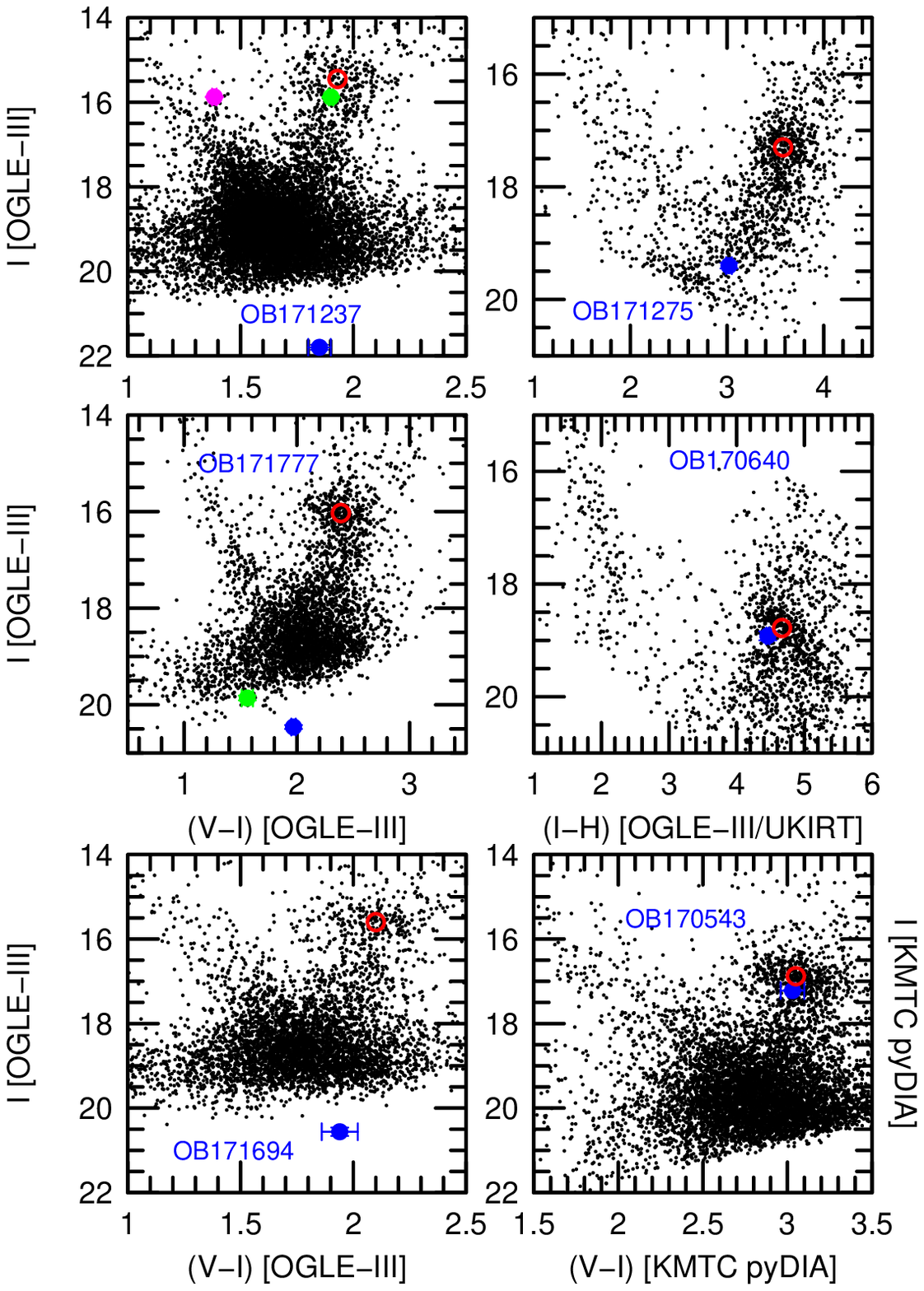}
\caption{CMDs for all 6 events, indicated by, e.g.,
  OB171275 for OGLE-2017-BLG-1275.  The red clump and
  the lens position are shown in red and blue, respectively.  Where
  relevant, the blended light is shown in green.  For
  OB171237, the bright variable at $1.4^{\prime\prime}$ is shown in
  magenta.  When there are multiple solutions, we only show the source
  and blend for the lowest-$\chi^2$ solution.  }
\label{fig:allcmd}
\end{figure}

Toward this end, we construct an $I$ vs.\ $I-H$ color-magnitude diagram (CMD)
by matching field star photometry from UKIRT and OGLE-III, which are both
calibrated.  See Figure~\ref{fig:allcmd}.

The $H$-band light curve is already on the same scale as the field stars.
We align the KMTC02 pySIS photometry (used in the fits) to KMTC02 pyDIA
by regression, and then to OGLE-III based on field stars.  The comparison
yields an offset $\Delta(I-H) = -0.60\pm 0.05$ of the source relative to
the clump.  Adopting $(I-H)_{0,\rm cl} = 1.29$ from \citet{bensby13} and
\citet{bb88}, this implies $(I-H)_{0,S}=0.69\pm 0.05$.  From its $I$-band
magnitude, $\Delta I=+2.10\,$mag below the clump, the source is a turnoff
star or subgiant.  In terms of surface gravity, these are much closer
to dwarfs than giants.  Therefore, we estimate the corresponding
$(V-I)$ color using the dwarf-star color-color relations of \citet{bb88},
finding $(V-I)_{0,S} = 0.65\pm 0.05$.  This is certainly possible, particularly
within the errors, but it is relatively blue for a star of this magnitude.

Therefore, we conduct an independent assessment using $V/I$ photometry.
Because of severe extinction, our normal procedure of evaluating $(V-I)_{0,S}$
from a single observatory-field combination yields statistical errors of
$\sigma(V-I)_s\sim 0.14\,$mag, which is too large to obtain a useful check.
We therefore combine four such measurements from KMTC02, KMTC42, KMTS02,
and KMTS42, finding $\Delta(V-I)$ (relative to the clump) of
$(-0.22 \pm 0.13)$,
$(-0.17 \pm 0.14)$,
$(-0.41 \pm 0.16)$, and
$(-0.24 \pm 0.14)$, respectively,
As these are are mutually consistent ($\chi^2=1.4$ for 3 dof), we combine them
to obtain $\Delta(V-I) = -0.25 \pm 0.07$, which is substantially redder than
the $H$-band based determination.  To proceed further, we convert this to
$(I-H)_{0,S} = 0.91\pm 0.09$ using the dwarf-star tables from \citet{bb88}.
Hence the two determinations are separated by $2.1\,\sigma$ and are therefore
in mild tension.  To reflect this tension, we adopt
their weighted average, $(I-H)_{0,S} = 0.74$, but also adopt a larger
error than the standard error of the mean (0.046 mag), namely
$\sigma(I-H)_{0,S} = 0.07$.  For purposes of homogeneous reporting
in Table~\ref{tab:cmd},
we note that this is equivalent to $(V-I)_{0,S}=0.69\pm 0.05$.
%0.62 1.32 0.045 0.655
%0.68 1.46 0.05  0.73
%0.71 1.53 0.05  0.77
%0.74            0.815
%0.75 1.64 0.06  0.83
%0,81            0.91
%0.88 1.96 0.08  1.00

\begin{figure}
\plotone{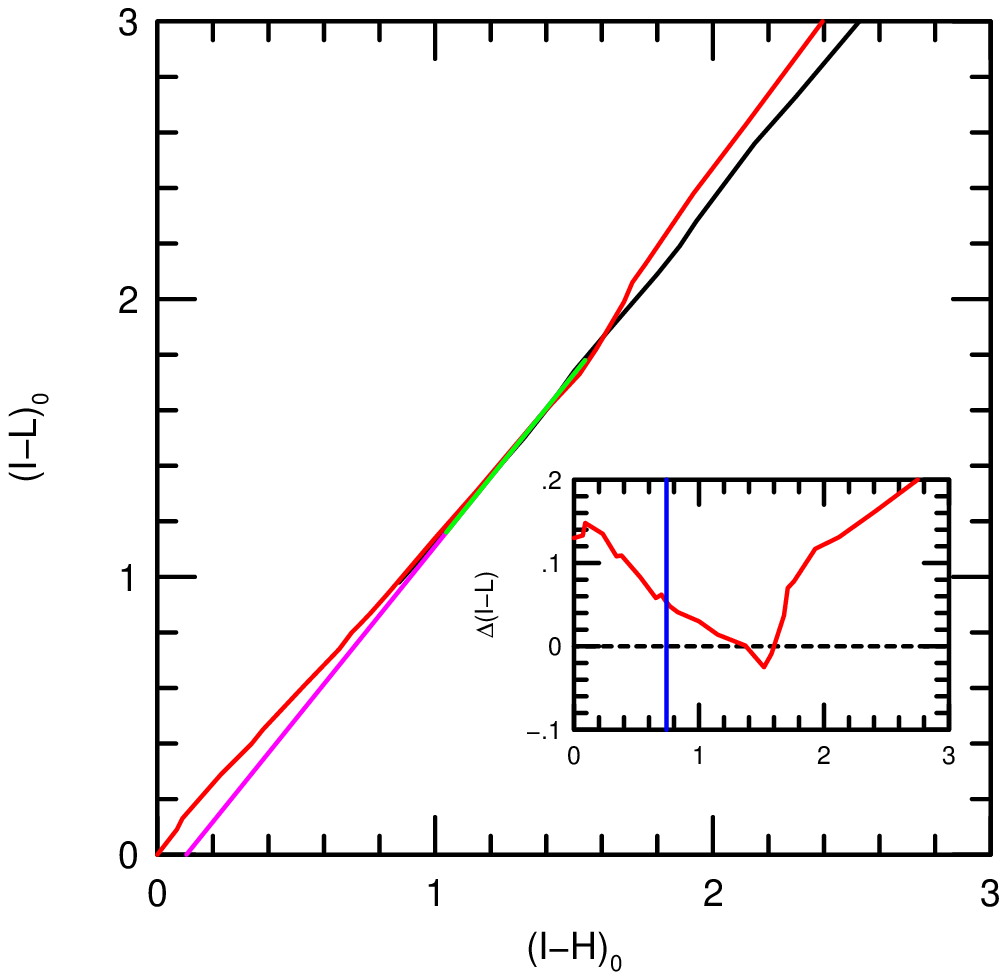}
\caption{Empirical $IHL$ color-color relation for giants (black) and
dwarfs (red) determined from local stars by \citet{bb88}.
The green line segment shows the slope of the relation that we derived
from bulge giants (whose range is represented by the length of the segment)
from OGLE-III, UKIRT, and {\it Spitzer} (OUS) data,
which has been transposed to the local relation, showing excellent
agreement with the slope of the black curve.  The magenta line is the
extrapolation of the green relation to a bluer range.  The inset shows
the offset between the magenta and red curves.  The $(I-L)$ color of the
dwarf (actually turnoff) source is determined by applying the extrapolated
OUS relation and then correcting this using the inset relation at the
color of the source (blue).
}
\label{fig:ihl}
\end{figure}

We now apply this $(I-H)_{0,S}$ color measurement to determine the
$(I-L)_S$ color constraint by applying the $IHL$ color-color relation.
To do so, we must take account of the fact that the $IHL$ color-color
relation is different for giants (which are the only stars bright
enough to enter the empirical determination based on field stars) and
dwarfs (including the object of interest, i.e., the microlensed
source).  This process is illustrated in Figure~\ref{fig:ihl}.  In the
main panel, the $IHL$ relations from \citet{bb88} are shown for dwarfs
and giants, in red and black, respectively.  The green line segment
shows the slope (1.24) of the empirical color-color relation derived
by matching field stars from OGLE-III, UKIRT, and {\it Spitzer}.  The
length of this segment represents the range of colors of the stars
that entered the determination.  Its height is set to match its center
to the black curve.  The key point is that the slopes of the black
curve and the green line are the same in this region.  If this were
not the case, it would mean that either \citet{bb88} had made a
serious error, or there was some problem with one of the 3 data sets
that we are using, or that the $IHL$ relation of bulge giants was
substantially different from local giants.  The magenta line
extrapolates the linear relation represented by the green segment to
bluer colors.  The inset shows the offset of the dwarf $IHL$ relation
from this extrapolation, while the blue vertical line indicates the
color of the microlensed source.

The empirical $IHL$ relation from field stars is
$(I-L) = 1.24[\Delta(I-H)] + 3.146$, where $\Delta(I-H)$ is the offset
from the observed clump centroid $(I-H)_{\rm cl}=3.58$ or $(I-H)_{0,\rm cl}=1.29$.
Finally, the color constraints must be applied to light-curve photometry,
e.g., $I_{\rm KMTC02,pysis}$, which is offset from calibrated OGLE-III CMD
photometry by $I_{\rm KMTC02,pysis}-I = -0.08$.  Combining all these terms,
\begin{equation}
(I_{\rm KMTC02,pysis}-L) = 1.24[\Delta(I-H)] + 3.146 + (I_{\rm KMTC02,pySIS}-I)
+\Delta(I-L),
  \label{eqn:colcol}
\end{equation}
where the last term is given by the inset in Figure~\ref{fig:ihl}.  This
yields $(I_{\rm KMTC02,pysis}-L)= 2.43 \pm 0.08$.

We also place constraints on the blended light.  In the light-curve fits,
only about 16\% of the baseline light is attributed to the blend.
Nevertheless, this may be partly due to the difficulty of carrying out
PSF photometry of the baseline object in crowded bulge fields.  Therefore,
we more conservatively adopt $I_B > 20.5$, which constrains the lens flux,
$I_L \geq I_B > 20.5$.  We expect this to provide only
a very weak constraint because the $\bpi_\e$ measurement already confines the
lens to be in or near the bulge, so this limit only implies
$M_{I,L} > I_B - I_{\rm cl} + M_{I,\rm cl} > 3.1$, which, among stars that
are in or near the bulge, would only rule out those that have already evolved
off the main sequence.  Moreover, to be ruled out, such stars would
need proper motions
$\mu_\rel \equiv \theta_\e/t_\e =\kappa M\pi_\e/t_\e \ga 10\,\masyr$,
which are themselves relatively rare.
Nevertheless, for completeness, we will include this constraint in the
Bayesian analysis of Section~\ref{sec:phys-ob171275}.  Note that the
blend is {\it not} displayed in Figure~\ref{fig:allcmd} because its color
is not known.

{\subsection{OGLE-2017-BLG-0640}% kb171726
  \label{sec:cmd-ob170640}}

Because this field is highly extincted and the source is intrinsically
red, $(V-I)_S$ cannot be measured from the KMT data.  Fortunately, this
event lies within the UKIRT Microlensing Survey \citep{ukirt17}, so we are
able to determine $\theta_*$ from an $I/H$ CMD.  See Figure~\ref{fig:allcmd}.
Although the CMD is based on a relatively small ($1^\prime$) circle centered
on the event, there is strong differential extinction.  That is, without
extinction, the mean $I$-band magnitude of clump stars would be approximately
independent of color, while the CMD shows that the bluer clump stars are,
on average, substantially brighter than the redder ones.  If this effect
were due to a pure gradient in extinction across the field.  The clump
center at the location of the event would simply be the center of this
tilted structure.  We have investigated the CMD on smaller angular scales
and find no evidence against the gradient hypothesis.  We therefore
adopt the center of the structure for the position of the clump (red dot).
The source is $\Delta(I-H)=-0.21\pm 0.03$ mag bluer than the clump.  To put
this on a homogeneous basis with other entries in Table~\ref{tab:cmd},
we use \citet{bb88}
to translate this offset to $\Delta(V-I)=-0.14\pm 0.02$, i.e.,
$(V-I)_{0,S}= 0.92\pm 0.02$.  Due to the fact that the source is a clump giant
suffering heavy extinction, we cannot place any useful constraint on blended
light.

%0.81   1.75    0.07    0.87
%0.91   2.05    0.08    1.06
%0.92                   1.08
%0.94   2.15    0.085   1.125
%0.94   2.16    0.085   
%1.00   2.31    0.095   
%1.08   2.50    0.10    

{\subsection{OGLE-2017-BLG-1237}% kb170422 
  \label{sec:cmd-ob171237}}

As in all but one of the events in this paper, we align our
measurements to the OGLE-III CMD.  To measure the source color, we
rely on KMTC03 pyDIA, for which there are two highly magnified
$V$-band points on the night of ${\rm HJD}^\prime=7935$ and four
moderately magnified points on ${\rm HJD}^\prime=7936$ and 7937.  The
relatively small scatter of the regression of these points (when
combined with the mean from the relatively unmagnified points)
confirms that the measurement is not strongly affected by the bright
variable at $1.4^{\prime\prime}$.  This is impressive, given the
variable is 6.4 mag brighter than the source in $V$ band.  However, as
mentioned in Section~\ref{sec:anal-ob171237}, the amplitude of this
double-mode pulsator is close to its minimum during these three days of
high and relatively-high magnification.

Combining the resulting $\theta_*=0.318\pm 0.024\,\muas$
measurement from Table~\ref{tab:cmd}
and the parameters from Table~\ref{tab:171237parms}, yields
\begin{equation}
  \theta_\e = {\theta_*\over \rho}=415\pm 37\,\muas;
  \qquad
  \mu_\rel = {\theta_*\over t_*} = 5.42\pm 0.41\,\masyr
  \label{eqn:thetaeval-1237}
\end{equation}

Given the proximity of the clump-giant blend at $\sim 0.5^{\prime\prime}$
and the bright variable at $\sim 1.4^{\prime\prime}$, we are unable to
place useful constraints on the light from the lens.

{\subsection{OGLE-2017-BLG-1777}% kb170282 
\label{sec:cmd-ob171777}}

Because of the central role of OGLE-IV photometry in the removal
of the complex trend (Appendix~\ref{sec:lc-corr}), we determine $I_{0,S}$
by calibrating OGLE-IV photometry to OGLE-III
(i.e., $I_{\rm OGLE-III} = I_{\rm OGLE-IV} + 0.064\pm 0.001$), but we still evaluate
$(V-I)_{0,S}$ using KMT data as described in the preamble to this section.
Indeed, we carry out these procedures twice, once using
KMTC42 data and the other time using KMTS42 data.
We thereby find
$(V-I)^{\rm KMTS42}_{S,OGLE-III}=2.005\pm 0.030$ and
$(V-I)^{\rm KMTC42}_{S,OGLE-III}=1.951\pm 0.025$. We adopt a calibrated value
$(V-I)_S=1.97\pm 0.02$, which then implies
$(V-I)_{0,S}=0.64\pm 0.03$, See Table~\ref{tab:cmd}.

The resulting $\theta_*=0.490\pm0.036\,\muas$ then implies for the
two solutions,
\begin{equation}
  \theta_\e = {63.9\pm 4.7\atop 72.1 \pm 5.3}\,\muas;
  \quad
  \mu_\rel = {358\pm 26\atop 382 \pm 28}\,\muas\,{\rm yr}^{-1},
  \quad
  {\rm (Local\ 1)\atop(Local\ 2)}.
  \label{eqn:thetaeval-1777}
\end{equation}

We note that two of the ``alarming'' features of these 2L1S models,
namely the very small $\theta_\e$ and $\mu_\rel$, are qualitatively similar for
the single-lens/xallarap model: $\theta_\e^{\rm 1L1S}=52.5\,\muas$ and
$\mu_\rel^{\rm 1L1S}=236\,\muas\,{\rm yr}^{-1}$.

Finally, we assess the limits on light from the lens based on both
KMTC42 and KMTS42 pyDIA astrometry and on a comparison
of KMT pyDIA astrometry with astrometry from OGLE-III and OGLE-IV.
We reference all positions to the bright giant because it is in all
four of these catalogs and because its position is likely to be well-determined.

We find that the offset of the lens position (determined from the difference
image) is $\Delta\btheta(E,N)_L = (825,100)\,\mas$ and $(821,111)\,\mas$
for KMTC42 and KMTS42, respectively, which confirms that both the
lens position and the giant-neighbor position are well determined.

Neither the KMTC42 nor the KMTS42 field-star catalogs contains any stars in the
neighborhood of the lens. However, the OGLE-III and OGLE-IV catalogs, which
are of higher quality, both have neighbors.  For OGLE-III, these are at
$\Delta\btheta(E,N)_{\rm N1}^{\rm O-III} = (880,130)\,\mas$ and
$\Delta\btheta(E,N)_{\rm N2}^{\rm O-III} = (530,90)\,\mas$ with $I$-band magnitudes
of 19.86 and 19.13, respectively.  For OGLE-IV, there is one at
$\Delta\btheta(E,N)_{\rm N2}^{\rm O-IV} = (570,110)\,\mas$ with $I=19.07$.
We consider, given the difficulty of measuring positions in the presence
of the bright neighbor, that N1 from OGLE-III is consistent with the lens
position but N2 (similar for both OGLE catalogs) is not.  We therefore
place an upper limit on the lens light $I_L> 19.7$.

{\subsection{OGLE-2017-BLG-0543}% kb170140 
\label{sec:cmd-ob170543}}

The characterization of the lens system is ambiguous, and so we
include the CMD analysis only for completeness.  The source photometry
and inferred $\theta_*$ are given in Table~\ref{tab:cmd}.  Among the
six events in this paper, OGLE-2017-BLG-0543 is the only one without
calibrated photometry.  However, because the determination of
$\theta_*$ requires only relative photometry, this is not a
fundamental limitation.  We note that the source is consistent with
being unblended.

{\subsection{OGLE-2017-BLG-1694}% kb172126 
\label{sec:cmd-ob171694}}

As for the case of OGLE-2017-BLG-0543, the characterization of the
lens system is ambiguous, and so we include the CMD analysis only for
completeness.  The source photometry and inferred $\theta_*$ are
given in Table~\ref{tab:cmd}.  By subtracting the source flux (derived
from fitting the pyDIA light curve to the best model and transforming
to the OGLE-III calibrated system) from the OGLE-III baseline flux, we
find blended flux with $I_B=20.25$.  However, we do not show this
blend in Figure~\ref{fig:allcmd} because we have no information about
its color.  That is, OGLE-III does not list a color for the baseline
object, while KMT pyDIA does not resolve this object.  We note that
the baseline object lies within about 200 mas of the source, but
we do not investigate the astrometry in detail.

\section{{Physical Parameters}
\label{sec:phys}}

To make Bayesian estimates of the lens properties, we follow the same
procedures as described in Section~5 of \citet{2018prime}.  We refer the
reader to that work for details.  Below, we repeat the text from
Section~5 of \citet{2018subprime} for the reader's convenience.

In Table~\ref{tab:physall},
we present the resulting Bayesian estimates
of the host mass $M_{\rm host}$, the planet mass $M_{\rm planet}$, 
the distance to the lens system $D_L$, and the planet-host projected
separation $a_\perp$.  For two of the four planetary
events, there are two or more competing solutions.  For these cases
(following \citealt{2018prime}),
we show the results of the Bayesian analysis for each solution separately,
and we then show the ``adopted'' values below these.  For $M_{\rm host}$,
$M_{\rm planet}$, and $D_L$, these are simply the weighted averages of the
separate solutions, where the weights are the product of the two
factors at the right side of each row.  The first factor is simply
the total weight from the Bayesian analysis.  The second is 
$\exp(-\Delta\chi^2/2)$ where $\Delta\chi^2$ is the $\chi^2$ difference
relative to the best solution.  For $a_\perp$,
we follow a similar approach provided that either the individual solutions
are strongly overlapping or that one solution is strongly dominant.  
If neither condition is met, we show the two values separately.

We present Bayesian analyses for 4 of the 6 events, but not for
OGLE-2017-BLG-0543, and OGLE-2017-BLG-1694, for which
we cannot distinguish between competing interpretations of the event.
See Sections~\ref{sec:anal-ob170543} and \ref{sec:anal-ob171694}.
Figure~\ref{fig:bayes} show histograms
for $M_{\rm host}$ and $D_L$ for these 4 events.

{\subsection{OGLE-2017-BLG-1275}% kb170314 
\label{sec:phys-ob171275}}

\begin{figure}
\plotone{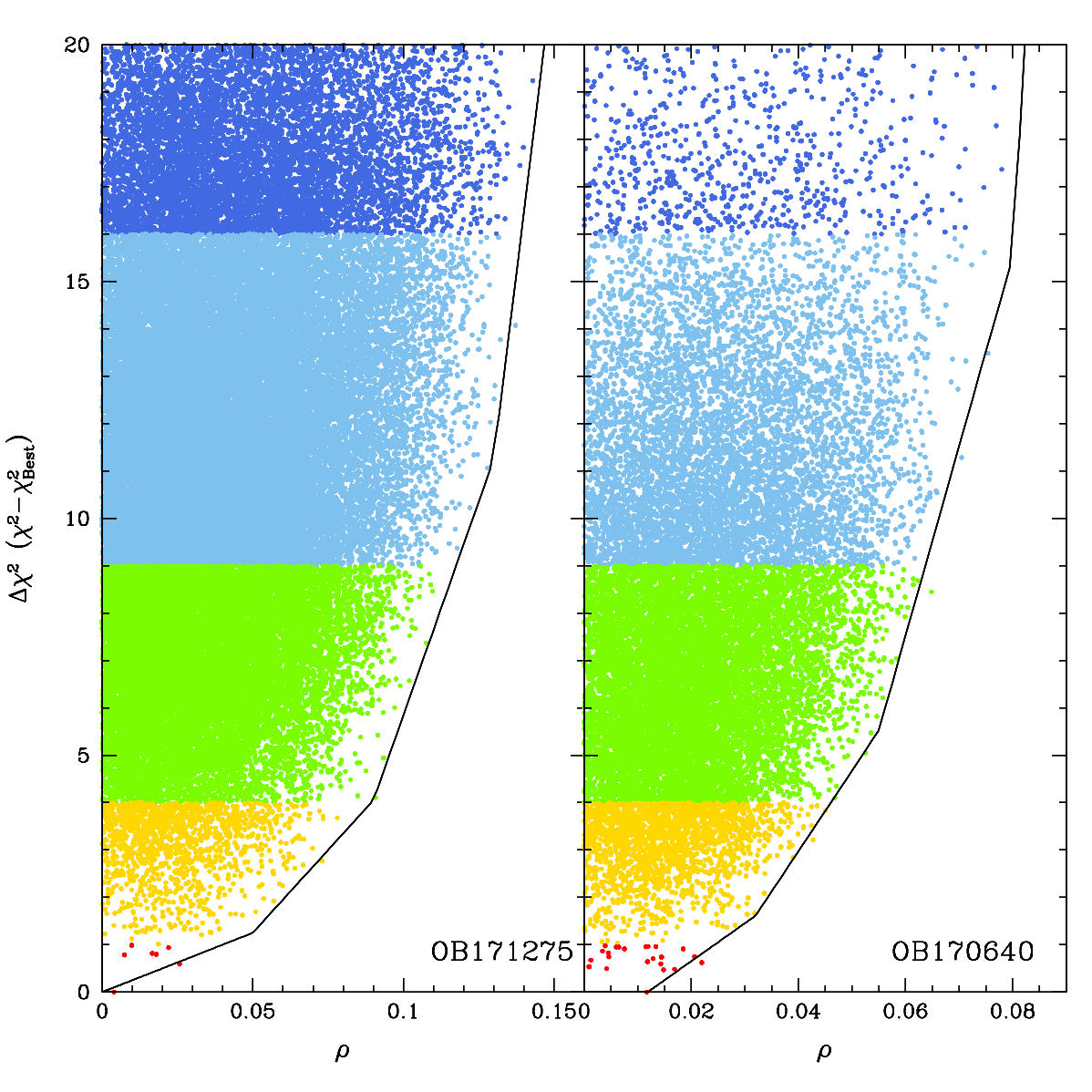}
\caption{Envelope functions (line segments) based on MCMC data (points),
  giving $\Delta\chi^2(\rho)$ for OGLE-2017-BLG-1275 and OGLE-2017-BLG-0640.
  Note that at, e.g., $2\,\sigma$, $\rho<0.09$ and $\rho<0.045$,
  so that $\mu_\rel = \theta_*/\rho t_\e$ is constrained to be
  $\mu_\rel \ga 0.46\,\masyr$, and $\mu_\rel \ga 2.80\,\masyr$, for
  OGLE-2017-BLG-1275 and OGLE-2017-BLG-0640, respectively.  Thus,
  the $\rho$ limit is completely unconstraining for the first case,
  and mildly constraining for the second.  See Equation~(\ref{eqn:probmu2}).
  %1.51/(0.09*13.38=0.46, 4.96/(0.045*14.38)=2.80
}
\label{fig:rho}
\end{figure}

\begin{figure}
\plotone{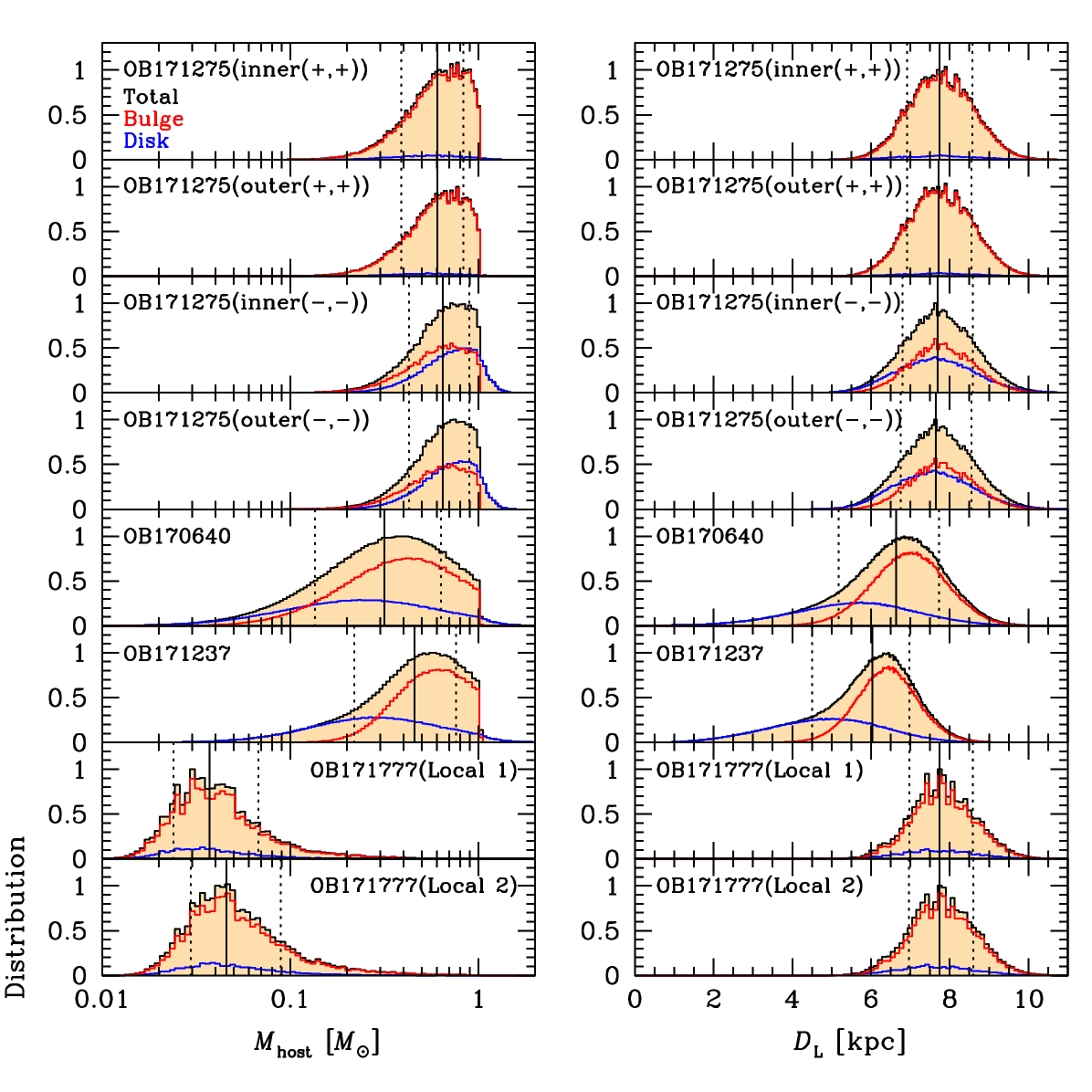}
\caption{Histograms of the host mass (left) and lens distance (right)
  for the three unambiguously planetary events, plus one very likely
  candidate (OB171777), as derived from the 
Bayesian analyses.  Disk (blue) and bulge (red) distributions are shown
separately, with their total shown in black.
}
\label{fig:bayes}
\end{figure}

OGLE-2017-BLG-1275 has four solutions,
${\rm inner}(++)$,
${\rm inner}(--)$,
${\rm outer}(++)$, and
${\rm outer}(--)$,
upon which there are a total of four constraints, i.e., on $t_\e$, $\rho$,
$\bpi_\e$, and $I_L$.  The first comes from Tables~\ref{tab:ob171275pp} and
\ref{tab:ob171275mm}, i.e.,
$\chi^2_{t_\e} = (t_{\e,\rm sim} - t_{\e,\rm table})^2/\sigma_{t_\e}^2$,
where $t_{\e,\rm sim}$ is the timescale of the simulated event.
The second is found by calculating $\rho_{\rm sim}=\theta_{\e,\rm sim}/\theta_*$
where $\theta_{\e,\rm sim}$ is the Einstein radius of the simulated event
and $\theta_*$ is from Table~\ref{tab:cmd}, and then applying the
$\chi^2(\rho)$ envelope function from Figure~\ref{fig:rho}.
We note that, even at the $1\,\sigma$ level, this constraint only implies
$\mu_\rel\ga 0.8\,\masyr$, which is hardly constraining, but we nevertheless
include this constraint for completeness.
For the third, we find
\begin{equation}
  \chi^2(\bpi_\e) = \sum_{i,j=1}^2 (a_i - a_{i,0})(a_j - a_{j,0})b_{i,j}
  \qquad b\equiv c^{-1},
  \label{eqn:chi2pie}
\end{equation}
where the $a_i$ are the $\bpi_\e$ components for the simulated event, while
$a_{i,0}$ and $c_{i,j}$ are the mean and covariance matrix of $\bpi_\e$
as derived from the MCMC.  These have very similar values and error bars
as the median-based values shown in Tables~\ref{tab:ob171275pp} and
\ref{tab:ob171275mm}, with correlation coefficients,
$+0.35$ ($\rm inner{(++)}$),
$-0.32$ ($\rm inner{(--)}$),
$+0.39$ ($\rm outer{(++)}$), and
$-0.40$ ($\rm outer{(--)}$).
The final constraint is $I_L >20.5$, as discussed
in Section~\ref{sec:cmd-ob171275}.

For all four solutions, the planet mass is peaked near
$M_{\rm planet}\sim 6\,M_{\rm Jup}$, while the host is near the
M-dwarf/K-dwarf boundary.  Note that the distance is $D_L\sim 7.7\,\kpc$,
i.e., close to the Galactocentric distance, regardless of whether it
belongs to the bulge or disk populations.  Also note that the latter is
prohibited by $\bpi_\e$ measurements for the $(+,+)$ solutions because
their directions are kinematically inconsistent with disk objects.
The projected planet-host separation is either about 1.2 au or 2.2 au.
By comparison, for an $M_{\rm host}\sim 0.6\,M_\odot$ host, the snow
line is about $a_{\rm snow} = 2.7\,\au (M_{\rm host}/M_\odot)=1.62\,\au$.  Hence,
this is a relatively rare case for microlensing, in which the planet could
be well inside the snow line, although because $a_\perp$ is bimodal,
it may well be beyond the snow line, even in projection.

Finally, we note that the posterior estimate of the proper motion is
$\mu_\rel = 6.7\pm 2.2\,\masyr$.  Based on this estimate, the lens
and source will be separated by $\Delta\theta\sim 87\pm 29\,\mas$, which
is a plausible date for first light on adaptive optics (A0) on extremely
large telescopes (ELTs).  Hence, it is likely that the lens mass
could be more precisely measured at AO first light \citep{masada}.

%a=sD_L thetaE, thetaE = a/D_L s, mu = a/D_L s tE
%tE = 13.4 (3%), D_L = 7.7 (10.6%), a/s = 1.9 (30%).
% mu = 6.7 +- 2.2 masyr
% a/s = (1.14 +- 0.36)/0.63=1.81+-0.57, (2.02 +- 0.63)/1.09=1.85+-0.58,
%(1.21 +- 0.36)/0.63=1.92+-0.57, (2.14 +- 0.63)/1.09=1.96+-0.58,

{\subsection{OGLE-2017-BLG-0640}% kb171726
\label{sec:phys-ob170640}}

OGLE-2017-BLG-0640 has one solution, upon which there are two
constraints, i.e., on $t_\e$ and $\rho$.  The first comes from
Table~\ref{tab:ob170640parms}, while the second comes from the
$\rho$-envelope function shown in Figure~\ref{fig:rho}, which is then combined
with the $\theta_*$ value from Table~\ref{tab:cmd} to yield a constraint
on $\theta_\e$.

The planet is of Jovian mass, and the host is an M dwarf, while the
system is most likely in the Galactic bulge.  The planet lies just
beyond the snow line in projection.

The posterior proper-motion estimate is $\mu_\rel = 7.2\pm 2.4\,\masyr$, which
would imply a similar estimate for the separation at ELT AO first light
as in the case of OGLE-2017-BLG-1275.  However, the contrast ratio is
expected to be much larger because the source is a giant (rather than
a subgiant), while the lens is predicted to be a middle M dwarf (rather
than near the boundary of M dwarfs and K dwarfs).  Hence, it may be more
challenging to resolve the lens and source separately for OGLE-2017-BLG-0640
than for OGLE-2017-BLG-1275.

%mu = a/D_L s = 1.14/6.63/0.61/14.38 = 7.2 +- 2.4

{\subsection{OGLE-2017-BLG-1237}% kb170422 
\label{sec:phys-ob171237}}

OGLE-2017-BLG-1237 has one solution, upon which there are two
constraints, i.e., on $t_\e$ and $\theta_\e$, which come from
Table~\ref{tab:171237parms} and Equation~(\ref{eqn:thetaeval-1237}),
respectively.  

The planet is estimated to be several Jupiter masses, while the host is
an early M dwarf.  Similarly to OGLE-2017-BLG-0640, the system most
likely lies in the bulge.  The planet lies well beyond the snow line.

In this case, the proper motion is directly measured (rather than
being estimated from the Bayesian analysis as in the previous two events):
$\mu_\rel=5.4\pm0.4\,\masyr$.  The source is faint, meaning that the
lens and source can certainly be resolved at EELT AO first light.
In fact, it may well be possible to resolve the lens earlier than that
using 10m class telescopes.  However, light from the clump-giant
blend at $0.5^{\prime\prime}$ may make this difficult.  Hence, this
issue should be carefully evaluated before undertaking such observations.

{\subsection{OGLE-2017-BLG-1777}% kb170282 
  \label{sec:phys-ob171777}}

OGLE-2017-BLG-1777 event has two solutions,
upon which there are a total of four constraints, i.e., on $t_\e$, $\rho$,
$\bpi_\e$, and $I_L$.  The first two come from Tables~\ref{tab:ob171777parms1}
and \ref{tab:ob171777parms2}.  The second then yields $\theta_\e=\theta_*/\rho$,
where $\theta_*$ is given in Table~\ref{tab:cmd}.
The third applies Equation~(\ref{eqn:chi2pie})
to the mean values and errors of $\bpi_\e$, which are similar to the values
in Tables~\ref{tab:ob171777parms1} and
\ref{tab:ob171777parms2}, with correlation
coefficients $-0.15$ and $+0.12$ for Locals 1 and 2, respectively.  The final
constraint is $I_L >19.7$, as discussed in Section~\ref{sec:cmd-ob171777}.

The lens system is composed of a bulge BD (or possibly VLM star)
orbited by a Neptune-class planet.
Insofar as the concept of ``snow line'' applies to such systems, the
planet lies well within the snow line.  As we have discussed, the
proper motion is extraordinarily low, $\mu_\rel=0.37\pm0.03\,\masyr$,
which would make it impossible to resolve the source and lens
(even if the latter proves to be luminous) using ELT AO for at least
several decades.  On the other hand, if the lens is luminous, the two
could conceivably by resolved using the VLTI GRAVITY interferometer
\citep{kojima1b}.
However, as we have emphasized, the first additional observation
that should be made to clarify the nature of this system is to measure
the RV variations induced by the putative BD (or VLM star) companion to
the source.

{\subsection{OGLE-2017-BLG-0543}% kb170140 
\label{sec:phys-ob170543}}

Because OGLE-2017-BLG-0543 has viable non-planetary explanations,
we do not carry out a Bayesian analysis.

{\subsection{OGLE-2017-BLG-1694}% kb172126 
\label{sec:phys-ob171694}}

Because OGLE-2017-BLG-1694 has viable non-planetary explanations,
we do not carry out a Bayesian analysis.

\section{{Discussion}
\label{sec:discussion}}
%2016 not in AF ob161850
%2016 11, 2017 12, 2018 19, 2019 13

In this paper, we have completed the analysis of planetary (and possibly
planetary) events from the 2017 KMT prime fields that were identified as
candidates by the KMT AnomalyFinder system.  The sample contains 11
unambiguous planets with unambiguous mass-ratio measurements, and one other
event for which we judge the planetary interpretation to be very likely.
In addition there are three unambiguous planets that have very
large uncertainties in $q$.
See Table~\ref{tab:2017events}.
While it is not the jurisdiction of this paper to decide which
planets will ultimately enter the statistical sample, for purposes of this
discussion we will adopt ``12''.  This can be compared to similar estimates
of this number
for three other AnomalyFinder prime-field years of 11 (2016), 19 (2018),
and 13 (2019).  The first of these estimates is derived from Table~12 of
\citet{2016prime}, while the latter two are derived from Figure~20 of
\citet{2019subprime}.  These numbers imply a mean and standard deviation
of $13.75\pm 3.59$.  By comparison, there are only two complete analyses
of AnomalyFinder subprime fields, i.e., 2018 and 2019, for which 
Figure~20 of \citet{2019subprime} indicates 14 and 12 planets, respectively.
Adopting Poisson errors (rather than errors derived from scatter), we
estimate an average rate (and standard error of the mean)
of $26.75\pm 3.15$ AnomalyFinder planets per year.
Thus, we can expect a statistical sample of $267\pm 33$
planets from 10 years of KMT data, 2016--2026 (excluding 2020 due to Covid-19),
which is the current horizon of the KMTNet project\footnote{For a
Poison process, the detection of $N_1$ events during a time $t_1$ implies
a mean and standard deviation of $N_2 = (t_2/t_1)[N_1 \pm \sqrt{2 N_1}]$
during a subsequent time $t_2$.  Hence, the uncertainty is
$\sqrt{2[(6/4)^2\times 55 + (8/2)^2\times 26]}= 33$.}.
%AF found 2016: 1991,1307?, 0506,
% not 1836?
Another relevant statistic is that three of the 12 planets in
Table~\ref{tab:2017events} were first identified by AnomalyFinder, whereas
the remainder were first identified by eye.  For the four years 2016--2019,
these ratios are 3/11, 3/12, 8/19, and 8/13.  The higher fractions for
the two most recent years may be random fluctuations, but they might also
be explained by the factor $\sim 2$ shorter interval between the time that
the events were discovered and the time that the AnomalyFinder search
was carried out.

Three of the four planets that we have reported call for special notice:
OGLE-2017-BLG-1275, OGLE-2017-BLG-1237, and OGLE-2017-BLG-1777.

OGLE-2017-BLG-1275 is the first planet with a {\it Spitzer} parallax
measurement that was originally discovered by AnomalyFinder.  This
discovery raises the question of whether there could be additional
planets with {\it Spitzer} parallax measurements that remain
``hidden'' in the data.  The {\it Spitzer} microlensing program ended
in 2019, so all such AnomalyFinder planets will be identified when the
2016-2017 subprime analyses are complete.  However, additional {\it
  Spitzer} planets that presently remain hidden could be discovered by
another channel.  The AnomalyFinder search is applied to the
end-of-year pySIS pipeline data, whereas all published KMT planets are
based on TLC pySIS rereductions.  These often have substantially
better $\Delta\chi^2 = \chi^2({\rm 1L1S}) - \chi^2({\rm 2L1S})$
compared to the pipeline pySIS.  Because TLC is labor intensive, it
would be prohibitive to rereduce all $\sim 1000$ {\it Spitzer} events,
but this would be feasible for promising subsets, such as all events
with $A_\max> 5$ and {\it Spitzer} coverage that might plausibly yield
a parallax measurement.

OGLE-2017-BLG-1237 is notable because it lies just $1.4^{\prime\prime}$ from
a variable, whose full amplitude of variations is about 15 times brighter
than the source flux.  Fortunately, the variable is sufficiently regular
that the impact of these variations on the event can be accurately modeled
based on data taken when the source is essentially unmagnified.

OGLE-2017-BLG-1777 is remarkable in several respects, some of which
are related to one another.  First, like OGLE-2017-BLG-1237, the
original light curve had to be corrected for effects from a
neighboring bright star (see Appendix~\ref{sec:lc-corr}).  Second,
both the Einstein radius $\theta_\e$ and the proper motion $\mu_\rel$
are unusually small, leading to a concern that both might be a product
of an incorrect measurement of $t_*$.  We investigate this issue in
Appendix~\ref{sec:rho} and conclude that this is not likely to be the
case.  Third, there is a $P\sim 13\,$day oscillation in the light
curve in the neighborhood of the peak.  After exploring a wide range of
models, we concluded that this feature could only be explained by
xallarap: orbit of the source star around an unseen companion.  The
implied mass of this companion is close to the boundary between BDs
and VLM stars.  We show in Appendix~\ref{sec:bd-xallarap} that
detection of such low-mass source companions is greatly enhanced for
events with low $\theta_\e$ and low $\mu_\rel$.  Finally, only by
including lens orbital motion in the microlens model are we able to
explain the final, otherwise discrepant, data point from the 2017
season.  Although this might seem to be an ad hoc solution for this
single point, the resulting orbital-motion parameters are strongly
peaked in the range that is expected for Kepler orbits.  The
concatenation of these unusual features gave us some pause regarding
this planetary candidate, but we finally concluded that it was very
likely to be real.  We argued that this complex solution could be
partially verified by RV measurements of the source, which would be
aimed at confirming the predicted $v\sin i\sim 6.5\,\kms$ amplitude of
source motion, with a period of $P\sim 12.7\,$day.

\acknowledgments
This research has made use of the KMTNet system operated by the Korea Astronomy 
and Space Science Institute (KASI) at three host sites of CTIO in Chile, SAAO in 
South Africa, and SSO in Australia. Data transfer from the host site to KASI was 
supported by the Korea Research Environment Open NETwork (KREONET).
Work by C.H. was supported by the grants of National Research Foundation 
of Korea (2020R1A4A2002885 and 2019R1A2C2085965).
%
%J.C.Y. acknowledges support from US NSF Grant No. AST-2108414 and JPL grant 1571564.
J.C.Y., S.-J. C., and I.-G.S. acknowledge support from US NSF Grant
No. AST-2108414, and J.C.Y. acknowledges support from JPL grant
1571564.
Y.S. acknowledges support from BSF Grant No. 2020740.
W.Zang and H.Y. acknowledge support by the National Science Foundation of China (Grant No. 12133005).
UKIRT is currently owned by the University of Hawaii (UH) and operated by the UH Institute for Astronomy; operations are enabled through the cooperation of the East Asian Observatory. When some of the 2017 data reported here were acquired, UKIRT was supported by NASA and operated under an agreement among the University of Hawaii, the University of Arizona, and Lockheed Martin Advanced Technology Center; operations were enabled through the cooperation of the East Asian Observatory. The collection of the 2017 data reported here was furthermore partially supported by NASA grants NNX17AD73G and NNG16PJ32C.
This paper makes use of data from the UKIRT microlensing surveys (Shvartzvald et al. 2017) provided by the UKIRT Microlensing Team and services at the NASA Exoplanet Archive, which is operated by the California Institute of Technology, under contract with the National Aeronautics and Space Administration under the Exoplanet Exploration Program.
The authors wish to recognize and acknowledge the very significant cultural role and reverence that the summit of Mauna Kea has always had within the indigenous Hawaiian community. We are most fortunate to have the opportunity to use data produced from observations conducted on this mountain.

\appendix
\section{Light-curve Correction For OGLE-2017-BLG-1777}
\label{sec:lc-corr}

\begin{figure}
\plotone{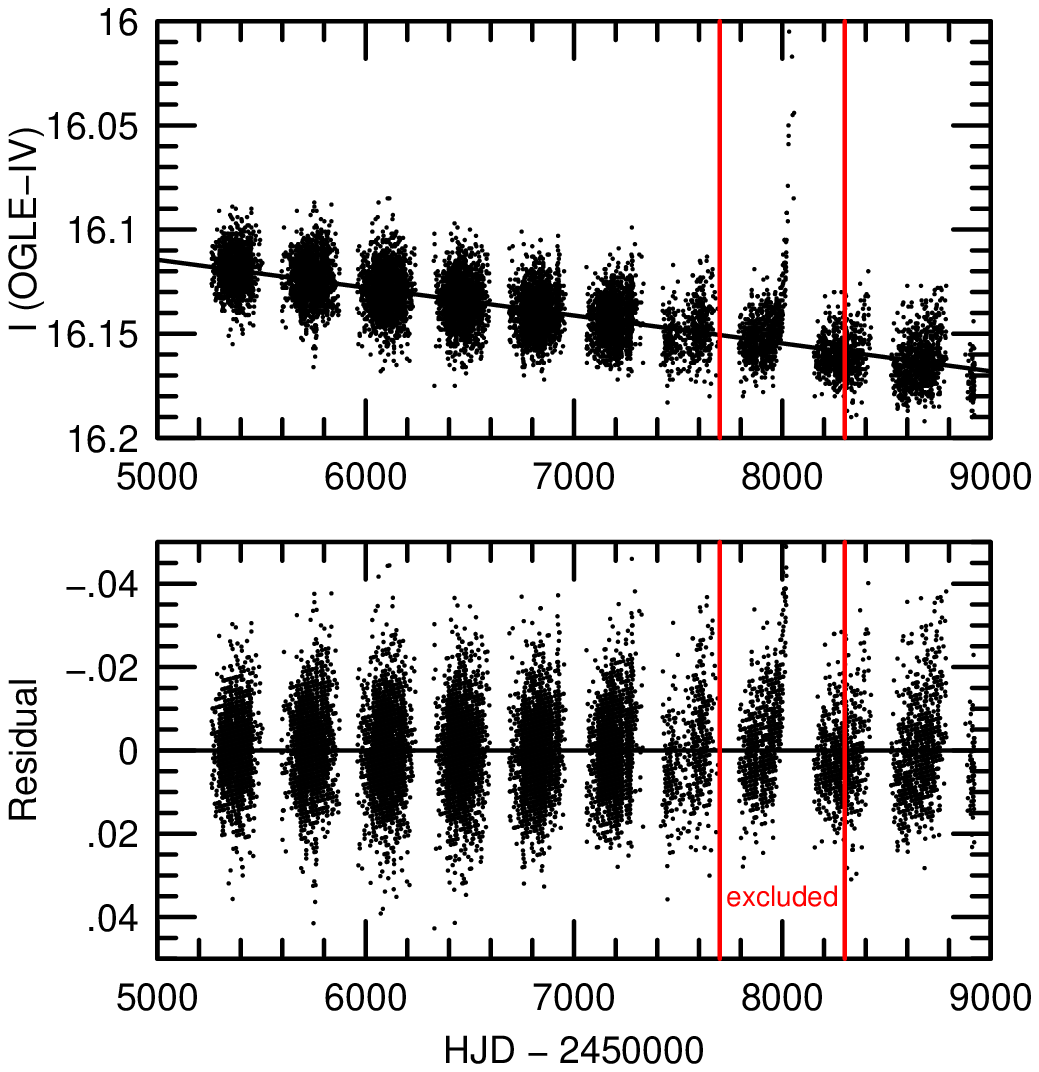}
\caption{Long-term linear trend in the OGLE-IV light curve of
  OGLE-2017-BLG-1777.  The magnitudes are constructed by first doing difference
  image photometry, then adding in the flux of the ``baseline object'', and
  finally converting to magnitudes.  The baseline object is unrelated to the
  microlensing event, but it accounts for the long-term trend seen here.
  The fit is a simple linear fit in magnitudes, with the region enclosed
  by red lines excluded from the fit.  More detailed investigations are
  carried out in flux units.  See Appendix~\ref{sec:lc-corr}.  
}
\label{fig:long}
\end{figure}

There is a long term, linear decline in the baseline flux of this event,
which is clearly manifest in the 10-year OGLE-IV light curve.
See Figure~\ref{fig:long}.
As we will show, the physical origin of this trend
is well-understood, and even if it were not, the trend itself is
precisely measured.  Hence, it could easily be removed.  However, a more
detailed analysis shows that while this ``fix'' may appear obvious, it is
also not correct.

Before discussing the origin of this effect, we start by noting that the
high baseline flux shown in Figure~\ref{fig:long} ($I\sim 16$) is an artifact,
which results from the fact that the OGLE alert was based on the apparent
light curve of a neighboring catalog star that is a giant of this magnitude.
The underlying OGLE light curves are simply DIA photometry, to which
OGLE adds the flux from the catalog star before converting to a magnitude
system.  This routine procedure has absolutely no impact on the light-curve
analysis, unless the modeled source flux exceeds this baseline flux, in
which case it might trigger concerns about ``negative blending'', which
is certainly does not apply to OGLE-2017-BLG-1777.

However, in the present case, the actual source is much closer to another
star from the OGLE-IV catalog (constructed in 2010),
which is 810 mas from the one that triggered the OGLE alert.
Hence, it was just by bad luck that OGLE triggered on the ``wrong star''.
This does happen from time to time (also with KMT triggers), but it does
not cause any substantive problems.  Thus, while it would not make any
difference to the modeling, it might give a more accurate impression of
possible blended light to add in the nearer ($I\sim 19$) catalog star,
rather than the giant.  Nevertheless, we retain the standard OGLE reduction
at this stage because this giant is the cause of both the linear slope
and the additional effects that we will uncover.

The underlying cause of the trend is the proper motion of this bright star
away from the source.  By comparing KMT difference images from the event in
2017 to reference images in the same year, we find that at $t_0$, this catalog
star lay 835 mas, approximately due west, of the source.  Using {\it Gaia} DR3
\citep{gaia16,gaia18}, we find that in the
frame of the Galactic bulge (which basically sets the astrometric alignment
of the images before they are subtracted) the source is moving at the
relatively high speed of $\mu_\parallel=5.5\,\masyr$ parallel to this separation
vector.  Hence, as time goes on, less and less of the flux from this star is
in the tapered point spread function (PSF) aperture that is used to
measure the flux.  Approximating the PSF as a Gaussian, one easily
finds that flux falling in the tapered aperture, $F_{\rm app}$, is related
to the flux of this contaminating star, $F_{\rm star}$, by
\begin{equation}
F_{\rm app}=F_{\rm star} \exp\biggl[-\ln 4\biggl({\theta\over \fwhm}\biggr)^2\biggr],
  \label{eqn:app1}
\end{equation}
where FWHM is the seeing.
To zeroth order, this excess flux disappears when the reference image
is subtracted from the current image prior to doing difference-image
photometry.  However, at next order, one measures the change with time,
which is most conveniently expressed in magnitudes relative to $F_{\rm star}$,
\begin{equation}
  {d I_{\rm star}\over d t} = \mu_\parallel{d I_{\rm star}\over d \theta} =
  {\mu_\parallel\over F_{\rm star}}\,{-2.5\over \ln 10}{d F_{\rm app}\over d\theta}=
    10\log 2{\mu_\parallel\theta_0\over \fwhm^2}
    \exp\biggl[-\ln 4\biggl({\theta_0\over \fwhm}\biggr)^2\biggr],
  \label{eqn:app2}
\end{equation}
where $\theta_0$ is the evaluation of $\theta$ at some reference time,
such as the peak of the event, $t_0$.
Adopting the median OGLE seeing $\fwhm = 1.32^{\prime\prime}$, this yields
$d I_{\rm star}/dt = 4.56\,{\rm mmag\,\rm yr^{-1}}$, which can be compared to
the empirical value shown in Figure~\ref{fig:long} of
$d I_{\rm star}/dt = 4.89\,{\rm mmag\,\rm yr^{-1}}$.  This qualitative agreement
shows that we have a reasonably good analytic understanding of this effect.
%3.01 * 2.64e-3 * 0.574 = 4.56e-3.

Unfortunately, this understanding is not as good as one might hope.
While it is plausible that the predicted slope should not be identical
to the empirical slope (because the
PSF is not actually Gaussian), Equation~(\ref{eqn:app2}) predicts that the
magnitude of the offset at fixed time should vary $\propto x\exp(-x)$ where
$x\equiv \ln 4(\theta_0/\fwhm)^2$.  However, when we include this dependence
on $\fwhm$ in the fit, $\chi^2$ does not improve, but on the contrary gets
slightly worse.

This means that we would not have been able to blindly apply the results
from the OGLE fit to the KMT data sets, i.e., scaling by $x\exp(-x)$
according to the median $\fwhm$.  We note that if we fit for the slope
parameters for each observatory (as part of a general fit to a
microlensing model), using KMT data from only 2017,
the errors are 30 to 50 times larger for KMT than for OGLE.  This is about
what one would expect based on an effective time baseline that is 15 times
shorter and sample sizes that are, on average, 8 times smaller, i.e.,
$15\times\sqrt{8}\sim 40$.  Nevertheless, despite these large errors,
we found that the slopes for the six KMT observatory/field combinations,
were much closer to zero than to the slopes predicted by scaling to the
extremely well-measured, long term OGLE slope.

Motivated, by this apparent conflict, we investigated whether the
OGLE data actually showed any evidence for an intra-season slope, or
whether the slope seen in Figure~\ref{fig:long}  was entirely due to discrete,
downward jumps between seasons, despite the fact that such a model would
appear to be extremely ad hoc.  Surprisingly, we found that the data are
consistent with this seemingly bizarre hypothesis.  We carried out a linear
fit of the form $F(t) = \sum_{i=1}^3 a_i f_i(t)$ where
$f_1(t) = 1$, $f_2(t)={\rm nint}(Y(t))$, $f_3(t) = Y(t)-{\rm nint}(Y(t))$,
and $Y(t)\equiv ({\rm HJD}^\prime - 7900)/365.25$.  That is, a constant,
plus an annual-jump term, plus an intra-season-slope term.  We found,
$a_i=(5.46658,-0.02708,-0.0065)\pm(0.00070,0.00016,0.00260)$.  That is,
the intra-season slope was constrained (at $1\,\sigma$) to be $>10$ times
smaller than the multi-year slope, $|a_3|\ll \sigma_3 = 0.096|a_2|$.

A plausible physical cause is differential refraction: as the field moves
toward the East over a season, the image of the neighboring bright star
moves closer to the source, causing the measured flux to increase.
We confirm that this explanation is consistent by fitting for
a continuous slope and an hour-angle term.  We find that the continuous
term is virtually identical to $a_2$ above, while the hour-angle term
has the opposite sign and effectively cancels the continuous term during each
season.  Assuming that this explanation is correct, the cancellation must
be accidental.  That is, the continuous term is proportional to the
neighbor brightness and the proper motion, while the differential refraction
term depends on the neighbor brightness, color, and position.  Hence,
for example, if the proper motion had been double, but the other parameters
remained the same, the continuous term would double and the in-season
slope would be half of the long-term slope, rather than zero.

Because we are less confident of the multi-year stability of KMT photometry,
we cannot make such a precise measurement of the intra-season
slope for the KMT observatories as for OGLE.  However, as mentioned above,
what triggered our detailed investigation of the structure of the OGLE
baseline light curve was that the intra-season KMT slope is consistent
with zero.  Therefore, we adopt the following treatment of the data.
\begin{enumerate}
\item{OGLE}
  \item[]{Use only 2016, 2017, 2018 data.}
  \item[]{Add $\Delta F(t)=-0.027[{\rm Year}-2017]$ to each measurement.}
\item{KMT}
  \item[]{Use only 2017 data.}
  \item[]{Do not modify flux measurements}
\end{enumerate}

Finally, we remove the zero point of the OGLE baseline flux,
$I_{\rm  base}=16.156$, and replace it with $I_{\rm base} =20.000$.  This is
set somewhat brighter than the best-fit source magnitude $I_S=20.4$
to avoid negative total fluxes for measurements with downward
statistical fluctuations, which would prohibit display in
magnitude-based figures.  As emphasized above, the choice of baseline
flux has absolutely no effect on the analysis.  However, choosing a
baseline that is close to the source flux enables clearer simultaneous
visualization of the light-curve evolution at all magnifications,
which span a factor of order 100.

\section{Xallarap Detections of Low-Mass Companions to the Source}
\label{sec:bd-xallarap}

To date, microlensing experiments have discovered and published of order
200 substellar objects, i.e., planets and brown dwarfs (BDs).  While a few
of these are isolated BDs (e.g., OGLE-2007-BLG-0224, \citealt{ob07224})
or possibly-isolated planets \citep{mroz17,ffp,moaffp},
the overwhelming majority are companions to stars.

However, shortly after \cite{mao91} first
proposed lensing systems as a new channel for finding planetary and
binary systems, \citet{griest-hu} suggested that microlensing could
also be used to find and characterize binary stars.  \citet{griest-hu}
mainly focused on single-lens binary-source (1L2S) events in which
the binary-source could be approximated as static and thus the companion
was detected entirely via its changing magnification during the event.
However, in their Figure~12, they already give an example that betrays
dramatic dynamical effects, i.e., periodic oscillations.  They argue
that such examples will be very rare, and they acknowledge
that they had to adopt extraordinarily improbable parameters to generate
the oscillations shown in their figure.  Moreover, while they do not
say so explicitly, the form of their figure shows that both components
of the binary are luminous, indeed of equal brightness.

\citet{han97} considered a range of flux ratios for the two sources and showed
(their Figure~3) that the light-curve oscillations would be dramatically larger
if the companion was effectively dark (Panel (c)) than if the two components
were equally luminous (Panel (a)).  When their Figure 3c
was shown at a meeting, Chris Stubbs shouted out from the audience
``xallarap'', and that name has stuck.  Prior to that meeting, this
neologism had only been used for the nemesis of the ``Green Lantern'' DC Comics
character.

Although \citet{han97} had carried out their study with the aim of
developing methods to better characterize microlensing events (and so,
in particular, the lenses), the main practical impact of
microlensing xallarap over the ensuing years was (similar to its
DC-character namesake) as a nuisance.  It was soon realized that
any microlensing parallax signal (light curve oscillations due to
Earth's annual motion, \citealt{gould92}) could be duplicated
with infinite precision by xallarap, provided that the source companion
had the same period, eccentricity, and phase as Earth, and that the
inclination of the binary orbit was the same as the ecliptic latitude of the
event.  Thus, in a large fraction of events in which the microlens
parallax is apparently detected, some effort is made to argue that the
alternative xallarap interpretation is unlikely.

The main method for doing this was introduced by \citet{poindexter05}.
If one does a systematic search for xallarap solutions (typically with
the eccentricity $e\equiv 0$ because the data normally do not support
higher-order models) and the solution's three xallarap parameters
$(P,\alpha,\delta)$, i.e., the period, phase, and inclination, are well
localized near those predicted from Earth's orbit, then it is
extremely improbable that the effects are due to xallarap.  On the other
hand, such xallarap searches do sometimes yield solutions with much
better $\chi^2$ than the parallax solution and with parameters far from those
of Earth, e.g., $P=0.5\,$yr.  In such cases, the parallax solution is rejected
in favor of xallarap.

To understand the challenges of detecting substellar and other low-mass
lenses using xallarap,
we begin by expressing the binary-source companion-host mass ratio $Q$ in
terms of observables.  Combining Kepler's Third Law with Newton's Third Law,
we obtain
\begin{equation}
  {Q^3\over (1+Q)^2} = Z \equiv {4\pi^2 a_1^3\over G M_S P^2}=
  {(a_1/\au)^3\over (P/{\rm yr})^2 (M_S/ M_\odot)}
  \label{eqn:qdef}
\end{equation}
where $a_1$ is the semi-major axis of the source orbit about the center of
mass,
\begin{equation}
a_1 = \xi\theta_\e D_S.
  \label{eqn:a1def}
\end{equation}

Provided that $\theta_\e$ can be measured and $D_S$ can be reliably estimated,
then $Z$ is an empirically determined quantity.  For the case of substellar
companions, i.e., $Q\ll 1$, the solution to Equation~(\ref{eqn:qdef})
is very well approximated by
\begin{equation}
Q = Z^{1/3} + {2\over 3}Z^{2/3},
  \label{eqn:qsol}
\end{equation}
but is adequately approximated for what follows by $Q\rightarrow Z^{1/3}$.
Further, to avoid clutter in what follows, we simply assume that the source
mass and distance are $M_S=1\,M_\odot$ and $D_S=8\,\kpc$.  We also note that
an important point of principle is that for $Z$ to be an ``observable'',
$\theta_\e$ must be measured, which requires that the lens transit the source,
i.e., $\rho\la u_0$.  Because, in almost all cases, $\theta_*\ll\theta_\e$,
this implies that $\rho\ll 1$, and therefore $u_0\ll 1$.  Hence, we can assume
that the xallarap features will take place at lens-source normalized
separations $u\la 0.5$, where the magnification scales $A\simeq u^{-1}$ and
hence small displacements $\eta\ll u$ within the Einstein ring induce
magnification changes $\Delta A \sim \eta A^2$, and thus fractional flux
changes of $\Delta A/A\sim \eta A$.  For definiteness, let us assume
that these can be reliably detected and characterized if they are a least
$\epsilon = 3\%$.  As a benchmark, we characterize the requirement of
``low-mass companion'' as $Q<0.08$.  Then, applying the various
approximations just described, we infer that to be detectable, $Q$ is
constrained by
\begin{equation}
  Q = {\xi\theta_\e D_S/\au\over (P/{\rm yr})^{2/3}}
  \ga {8\epsilon\theta_\e/\mas \over A(P/{\rm yr})^{2/3}}
  \label{eqn:bigqeval}
\end{equation}
where $A$ is the magnification at which the xallarap effect is effectively
being detected, i.e., when the companion is at maximum projected separation,
and where we have made use of the identity $\mas=\au/\kpc$.
This will occur when $|t-t_0| = P/4$, i.e., $A = t_\e/|t-t_0| = 4t_\e/P$.
Noting that $\mu_\rel = \theta_\e/t_\e = 4\theta_\e/PA$, we obtain
\begin{equation}
  Q \ga 0.044{\epsilon\over 3\%}\biggl[{1\over A}
  \biggl({\theta_\e\over 0.1\mas}\biggr)
  \biggl({\mu_\rel\over \masyr}\biggr)^2\biggr]^{1/3}
  \label{eqn:qeval2}
\end{equation}

%8*eps*thE/mas/[A^{1/3}*(4*tE/P)^{2/3}*(P/yr)^{2/3}] =
%8/4^{2/3}*eps*[(thE/mas)*(mu/masyr)^2/A]^{1/3}
%8/4^{2/3}*eps--> 0.095: then with mas -> 0.1 mas => 0.095/10^{1/3} = 0.044

If we ignore for the moment the factor $A^{1/3}$, then
Equation~(\ref{eqn:qeval2}) implies that it is difficult to probe into
the low-mass regime and extremely difficult to probe the planetary
regime (nominally $Q\la 0.012$) regime.  This is because
events with either $\theta_\e\la 0.1\,\mas$ or $\mu_\rel\la 1\,\masyr$
are each rare, so the combination is very rare.  Moreover, only events
with $\theta_\e$ measurements can yield $Q$ determinations, and
these require $u_0 \la \rho$, which occurs with probability
$p\sim\theta_*/\theta_\e$ or typically $p\sim 0.6\% (\theta_\e/0.1\,\mas)$
for dwarf sources and $p\sim 6\% (\theta_\e/0.1\,\mas)$ for giant sources.
Finally, there must actually be low-mass companions in these relatively
short $P$ orbits.

To understand the potential for enhanced sensitivity from possible high
magnification, we restrict consideration to companions that are no closer than
3 stellar radii on the grounds that nearer ones are likely to be rare, i.e.,
$A<\theta_\e/3\theta_*$.  We can then rewrite Equation~(\ref{eqn:qeval2}) as
\begin{equation}
  Q \ga 0.025{\epsilon\over 3\%}\biggl[
  {\theta_*\over 6\,\muas}
  \biggl({\mu_\rel\over \masyr}\biggr)^2\biggr]^{1/3},
  \label{eqn:qeval3}
\end{equation}
where we have normalized the source radius to that of a typical first-ascent
giant\footnote{Note, however, that clump giants must be excluded because they
would have swallowed any such companions prior to the helium flash.}.
Thus, under the assumptions of this calculation, giant sources cannot
probe the planetary regime at all, while dwarf sources, being about 10 times
smaller, could marginally probe it.

%(thE/0.1mas)/A > 3 th*/100 uas = 0.18*(th*/6 uas) : 0.044*(0.18)^{1/3} = 0.025.

Thus, this method would not appear to be a promising one for learning about
hot low-mass companions.  And these poor prospects may have contributed
to the paucity of systematic investigations of the method.

Nevertheless, OGLE-2017-BLG-1777 appears to contain such a low-mass source
companion.  Adopting (for the moment) $D_S=8\,\kpc$ and $M_S=1\,M_\odot$, and
inserting $(P,\xi,\rho)$ from Table~\ref{tab:ob171777parms2} (and $\theta_*$
from Section~\ref{sec:cmd-ob171777}) into Equations~(\ref{eqn:qdef}) and
(\ref{eqn:a1def}), we obtain 
\begin{equation}
  Z^{1/3} = 0.0687\pm 0.0084;\qquad
  Q = 0.0718\pm 0.0088 \qquad ({\rm Local\ 1})
  \label{eqn:qevalsol1}
\end{equation}
and
\begin{equation}
  Z^{1/3} = 0.0769\pm 0.0120;\qquad
  Q = 0.0808\pm 0.0126 \qquad ({\rm Local\ 2})
  \label{eqn:qevalsol2}
\end{equation}
for the two solutions shown in Table~\ref{tab:ob171777parms2}.  We then
reinsert the dependence on $D_S$ and $M_S$ to find,
\begin{equation}
  {M_{\rm comp}({\rm Local\ 1})\atop M_{\rm comp}({\rm Local\ 2})} =
  {(0.0718\pm 0.0084)\atop(0.0808\pm 0.0126)}M_\odot\,{D_S\over 8\,\kpc}
  \biggr({M_S\over M_\odot}\biggr)^{2/3}.
  \label{eqn:mcompeval}
\end{equation}
That is, both solutions indicate source companions that
are very close to the star/BD boundary.

The predicted semi-amplitude of the RV of the source due to its companion is
$v\sin i = 2\pi (a_1/P)\cos(\delta_S)/\sqrt{1-e^2}\rightarrow 2\pi (a_1/P)$,
where we have suppressed the extremely small (and oppositely signed) 
corrections due to inclination and eccentricity.  Explicitly,
\begin{equation}
  v\sin i \simeq 2\pi{a_1\over P} =
  {(6.28\pm 0.77)\atop(7.03\pm 1.10)}\kms\,{D_S\over 8\,\kpc}, \qquad
  {({\rm Local\ 1})\atop ({\rm Local\ 2})}.
  \label{eqn:vsinieval}
\end{equation}
%    1e-5         1e-5         1e-5         1e-5
% xi=142+-010; rho=767+-63; thE=638+-64; a1=725+- 89 P=12.55; z13=687+- 84 
% xi=141+-013; rho=679+-74; thE=722+-91; a1=814+-127 P=12.59; z13=769+-120

% xi*A = dA/A > eps => xi > eps/A

%1/u - 1/(u + eta) = 1/u - 1/u(1 + eta/u) = A(1 - (1 - eta/u)= A^2 eta

\section{Are The Low Values of $\theta_\e$ and $\mu_\rel$ for OGLE-2017-BLG-1777 Connected?}
\label{sec:rho}

Because the detection of VLM stars and substellar objects via xallarap requires
a combination of very improbable characteristics, i.e., very small $\theta_\e$,
very small $\mu_\rel$, $\rho\la u_0$, as well as the presence of a close-in
VLM-star substellar companion to the source, one must be concerned that some
rare effect is artificially generating the appearance of one or more of these
characteristics.  In this Appendix, we investigate the circumstances that
could lead to simultaneous underestimates of $\theta_\e$ and $\mu_\rel$.

We begin by pointing out that the
parameter combination $\theta_\e\mu_\rel$ appears to be an ``invariant'', i.e.,
it seems to depend only on directly observable characteristics of the light
curve.  If correct, this would make the identification of such exceptional
events relatively immune to misinterpretation in that, even if $\theta_\e$
were systematically underestimated, it would only mean that $\mu_\rel$
was overestimated.  The formulation of this invariance will then allow
us to explore potential Achilles Heels to this comforting argument and to
investigate whether these apply to OGLE-2017-BLG-1777 in particular.

Although it is not essential, we assume that the photometry
has been transformed to a calibrated system.  This will avoid unnecessary
complications, and in any case, it is true of the event under examination.
We can then write,
%$\Delta F_\max\equiv F_{\rm peak}-F_{\rm base} = (A-1)f_S \simeq f_s/u_0$, and can therefore write
\begin{equation}
  \theta_\e\mu_\rel = \biggl({\theta_*\over t_*}\biggr)^2 t_\e =
        {f_{0,S} t_\e\over\pi S t_*^2}\simeq 10^{0.4\,A_I}
        {f_S t_\e\over \pi S t_*^2},
  \label{eqn:invariant}
\end{equation}
where $S=f_{0,S}/\pi\theta_*^2$ is the surface brightness, which depends only
on the dereddened source color, $(V-I)_{0,S} = (V-I)_S - E(V-I)$.
We examine the five parameters $(A_I,E(V-I),(V-I)_S,f_S t_\e, t_*)$
from the standpoint of ``what could go wrong'', keeping in mind that
pathological cases must be considered.

The extinction $A_I$ and reddening $E(V-I)$), are model-independent
because they are evaluated directly from the red-clump position on the
CMD, without reference to the microlensing data.  In principle, it is
possible for this evaluation to fail catastrophically if there is a
small dense cloud directly in front of the source, so that its
extinction is very different from the clump stars in its neighborhood.
This would be extremely rare, but the possibility must be considered
when the alternative is an extremely unlikely combination of
$\theta_\e$ and $\mu_\rel$.

The source color $(V-I)_S$ is model-independent because it can be derived
from regression of the $V$-band flux on the $I$-band flux.  Hence, one must
carefully check whether $(V-I)_S$ has been properly measured,
but this concern is not related to issues of a possibly incorrect model.

The parameter combination $f_S t_\e$ is an invariant for high-magnification
events, $A_\max \gg 1$.  We now show that this is because,
like $(V-I)_S$, it can be derived from model-independent regression of the data.
At intermediate magnification, $1\ll A\ll A_\max$,
(and excluding regions of the light curve that are impacted by short anomalies),
we have $u=\sqrt{u_0^2 + (t-t_0)^2/t_\e^2}\rightarrow |t-t_0|/t_\e$ and
$A\rightarrow 1/u\rightarrow t_\e/|t-t_0|$.  Therefore, the time evolution
of the observed flux relative to the baseline flux is:
$F(t_i) - f_{\rm base}=[A(t_i)-1]f_S \simeq A(t_i) f_S=f_St_\e/|t_i-t_0|$.
Thus, because
$F(t)$, $f_{\rm base}$, $t_i$ and $t_0$ are all direct observables, $f_S t_\e$
is tightly constrained.  Stated otherwise, $f_S t_\e$ can essentially be
evaluated as a regression of $F(t_i)$ on $|t_i - t_0|^{-1}$ at intermediate
magnifications.  While $f_S t_\e$ is never measured this way,
any model that fits the data will obey it.  Hence, $f_S t_\e$ is a robust,
model-independent invariant.

The source self-crossing time, $t_*=\rho t_\e$ is often called an ``invariant''
because it changes very little as $t_\e$ is held fixed at various different
values.  That is, within such a restricted class of models, the duration of
the ``bump'', $t_{\rm bump}$ that is caused by the source transiting a
caustic is related to $t_*$ by some definite factor,
$t_* = k t_{\rm bump}$, which is determined by the overall geometry.  Then,
because $t_{\rm bump}$ is given directly by the data, $t_*$ is invariant.
However, the factor $k$ can be different from one class of models to
another.  For example, if the form of the bump is that of a generic
caustic crossing (e.g., Figure~1 from \citealt{gouldandronov99}),
then $k=(\sin \psi)/2$, where $\psi$ is the angle of the source trajectory
relative to the caustic.  If the event has been modeled with a source
trajectory of $\psi=90^\circ$, but the true caustic crossing has
$\psi=10^\circ$, i.e., the long duration of $t_{\rm bump}$ has been attributed
to a slow source rather than an acute crossing, then $t_*$ would be
overestimated by a factor $\sin(90^\circ)/\sin(10^\circ)=5.8$.

We now examine how well each of these factors are constrained for the
case of OGLE-2017-BLG-1777.  The absolute and relative extinction, $A_I$
and $E(V-I)$ are not major concerns.  If there were a small dust cloud
in front of the source, then it would be intrinsically bluer than we have
inferred.  However, it is already near the limit of
the observed color range for bulge
dwarfs.  Moreover, unless this cloud had very unusual properties, the
net effect of making the source brighter and bluer would be reduce $\theta_*$,
relative to our estimate, which is the opposite of what would need to
explain the small value of $\theta_\e\mu_\rel$.  In principle, there could
be a hole in the dust, but this is extraordinarily unlikely because it
would require a strongly correlated behavior among physically independent
dust clouds along the line of sight.

The $(V-I)_S$ source color is measured from two different sites, each
to good statistical precision based on several magnified $V$-band points,
and in good agreement with each other.

The argument for the invariance of $f_S t_\e$ could in principle be
undermined for cases of strong parallax because the approximation
$A\rightarrow 1/u\rightarrow t_\e/|t-t_0|$ would no longer be strictly valid.
See, e.g., \citet{smp03}.  Nevertheless, we note that for the collection of
very different physical models in Tables~\ref{tab:ob171777parms1} and
\ref{tab:ob171777parms2}, some with and some without
parallax, we have $f_S t_\e = (6.99,6.78,7.19,7.48)\,$day, where we have
expressed the source flux $f_S$ on an $I=18$ scale.  The full range
of this variation is only $\sim 10\%$, despite the large parallax of one
solution.

Hence, the major potential issue is $t_*$, for which the concern is
heightened by the fact that the light curve does not exhibit either of
the two classical forms, i.e., those associated with fold caustics
(e.g., Figure 1 from \citealt{gouldandronov99}) or with point caustics
(e.g., Figure 1 from \citealt{ob07224}).  Hence, one should first be
concerned that finite-source effects are detected at all.  One
piece of evidence that they are is that $\rho$ is measured to $\la 10\%$
in all the models.  We further explored the possibility that the $\rho$
measurements were spurious by fitting to $\rho=0$ models.  However, these
models completely failed to match the observed light curve near the peak.

An additional concern for xallarap models is that the ``directly observable''
duration of the finite-source effects can be affected by internal motion
of the source relative to its center of mass.  In the present case, we
also have, for the final model, orbital motion of the lens, which
can also produce this effect.  As an overall check we can compare the $t_*$
values for the various models, which are shown in
Tables~\ref{tab:ob171777parms1} and
\ref{tab:ob171777parms2}.  Indeed,
we find a full range of variation of about a factor 1.6.  The largest
difference is between models that include or exclude xallarap, showing
that internal motion can be a big effect.  There is also a factor 1.4
difference between 2L1S and 1L1S models that both allow for xallarap.
Hence, the difference largely stems from the ``cusp entrance'' form of the
first, versus the ``point caustic'' form of the second.

Thus, as expected, among the five factors, the one that deviates most
from ``invariance'' is $t_*$.  Nevertheless, its range of variation, allowing
for very different physical models, is still modest compared to the
extreme values of $\theta_*$ and $\mu_\rel$.

We conclude that these extreme values are unlikely to have been generated
by problems in the modeling.

\input tabnames

\input ob1275parms.tex

\input ob1275spitzer.tex

\input ob1275pp.tex

\input ob1275mm.tex

\input ob0640parms.tex

\input ob1237parms.tex

\input ob1777parms1.tex

\input ob1777parms2.tex

\input ob0543parms.tex

\input ob1694parmsplan.tex

\input ob1694parmsbd.tex

\input ob1694parmsstar.tex

\input tabcmd

\input tabphysall

\input tab2017

\clearpage

\end{document}

%% file: author.tex
\author{\textsc{
Yoon-Hyun Ryu$^{1}$, 
Andrzej Udalski$^{2}$,
Jennifer C. Yee$^{3}$, 
Weicheng Zang$^{3,4}$,
Yossi Shvartzvald$^{5}$,
Cheongho Han$^{6}$
Andrew Gould$^{7,8}$\\ 
(Lead Authors)\\
Michael D. Albrow$^{9}$, 
Sun-Ju Chung$^{1,3}$, 
Kyu-Ha Hwang$^{1}$, 
Youn Kil Jung$^{1,11}$, 
In-Gu Shin$^{3}$, 
Hongjing Yang$^{4}$, 
Sang-Mok Cha$^{1,10}$, 
Dong-Jin Kim$^{1}$,
Seung-Lee Kim$^{1}$, 
Chung-Uk Lee$^{1}$,
Dong-Joo Lee$^{1}$,
Yongseok Lee$^{1,10}$, 
Byeong-Gon Park$^{1,11}$, 
Richard W. Pogge$^{8}$,
Hanyue Wang$^{3}$  \\
(The KMTNet Collaboration)\\
Przemek Mr{\'o}z$^{2}$,
Micha{\l} K. Szyma{\'n}ski$^{2}$,
Jan Skowron$^{2}$,
Radek Poleski$^{2}$,
Igor Soszy{\'n}ski$^{2}$,  
Pawe{\l} Pietrukowicz$^{2}$,
Szymon Koz{\l}owski$^{2}$,
Krzysztof Ulaczyk$^{12}$,
Krzysztof A. Rybicki$^{2,5}$,
Patryk Iwanek$^{2}$,
Marcin Wrona$^{2}$\\
(The OGLE Collaboration)\\
Charles Beichman$^{13}$,
Geoffry Bryden$^{14}$,
Sean Carey$^{13}$,
Calen B.~Henderson$^{13}$,
Sebastiano Calchi Novati$^{13}$,
Wei Zhu$^{4}$\\
(The Spitzer Team)\\
Savannah Jacklin$^{15}$,
Matthew T.~Penny$^{16}$\\
(The UKIRT Team)}}

%----------------------------------------------------------------
\affil{$^{1}$Korea Astronomy and Space Science Institute, Daejon
34055, Republic of Korea}

\affil{$^{2}$Astronomical Observatory, University of Warsaw, 
Al.~Ujazdowskie~4, 00-478~Warszawa, Poland}

\affil{$^{3}$ Center for Astrophysics $|$ Harvard \& Smithsonian, 60 Garden
St., Cambridge, MA 02138, USA}

\affil{$^{4}$ Department of Astronomy,
Tsinghua University, Beijing 100084, China}

\affil{$^{5}$Department of Particle Physics and Astrophysics, 
Weizmann Institute of Science, Rehovot 76100, Israel}

\affil{$^{6}$Department of Physics, Chungbuk National University,
Cheongju 28644, Republic of Korea}

\affil{$^{7}$Max-Planck-Institute for Astronomy, K\"{o}nigstuhl 17,
69117 Heidelberg, Germany}

\affil{$^{8}$Department of Astronomy, Ohio State University, 140 W.
18th Ave., Columbus, OH 43210, USA}

\affil{$^{9}$University of Canterbury, Department of Physics and
Astronomy, Private Bag 4800, Christchurch 8020, New Zealand}

\affil{$^{10}$School of Space Research, Kyung Hee University,
Yongin, Kyeonggi 17104, Republic of Korea}

\affil{$^{11}$Korea University of Science and Technology, Korea, 
(UST), 217 Gajeong-ro, Yuseong-gu, Daejeon, 34113, Republic of Korea}

\affil{$^{12}$Department of Physics, University of Warwick, 
Gibbet Hill Road, Coventry, CV4~7AL,~UK}

\affil{$^{13}$IPAC, Mail Code 100-22, Caltech, 1200 E. California Blvd., Pasadena, CA 91125}

\affil{$^{14}$Jet Propulsion Laboratory, California Institute of Technology, 4800 Oak Grove Drive, Pasadena, CA 91109, USA}

\affil{$^{15}$Vanderbilt University, Department of Physics \& Astronomy, Nashville, TN 37235, USA}

\affil{$^{16}$Department of Physics and Astronomy, Louisiana State University,
Baton Rouge, LA 70803, USA}

%\affil{$^{SS}$School of Physics and Astronomy, Tel-Aviv University, Tel-Aviv 6997801, Israel}

%% file: tabnames.tex
 \begin{deluxetable}{llrrrrr}
 \tablecolumns{7} \tablewidth{0pc}
 \tablecaption{\textsc{Event Names, Cadences, Alerts, and Locations}}
 \tablehead{\colhead{Name} & 
\colhead{$\Gamma\,({\rm hr}^{-1})$} &
\colhead{Alert Date} &
\colhead{RA$_{\rm J2000}$} &
\colhead{Dec$_{\rm J2000}$} &
\colhead{$l$} &
\colhead{$b$} }
%\hline
 \startdata
\hline
OGLE-2017-BLG-1275& 0.8 & 10 Jul 2017 & 17:51:39.70 & $-29$:29:52.69  & $+0.20$  & $-1.42$  \\  % nataf 14.43 NO RHO, YES OIII
KMT-2017-BLG-0314 & 4.0 \\ % A_I=2.81
\hline
OGLE-2017-BLG-0640& 0.8 & 13 May 2017 & 17:49:51.63 & $-29$:27:41.40  & $+0.03$  & $-1.06$  \\ % nataf 14.44 NO RHO, YES OIII, but not useful, use UKIRT
KMT-2017-BLG-1726 & 4.0 \\ % A_I=4.69 
\hline
OGLE-2017-BLG-1237& 0.35 & 01 Jul 2017 & 18:06:35.99 & $-28$:19:43.79  & $+2.84$  & $-3.67$  \\ % nataf 14.35  YES RHO, YES OIII
KMT-2017-BLG-0422 & 2.4 \\ % A_I=1.01 "by-eye" "class: binary"
\hline
OGLE-2017-BLG-1777& 0.8 & 14 Oct 2017 & 17:53:23.25 & $-29$:54:36.00  & $+0.04$  & $-1.95$  \\ % nataf 14.44  YES RHO, YES OIII
KMT-2017-BLG-0282 & 4.0 \\ % A_I=1.54 "by-eye" "class: binary"
\hline
OGLE-2017-BLG-0543& 0.13 & 13 Apr 2017 & 17:51:19.96 & $-32$:02:29.69  & $-2.02$  & $-2.66$  \\% nataf 14.55 NO RHO, NO OIII
KMT-2017-BLG-0140 & 4.0 \\ % A_I=2.24 "possible planet"
\hline
OGLE-2017-BLG-1694& 3.0& 06 Sep 2017 & 17:59.14.60 & $-28$:34:48.00  & $-1.83$  & $-2.39$  \\ % nataf \14.55  NO RHO YES OIII
KMT-2017-BLG-2126 & 4.0 \\ % A_I=1.30 "possible planet"

\hline
 \enddata
% \tablecomments{a: Manually alerted by MOA, based on OGLE alert}
 \label{tab:names}
 \end{deluxetable}

%% file: ob1275parms.tex
\begin{deluxetable}{lccc}
\tablecolumns{4} \tablewidth{0pc} \tablecaption{\textsc{Ground-only Parameters for OGLE-2017-BLG-1275}} \tablehead{\colhead{Parameters} & \colhead{inner} & \colhead{outer} & \colhead{wide} } \startdata
  $\chi^2/\rm{dof}$             &9823.68/9824              &9823.67/9824               &9858.66/9824\\
  $t_0-2457940$                 &9.962 $\pm$ 0.017          &9.950 $\pm$ 0.017           &9.918 $\pm$ 0.017  \\
  $u_0$                         &0.395 $\pm$ 0.020          &0.389 $\pm$ 0.019           &0.399 $\pm$ 0.017      \\
  $t_{\rm E}$ $(\rm{days})$     &13.41 $\pm$ 0.42         &13.51 $\pm$ 0.42          &13.00 $\pm$ 0.34  \\
  $s$                           &0.631 $\pm$ 0.014          &1.092 $\pm$ 0.040           &1.254 $\pm$ 0.009       \\
  $q$ $(10^{-3})$               &8.25 $\pm$ 1.31          &9.13 $\pm$ 1.69           &1.30 $\pm$ 0.23       \\
  log $q$ (mean)                &-2.084 $\pm$ 0.069         &-2.041 $\pm$ 0.080          &-2.889 $\pm$ 0.076   \\
  $\alpha$ $(\rm{rad})$         &4.758 $\pm$ 0.014          &4.755 $\pm$ 0.013           &1.718 $\pm$ 0.005   \\
  $\rho$ $(10^{-2})$            &$3.4_{-2.1}^{+2.7}$  &$3.9_{-2.5}^{+3.2}$   &$0.17_{-0.11}^{+0.17}$  \\
  $I_S$ [OGLE]            &19.358 $\pm$ 0.074         &19.373 $\pm$ 0.074          &19.277 $\pm$ 0.061    \\
  $I_B$ [OGLE]            &$21.09_{-0.31}^{+0.46}$ &$21.02_{-0.28}^{+0.42}$  &$21.62_{-0.42}^{+0.73}$            \\
%  $t_*$ $(\rm{hours})$          &$11.0_{-6.9}^{+8.7}$ &$12.6_{-8.0}^{+10.3}$ &$0.54_{-0.36}^{+0.52}$   \\
\enddata
%\tablecomments{}
\label{tab:ob171275parms}
\end{deluxetable}

%% file: ob1275spitzer.tex
\begin{deluxetable}{lcccc}
  \tablecolumns{5} \tablewidth{0pc} \tablecaption{\textsc{Spitzer-``only'' Parameters for OGLE-2017-BLG-1275}}
  \tablehead{\colhead{Parameters} & \colhead{inner (+,+)} & \colhead{outer (+,+)} & \colhead{inner (-,-)} & \colhead{outer (-,-)}} \startdata
  $\chi^2/\rm{dof}$         &19.71/13                  &13.00/13                  &19.75/13                  &13.34/13\\
  $\pi_{\rm{E},\it{N}}$     &-0.059 $\pm$ 0.029         &-0.084 $\pm$ 0.022         &0.062 $\pm$ 0.028          &0.089 $\pm$ 0.022\\
  $\pi_{\rm{E},\it{E}}$     &0.027 $\pm$ 0.015          &0.022 $\pm$ 0.009          &0.022 $\pm$ 0.017          &0.013 $\pm$ 0.011\\
  $L_S$ [Spitzer]           &17.17 $\pm$ 0.20         &17.30 $\pm$ 0.16         &17.17 $\pm$ 0.20         &17.33 $\pm$ 0.17 \\
  $L_B$ [Spitzer]           &$16.60_{-0.14}^{+0.20}$ &$16.52_{-0.11}^{+0.14}$ &$16.60_{-0.13}^{+0.20}$ &$16.50_{-0.10}^{+0.14}$\\
  \enddata
%\tablecomments{}
\label{tab:1275spitzer}
\end{deluxetable}

%% file: ob1275pp.tex
\begin{deluxetable}{lcccc}
  \tablecolumns{5} \tablewidth{0pc} \tablecaption{\textsc{OGLE-2017-BLG-1275 (+,+) Parameters for Ground+Spitzer Data}}
  \tablehead{\colhead{Parameters} & \colhead{inner (+,+)} & \colhead{outer (+,+)} & \colhead{inner-FREE (+,+)} & \colhead{outer-FREE (+,+)} } \startdata
  $\chi^2/\rm{dof}$             &9847.41/9838              &9844.09/9838               &9843.13/9837              &9836.90/9837\\
  $t_0-2457940$                 &9.962 $\pm$ 0.017          &9.951 $\pm$ 0.017           &9.961 $\pm$ 0.018          &9.949 $\pm$ 0.017\\
  $u_0$                         &0.395 $\pm$ 0.020          &0.390 $\pm$ 0.019           &0.397 $\pm$ 0.020          &0.392 $\pm$ 0.020\\
  $t_{\rm E}$ $(\rm{days})$     &13.41 $\pm$ 0.42         &13.50 $\pm$ 0.42          &13.38 $\pm$ 0.42         &13.44 $\pm$ 0.42\\
  $s$                           &0.630 $\pm$ 0.013          &1.089 $\pm$ 0.040           &0.630 $\pm$ 0.013          &1.089 $\pm$ 0.039\\
  $q$ $(10^{-3})$               &8.36 $\pm$ 1.34          &9.11 $\pm$ 1.63           &8.35 $\pm$ 1.28          &9.08 $\pm$ 1.60\\
  log $q$ (mean)                &-2.079 $\pm$ 0.069         &-2.041 $\pm$ 0.078          &-2.080 $\pm$ 0.067         &-2.043 $\pm$ 0.077\\
  $\alpha$ $(\rm{rad})$         &4.755 $\pm$ 0.014          &4.753 $\pm$ 0.013           &4.756 $\pm$ 0.013          &4.754 $\pm$ 0.013\\
  $\rho$ $(10^{-2})$            &$3.0_{-1.9}^{+2.7}$  &$3.8_{-2.4}^{+3.3}$   &$2.7_{-1.8}^{+2.5}$  &$3.4_{-2.1}^{+2.9}$\\
  $\pi_{\rm{E},\it{N}}$        &$-0.025$ $\pm$ 0.015       &$-0.032$ $\pm$ 0.016       &$-0.054$ $\pm$ 0.032       &$-0.083$ $\pm$ 0.024\\
  $\pi_{\rm{E},\it{E}}$         &0.036 $\pm$ 0.017          &0.032 $\pm$ 0.014           &0.029 $\pm$ 0.017          &0.023 $\pm$ 0.011\\
  $L_S$ [Spitzer]               &16.92 $\pm$ 0.11         &16.96 $\pm$ 0.11          &17.11 $\pm$ 0.23         &17.31 $\pm$ 0.19\\
  $L_B$ [Spitzer]               &16.84 $\pm$ 0.12         &16.81 $\pm$ 0.12          &$16.64_{-0.16}^{+0.23}$ &$16.51_{-0.12}^{+0.15}$\\
  $I_S$ [OGLE]   &19.359 $\pm$ 0.074         &19.371 $\pm$ 0.073          &19.352 $\pm$ 0.076         &19.362 $\pm$ 0.074\\
  $I_B$ [OGLE]        &$21.08_{-0.30}^{+0.46}$ &$21.03_{-0.28}^{+0.42}$  &$21.12_{-0.32}^{+0.48}$ &$21.07_{-0.29}^{+0.45}$\\
%  $t_*$ $(\rm{hours})$      &$9.718_{-6.083}^{+8.485}$  &$12.279_{-7.785}^{+10.535}$ &$8.645_{-5.651}^{+8.111}$  &$11.095_{-6.860}^{+9.126}$\\
\enddata
%\tablecomments{}
\label{tab:ob171275pp}
\end{deluxetable}

%% file: ob1275mm.tex
\begin{deluxetable}{lcccc}
  \tablecolumns{5} \tablewidth{0pc} \tablecaption{\textsc{OGLE-2017-BLG-1275 $(-,-)$ Parameters for Ground+Spitzer Data}}
  \tablehead{\colhead{Parameters} & \colhead{inner $(-,-)$} & \colhead{outer $(-,-)$} & \colhead{inner-FREE $(-,-)$} & \colhead{outer-FREE $(-,-)$} } \startdata
  $\chi^2/\rm{dof}$             &9847.18/9838              &9845.24/9838               &9843.58/9837              &9836.81/9837\\
  $t_0-2457940$                 &9.963 $\pm$ 0.018          &9.951 $\pm$ 0.017           &9.962 $\pm$ 0.018          &9.950 $\pm$ 0.017\\
  $u_0$                         &$-0.396$ $\pm$ 0.021         &$-0.390$ $\pm$ 0.020          &$-0.397$ $\pm$ 0.020      &$-0.392$ $\pm$ 0.020\\
  $t_{\rm E}$ $(\rm{days})$     &13.38 $\pm$ 0.42         &13.48 $\pm$ 0.42          &13.37 $\pm$ 0.41         &13.44 $\pm$ 0.42\\
  $s$                           &0.631 $\pm$ 0.014          &1.089 $\pm$ 0.039           &0.630 $\pm$ 0.013          &1.089 $\pm$ 0.038\\
  $q$ $(10^{-3})$               &8.29 $\pm$ 1.30          &9.12 $\pm$ 1.60           &8.29 $\pm$ 1.27          &9.08 $\pm$ 1.58\\
  log $q$ (mean)                &$-2.082$ $\pm$ 0.068         &$-2.041$ $\pm$ 0.076          &$-2.082$ $\pm$ 0.067     &$-2.043$ $\pm$ 0.076\\
  $\alpha$ $(\rm{rad})$         &$-4.755$ $\pm$ 0.013      &$-4.752$ $\pm$ 0.013        &$-4.756$ $\pm$ 0.013       &$-4.754$ $\pm$ 0.013\\
  $\rho$ $(10^{-2})$            &$3.3_{-2.1}^{+2.7}$  &$3.8_{-2.4}^{+3.2}$   &$2.9_{-1.8}^{+2.6}$  &$3.5_{-2.2}^{+2.8}$\\
  $\pi_{\rm{E},\it{N}}$         &0.027 $\pm$ 0.015          &0.034 $\pm$ 0.016           &0.060 $\pm$ 0.030          &0.087 $\pm$ 0.024\\
  $\pi_{\rm{E},\it{E}}$         &0.034 $\pm$ 0.017          &0.029 $\pm$ 0.015           &0.022 $\pm$ 0.018          &0.014 $\pm$ 0.012\\
  $L_S$ [Spitzer]               &16.91 $\pm$ 0.11         &16.95 $\pm$ 0.11          &17.14 $\pm$ 0.22         &17.31 $\pm$ 0.19\\
  $L_B$ [Spitzer]               &16.85 $\pm$ 0.13         &16.82 $\pm$ 0.12          &$16.63_{-0.15}^{+0.23}$ &$16.51_{-0.12}^{+0.15}$\\
  $I_S$ [OGLE]          &19.354 $\pm$ 0.075         &19.369 $\pm$ 0.074          &19.352 $\pm$ 0.073         &19.361 $\pm$ 0.074\\
  $I_B$ [OGLE]        &$21.11_{-0.31}^{+0.48}$ &$21.04_{-0.29}^{+0.43}$  &$21.12_{-0.31}^{+0.48}$ &$21.07_{-0.30}^{+0.45}$\\
%  $t_*$ $(\rm{hours})$   &$10.708_{-6.834}^{+8.750}$ &$12.347_{-7.770}^{+10.244}$ &$9.244_{-5.831}^{+8.457}$  &$11.296_{-7.136}^{+8.972}$\\
\enddata
%\tablecomments{}
\label{tab:ob171275mm}
\end{deluxetable}

%% file: ob0640parms.tex
\begin{deluxetable}{lc}
\tablecolumns{2} \tablewidth{0pc} \tablecaption{\textsc{Microlens Parameters for OGLE-2017-BLG-0640}} \tablehead{\colhead{Parameters} & \colhead{2L1S}  } \startdata
  $\chi^2/\rm{dof}$             &10737.638/10738              \\
  $t_0-2457880$                 &7.918 $\pm$ 0.029          \\
  $u_0$                         &0.658 $\pm$ 0.034          \\
  $t_{\rm E}$ $(\rm{days})$     &14.382 $\pm$ 0.471         \\
  $s$                           &0.607 $\pm$ 0.010          \\
  $q$ $(10^{-3})$               &4.884 $\pm$ 0.266          \\
  log $q$ (mean)                &-2.312 $\pm$ 0.024         \\
  $\alpha$ $(\rm{rad})$         &4.136 $\pm$ 0.010          \\
  $\rho$ $(10^{-2})$            &$2.044_{-1.284}^{+1.679}$  \\
  $I_S$ [OGLE]            &19.587 $\pm$ 0.094         \\
  $I_B$ [OGLE]            &$22.14_{-0.59}^{+1.09}$ \\
%  $t_*$ $(\rm{hours})$          &$7.045_{-4.418}^{+5.794}$ \\
\enddata
%\tablecomments{}
\label{tab:ob170640parms}
\end{deluxetable}

%% file: ob1237parms.tex
\begin{deluxetable}{lc}
\tablecolumns{2} \tablewidth{0pc} \tablecaption{\textsc{Microlens Parameters for OGLE-2017-BLG-1237}} \tablehead{\colhead{Parameters} & \colhead{2L1S}  } \startdata
  $\chi^2/\rm{dof}$             &6616.58/6617              \\
  $t_0-2457930$                 &5.888 $\pm$ 0.004          \\
  $u_0$ $(10^{-2})$             &$1.766_{-0.108}^{+0.075}$          \\
  $t_{\rm E}$ $(\rm{days})$     &$27.94_{-1.13}^{+1.74}$         \\
  $s$                           &1.041 $\pm$ 0.001          \\
  $q$ $(10^{-3})$               &$7.93_{-0.46}^{+0.35}$          \\
  log $q$ (mean)                &-2.105 $\pm$ 0.024         \\
  $\alpha$ $(\rm{rad})$         &4.075 $\pm$ 0.005          \\
  $\rho$ $(10^{-4})$            &$7.66_{-0.47}^{+0.35}$  \\
  $I_S$ [OGLE]                  &$21.865_{-0.046}^{+0.068}$         \\
  $I_B$ [OGLE]                  &15.958 $\pm$ 0.001 \\
  $t_*$ $(\rm{hours})$          &0.514 $\pm$ 0.007 \\
\enddata
%\tablecomments{tab:171237parms}
\label{tab:171237parms}
\end{deluxetable}

%% file: ob1777parms1.tex
\begin{deluxetable}{lcccc}
\tablecolumns{5} \tablewidth{0pc} \tablecaption{\textsc{Intermediate-Model Parameters for OGLE-2017-BLG-1777}} \tablehead{\colhead{}
                     &\colhead{2L1S}   & \colhead{2L1S}     & \colhead{1L1S}     &\colhead{2L1S} \\
\colhead{Parameters} &\colhead{Static} &\colhead{Parallax}  &\colhead{Xallarap}  &\colhead{Xallarap}} \startdata
  $\chi^2/\rm{dof}$             &12777.22/11900      &12453.50/11898      &11922.83/11896              &11901.61/11893  \\
  $t_0-2458040$                 &1.184 $\pm$ 0.002    &1.169 $\pm$ 0.002    &1.352 $\pm$ 0.003            &$1.255_{-0.025}^{+0.013}$ \\
  $u_0$ $(10^{-2})$             &1.016 $\pm$ 0.002    &1.987 $\pm$ 0.046    &0.814 $\pm$ 0.025            &0.714 $\pm$ 0.023  \\
  $t_{\rm E}$ (days)            &60.40 $\pm$ 0.91   &31.74 $\pm$ 0.53   &81.36 $\pm$ 2.40           &85.89 $\pm$ 3.03  \\
  $s$                           &0.796 $\pm$ 0.004    &0.801 $\pm$ 0.003    &-                            &$0.744_{-0.056}^{+0.041}$ \\
  $q$ $(10^{-3})$               &6.08 $\pm$ 0.14    &11.93 $\pm$ 0.29   &-                            &$0.86_{-0.20}^{+0.38}$ \\
  log $q$ (mean)                &-2.216 $\pm$ 0.010   &-1.923 $\pm$ 0.010   &-                            &-3.045 $\pm$ 0.129 \\
  $\alpha$ $(\rm{rad})$         &2.983 $\pm$ 0.001    &2.949 $\pm$ 0.003    &-                            &3.003 $\pm$ 0.025   \\
  $\rho$ $(10^{-3})$            &13.33 $\pm$ 0.21   &24.63 $\pm$ 0.37   &8.24 $\pm$ 0.27            &5.78 $\pm$ 032  \\
  $\pi_{\rm{E},\it{N}}$         &-                    &1.371 $\pm$ 0.060    &-                            &-  \\
  $\pi_{\rm{E},\it{E}}$         &-                    &1.835 $\pm$ 0.078    &-                            &-   \\
  $\xi_{\it{N}}$ $(10^{-2})$    &-                    &-                    &-0.829 $\pm$ 0.151           &$-0.267_{-0.191}^{+0.289}$ \\
  $\xi_{\it{E}}$ $(10^{-2})$    &-                    &-                    &1.348 $\pm$ 0.045            &1.098 $\pm$ 0.065 \\
  $\alpha_S$                    &-                    &-                    &305.0 $\pm$ 23.66         &$334.5_{-25.1}^{+14.8}$  \\
  $\delta_S$                    &-                    &-                    &$-15.4_{-7.2}^{+5.4}$  &$6.1_{-11.6}^{+4.9}$  \\
  ellipticity                   &-                    &-                    &0.055 $\pm$ 0.026            &$0.114_{-0.040}^{+0.050}$  \\
  phase                         &-                    &-                    &0.400 $\pm$ 0.071            &$0.367_{-0.041}^{+0.073}$  \\
  period (days)                 &-                    &-                    &12.75 $\pm$ 0.14           &12.62 $\pm$ 0.16   \\
  $I_S$ [OGLE]                  &20.390 $\pm$ 0.018   &19.674 $\pm$ 0.021   &20.665 $\pm$ 0.032           &$20.695_{-0.031}^{+0.033}$  \\
  $t_*$ $(\rm{hours})$          &19.32 $\pm$ 0.10   &18.77 $\pm$ 0.13   &16.06 $\pm$ 0.21           &11.93 $\pm$ 0.42  \\
\enddata
%\tablecomments{}
\label{tab:ob171777parms1}
\end{deluxetable}

%% file: ob1777parms2.tex
\begin{deluxetable}{lcc}
  \tablecolumns{3} \tablewidth{0pc} \tablecaption{\textsc{Final-Model Parameters for OGLE-2017-BLG-1777}}
  \tablehead{\colhead{Parameters} & \colhead{2L1S (Local 1)} & \colhead{2L1S (Local 2)}}% \\
%\colhead{} &\colhead{Xallarap} &\colhead{Xallarap}\\
%\colhead{} &\colhead{Parallax} &\colhead{Parallax}\\
%\colhead{Parameters} &\colhead{Orbital motion} &\colhead{Orbital motion}}
\startdata
  $\chi^2/\rm{dof}$             &11886.17/11889             &11886.25/11889\\
  $t_0-2458040$                 &$1.251_{-0.022}^{+0.015}$   &1.224 $\pm$ 0.018\\
  $u_0$ $(10^{-2})$             &0.945 $\pm$ 0.077           &0.924 $\pm$ 0.072\\
  $t_{\rm E}$ (days)            &65.3 $\pm$ 5.0          &68.0 $\pm$ 5.3\\
  $s$                           &0.732 $\pm$ 0.031           &0.704 $\pm$ 0.030\\
  $q$ $(10^{-3})$               &$1.17_{-0.27}^{+0.40}$   &1.63 $\pm$ 0.44\\
  $\log q$ (mean)                &-2.93 $\pm$ 0.12          &-2.79 $\pm$ 0.12 \\
  $\alpha$ $(\rm{rad})$         &3.015 $\pm$ 0.017           &3.008 $\pm$ 0.014\\
  $\rho$ $(10^{-3})$            &7.67 $\pm$ 0.63           &6.97 $\pm$ 0.62 \\
  $\pi_{\rm{E},\it{N}}$         &0.09 $\pm$ 0.29           &0.02 $\pm$ 0.25 \\
  $\pi_{\rm{E},\it{E}}$         &$0.246_{-0.080}^{+0.120}$   &0.230 $\pm$ 0.090 \\
  $ds/dt$ $({\rm yr}^{-1})$      &$3.15_{-0.46}^{+0.35}$   &$3.63_{-0.67}^{+0.50}$ \\
  $d\alpha/dt$ $({\rm yr}^{-1})$ &0.45 $\pm$ 0.57           &$-1.20_{-0.56}^{+0.38}$\\
  $\xi_{\it{N}}$ $(10^{-2})$    &$-0.259_{-0.132}^{+0.205}$  &-0.074 $\pm$ 0.199\\
  $\xi_{\it{E}}$ $(10^{-2})$    &1.395 $\pm$ 0.092           &1.416 $\pm$ 0.096\\
  $\alpha_S$                    &324.0 $\pm$ 21.6        &336.0 $\pm$ 11.6\\
  $\delta_S$                    &$-4.5_{-6.7}^{+8.3}$  &$7.2_{-6.1}^{+3.8}$\\
  ellipticity                   &0.112 $\pm$ 0.038           &$0.098_{-0.032}^{+0.039}$\\
  phase                         &0.386 $\pm$ 0.056           &0.363 $\pm$ 0.032\\
  period (days)                 &12.55 $\pm$ 0.15          &12.58 $\pm$ 0.16\\
  $I_S$ [OGLE-IV]                  &20.394 $\pm$ 0.090          &20.407 $\pm$ 0.086\\
  $t_*$ $({\rm hours})$         & 12.00 $\pm$ 0.51          &11.40 $\pm$ 0.48\\
  $\beta$                       & $0.35_{-0.11}^{+0.14}$  & $0.51_{-0.20}^{+0.16}$ \\
\enddata
%\tablecomments{}
\label{tab:ob171777parms2}
\end{deluxetable}

%% file: ob0543parms.tex
\begin{deluxetable}{lccc}
	\tablecolumns{4} \tablewidth{0pc} \tablecaption{\textsc{Microlens Parameters for OGLE-2017-BLG-0543}} \tablehead{\colhead{Parameters}
		& \colhead{outer}  & \colhead{inner} & \colhead{1L2S} }\startdata
	$\chi^2/\rm{dof}$             &8614.517/8615        &8631.408/8615                &8617.388/8614              \\
	$t_0-2457870$                 &0.882 $\pm$ 0.050    &1.040 $\pm$ 0.042            &0.791 $\pm$ 0.057          \\
	$u_0$                         &1.251 $\pm$ 0.003    &1.242 $\pm$ 0.003            &$1.276_{-0.006}^{+0.007}$  \\
	$t_{\rm E}$ $(\rm{days})$     &12.080 $\pm$ 0.061   &12.083 $\pm$ 0.061           &12.104 $\pm$ 0.077         \\
	$s$                           &1.518 $\pm$ 0.027    &2.185 $\pm$ 0.037            &-          \\
	$q$ $(10^{-3})$               &4.250 $\pm$ 0.727    &4.430 $\pm$ 0.672            &-          \\
	log $q$ (mean)                &-2.373 $\pm$ 0.074   &-2.354 $\pm$ 0.066           &-          \\
	$\alpha$ $(\rm{rad})$         &1.373 $\pm$ 0.007    &1.389 $\pm$ 0.006            &-          \\
	$\rho$                        &0.079 $\pm$ 0.052    &$0.061_{-0.038}^{+0.049}$    &$0.114_{-0.072}^{+0.093}$  \\
    $t_{0,2}-2457870$             &-                    &-                            &3.802 $\pm$ 0.080          \\
    $u_{0,2}$                     &-                    &-                            &$0.093_{-0.024}^{+0.018}$          \\
    $\rho_2$                      &-                    &-                            &$0.086_{-0.054}^{+0.044}$  \\
    $q_F$ $(10^{-3})$             &-                    &-                            &3.656 $\pm$ 0.643          \\
    $I_S$ [KMTC]                  &17.336 $\pm$ 0.001   &17.336 $\pm$ 0.001           &17.336 $\pm$ 0.001         \\
    $I_B$ [KMTC]                  &-                    &-                            &-                       \\
    $I_{S,2}$ [KMTC]              &-                    &-                            &23.433 $\pm$ 0.192         \\
%	$t_*$ $(\rm{hours})$          &22.758 $\pm$ 15.063  &$17.538_{-10.911}^{+14.361}$ &$33.299_{-21.018}^{+26.750}$   \\
%	$t_{*,2}$ $(\rm{hours})$      &-                    &-                            &$25.119_{-15.544}^{+12.624}$   \\
	\enddata
	%\tablecomments{}
	\label{tab:ob170543parms}
\end{deluxetable}

%% file: ob1694parmsplan.tex
\begin{deluxetable}{lcccc}
	\tablecolumns{5} \tablewidth{0pc} \tablecaption{\textsc{Planet models for OGLE-2017-BLG-1694}} \tablehead{\colhead{Parameters}
		& \colhead{Close}  & \colhead{Wide} & \colhead{Resonant 1}& \colhead{Resonant 2}} \startdata
	$\chi^2/\rm{dof}$             &10193.24/10193           &10194.72/10193           &10193.56/10193           &10215.75/10193   \\
	$t_0-2458000$                 &3.349 $\pm$ 0.032         &3.359 $\pm$ 0.039         &3.390 $\pm$ 0.031         &3.286 $\pm$ 0.069 \\
	$u_0$                         &0.193 $\pm$ 0.016         &0.204 $\pm$ 0.017         &0.222 $\pm$ 0.014         &0.322 $\pm$ 0.028 \\
	$t_{\rm E}$ $(\rm{days})$     &15.77 $\pm$ 0.74        &15.65 $\pm$ 0.80        &14.93 $\pm$ 0.62        &13.07 $\pm$ 0.65 \\
	$s$                           &0.637 $\pm$ 0.077         &1.926 $\pm$ 0.169         &1.164 $\pm$ 0.017         &1.104 $\pm$ 0.040 \\
	$q$ $(10^{-2})$               &$2.321_{-0.909}^{+1.625}$ &$1.915_{-0.624}^{+1.000}$ &$0.156_{0.036}^{+0.049}$  &3.409 $\pm$ 0.715\\
	log $q$ (mean)           &$-1.63$ $\pm$ 0.23     &$-1.71$ $\pm$ 0.19        &$-2.80$ $\pm$ 0.12   &$-1.47$ $\pm$ 0.09 \\
	$\alpha$ $(\rm{rad})$         &0.965 $\pm$ 0.052         &1.022 $\pm$ 0.037         &1.095 $\pm$ 0.019         &3.548 $\pm$ 0.045 \\
	$\rho$                        &0.035 $\pm$ 0.022         &$0.025_{-0.016}^{+0.019}$ &0.068 $\pm$ 0.006         &$0.088_{-0.036}^{+0.024}$\\
    $I_S$ [KMTC]                  &20.429 $\pm$ 0.096        &20.363 $\pm$ 0.097        &20.293 $\pm$ 0.079        &19.877 $\pm$ 0.118 \\
    $I_B$ [KMTC]                  &19.678 $\pm$ 0.047        &19.712 $\pm$ 0.053        &19.753 $\pm$ 0.047        &$20.114_{-0.127}^{+0.171}$\\
%	$t_*$ $(\rm{hours})$          &13.17 $\pm$ 8.21        &9.38 $\pm$ 6.53         &24.17 $\pm$ 1.66        &$27.71_{-11.67}^{+7.21}$\\
	\enddata
	%\tablecomments{}
	\label{tab:ob171694parmsplan}
\end{deluxetable}

%% file: ob1694parmsbd.tex
\begin{deluxetable}{lcc}
  \tablecolumns{3} \tablewidth{0pc} \tablecaption{\textsc{Brown Dwarf models for OGLE-2017-BLG-1694}}
  \tablehead{\colhead{Parameters}
		& \colhead{Close}  & \colhead{Wide} } \startdata
	$\chi^2/\rm{dof}$             &10201.45/10193           &10201.27/10193           \\
	$t_0-2458000$                 &3.410 $\pm$ 0.031         &3.386 $\pm$ 0.032         \\
	$u_0$                         &0.134 $\pm$ 0.009         &0.136 $\pm$ 0.010         \\
	$t_{\rm E}$ $(\rm{days})$     &18.80 $\pm$ 0.87        &19.24 $\pm$ 0.92        \\
	$s$                           &0.492 $\pm$ 0.016         &1.892 $\pm$ 0.076         \\
	$q$ $(10^{-2})$               &5.05 $\pm$ 0.84         &6.20 $\pm$ 1.12 \\
	log $q$ (mean)                &$-1.300$ $\pm$ 0.074        &$-1.209$ $\pm$ 0.081        \\
	$\alpha$ $(\rm{rad})$         &4.897 $\pm$ 0.019         &4.920 $\pm$ 0.025         \\
	$\rho$                        &$0.021_{-0.013}^{+0.018}$ &$0.023_{-0.015}^{+0.019}$ \\
    $I_S$ [KMTC]                  &20.820 $\pm$ 0.079        &20.785 $\pm$ 0.084        \\
    $I_B$ [KMTC]                  &19.526 $\pm$ 0.024        &19.536 $\pm$ 0.026        \\
%	$t_*$ $(\rm{hours})$          &$9.397_{-5.909}^{+8.192}$ &$10.784_{-6.880}^{+8.982}$         \\
	\enddata
	%\tablecomments{}
	\label{tab:ob171694parmsbd}
\end{deluxetable}

%% file: ob1694parmsstar.tex
\begin{deluxetable}{lcccc} 
  \tablecolumns{5} \tablewidth{0pc} \tablecaption{\textsc{Star models for OGLE-2017-BLG-1694}}\tablehead{\colhead{Parameters}
		& \colhead{Close 1}  & \colhead{Wide 1} & \colhead{Close 2}  & \colhead{Wide 2}} \startdata
	$\chi^2/\rm{dof}$             &10198.91/10193              &10196.59/10193              &10194.32/10193            &10198.98/10193   \\
	$t_0-2458000$                 &3.596 $\pm$ 0.039            &3.226 $\pm$ 0.033            &3.063 $\pm$ 0.062          &3.667 $\pm$ 0.036 \\
	$u_0$                         &0.163 $\pm$ 0.011            &0.130 $\pm$ 0.014            &0.203 $\pm$ 0.015          &0.117 $\pm$ 0.014 \\
	$t_{\rm E}$ $(\rm{days})$     &17.02 $\pm$ 0.78           &$21.50_{-1.84}^{+2.66}$   &15.80 $\pm$ 0.79         &$26.62_{-2.54}^{+3.39}$ \\
	$s$                           &0.329 $\pm$ 0.012            &4.228 $\pm$ 0.652            &0.389 $\pm$ 0.021          &4.397 $\pm$ 0.324 \\
	$q$ $(10^{-2})$               &$46.5_{-9.0}^{+13.2}$  &$43.1_{-21.1}^{+41.2}$ &$36.1_{-4.8}^{+8.6}$ &$170.2_{-50.2}^{+77.1}$\\
	log $q$ (mean)                &$-0.326$ $\pm$ 0.100           &$-0.356$ $\pm$ 0.303     &$-0.426$ $\pm$ 0.087         &0.236 $\pm$ 0.156 \\
	$\alpha$ $(\rm{rad})$         &0.895 $\pm$ 0.033            &0.848 $\pm$ 0.025            &$5.416_{-0.049}^{+0.065}$  &5.613 $\pm$ 0.039 \\
	$\rho$                        &$0.040_{-0.025}^{+0.032}$    &$0.015_{-0.010}^{+0.015}$    &$0.026_{-0.017}^{+0.026}$  &$0.020_{-0.013}^{+0.016}$\\
    $I_S$ [KMTC]                  &20.622 $\pm$ 0.079           &20.669 $\pm$ 0.078           &20.392 $\pm$ 0.092         &20.450 $\pm$ 0.075 \\
    $I_B$ [KMTC]                  &19.593 $\pm$ 0.030           &19.576 $\pm$ 0.028           &19.697 $\pm$ 0.048         &19.668 $\pm$ 0.037\\
%	$t_*$ $(\rm{hours})$     &$16.270_{-10.210}^{+12.725}$ &$8.058_{-5.184}^{+7.722}$    &$10.001_{-6.446}^{+9.724}$ &$13.175_{-8.408}^{+10.863}$\\
	\enddata
	%\tablecomments{}
	\label{tab:ob171694parmsstar}
\end{deluxetable}

%% file: tabcmd.tex
 \begin{deluxetable}{lrrrrrrrr}
 \tablecolumns{9} \rotate \tablewidth{0pc}
 \tablecaption{\textsc {CMD Parameters}}
 \tablehead{\colhead{Name} & 
\colhead{$(V-I)_{\rm S}$} &
\colhead{$(V-I)_{\rm cl}$} &
\colhead{$(V-I)_{\rm S,0}$} &
\colhead{$I_{\rm S}$} &
\colhead{$I_{\rm cl}$} &
\colhead{$I_{\rm cl,0}$} &
\colhead{$I_{\rm S,0}$} &
\colhead{$\theta_*\ (\muas)$} }
%\hline
 \startdata
 OGLE-2017-BLG-1275& N.A.          & 3.65$\pm$0.05 & 0.69$\pm$0.05 & 19.40$\pm$0.05 & 17.30$\pm$0.03 & 14.43 & 16.53$\pm$0.06 & 1.516$\pm$0.121 \\
 OGLE-2017-BLG-0640& N.A.          & N.A.          & 0.92$\pm$0.02 & 18.92$\pm$0.08 & 18.78$\pm$0.10 & 14.44 & 14.58$\pm$0.13 & 4.961$\pm$0.417 \\
 OGLE-2017-BLG-1237& 1.85$\pm$0.03 & 1.93$\pm$0.03 & 0.98$\pm$0.04 & 21.80$\pm$0.06 & 15.45$\pm$0.05 & 14.35 & 20.70$\pm$0.08 & 0.318$\pm$0.024 \\
 OGLE-2017-BLG-1777& 1.97$\pm$0.02 & 2.39$\pm$0.02 & 0.64$\pm$0.03 & 20.46$\pm$0.09 & 16.03$\pm$0.04 & 14.44 & 18.87$\pm$0.10 & 0.490$\pm$0.036 \\
 OGLE-2017-BLG-0543& 3.03$\pm$0.07 & 3.05$\pm$0.03 & 1.04$\pm$0.08 & 17.22$\pm$0.01 & 16.88$\pm$0.04 & 14.55 & 14.89$\pm$0.04 & 4.959$\pm$0.477 \\
 OGLE-2017-BLG-1694& 1.94$\pm$0.08 & 2.10$\pm$0.04 & 0.90$\pm$0.11 & 20.56$\pm$0.10 & 15.60$\pm$0.06 & 14.55 & 19.51$\pm$0.12 & 0.497$\pm$0.0.079 \\
\hline
 \enddata
 \tablecomments{$(V-I)_{\rm cl,0}=1.06$}
 \label{tab:cmd}
 \end{deluxetable}

%OGLE-2018-BLG-0866& 2.50$\pm$0.07 & 2.38$\pm$0.02 & 1.18$\pm$0.07 & $\pm$ & 16.15$\pm$0.04 & 14.46 & $\pm$ & $\pm$ \\

%% file: tabphysall.tex
\begin{deluxetable}{lccccccc}
\tablecolumns{8} 
\tablewidth{0pc}\tablecaption{\textsc{Physical properties}} 
\tablehead{\colhead{Event} & \multicolumn{4}{c}{Physical Properties} & \colhead{} &
\multicolumn{2}{c}{Relative Weights}\\
\cline{7-8} \colhead{Models}&\colhead{$M_{\rm host}$ $[M_\sun]$} &\colhead{$M_{\rm planet}$ $[M_J]$} 
&\colhead{$D_{\rm L}$ [kpc]} &\colhead{$a_\bot$ [au]} &  &
\colhead{Gal.Mod.} & \colhead{$\chi^2$}} \startdata
 OB171275  \\
 inner$(+,+)$ &0.60 $\pm$ 0.22  &5.30 $\pm$ 1.96  &7.74 $\pm$ 0.82  &1.14 $\pm$ 0.36 &&0.502 &0.190\\
 outer$(+,+)$ &0.60 $\pm$ 0.22  &5.75 $\pm$ 2.10  &7.72 $\pm$ 0.82  &2.02 $\pm$ 0.63 &&0.532 &1.000\\
 inner$(-,-)$ &0.65 $\pm$ 0.24  &5.67 $\pm$ 2.05  &7.69 $\pm$ 0.89  &1.21 $\pm$ 0.36 &&0.903 &0.213\\
 outer$(-,-)$ &0.65 $\pm$ 0.23  &6.23 $\pm$ 2.21  &7.65 $\pm$ 0.90  &2.14 $\pm$ 0.63 &&1.000 &0.563\\
% Adopted      &0.63 $\pm$ 0.23  &5.90 $\pm$ 2.20  &7.69 $\pm$ 0.90  & bimodal && &\\
              &                 &                 &                 &1.19 $\pm$ 0.36 && &\\[-1ex] 
 \raisebox{1.5ex}{Adopted}&\raisebox{1.5ex}{0.63 $\pm$ 0.23}&\raisebox{1.5ex}{5.90 $\pm$ 2.20}&\raisebox{1.5ex}{7.69 $\pm$ 0.90}  &2.09 $\pm$ 0.63 && &\\[1ex]
 \cline{1-8}\\[-1ex]
OB170640     &$0.32_{-0.18}^{+0.32}$  &$1.62_{-0.94}^{+1.64}$ &$6.63_{-1.45}^{+1.09}$ &1.14 $\pm$ 0.38 && &\\[1ex]
\cline{1-8}\\[-1ex]
OB171237     &$0.46_{-0.24}^{+0.30}$  &$3.80_{-1.99}^{+2.49}$ &$6.03_{-1.53}^{+0.94}$ &$2.53_{-0.64}^{+0.50}$ && &\\[1ex]
\cline{1-8}\\[-1ex]         
 OB171777  \\
2L1S (Local 1) &$0.038_{-0.014}^{+0.030}$  &$0.046_{-0.017}^{+0.037}$  &7.74 $\pm$ 0.81  &0.39 $\pm$ 0.06        &&0.539 &1.000\\
2L1S (Local 2) &$0.046_{-0.017}^{+0.042}$  &$0.078_{-0.028}^{+0.073}$  &7.74 $\pm$ 0.82  &$0.43_{-0.07}^{+0.08}$ &&1.000 &0.961\\
Adopted        &$0.044_{-0.017}^{+0.042}$  &$0.067_{-0.030}^{+0.073}$  &7.74 $\pm$ 0.82  &$0.41 \pm 0.08$ && &\\
 \enddata
 %\tablecomments{}
 \label{tab:physall}
\end{deluxetable}

%% file: tab2017.tex
\begin{deluxetable}{llccl}
\tablecolumns{5} \tablewidth{0pc}
\tablecaption{\textsc{AnomalyFinder Planets in KMT Prime Fields for 2017}}
\tablehead{\colhead{Event Name} &
\colhead{KMT Name} &
\colhead{$\log q$} &
\colhead{$s$} &
\colhead{Reference} }
\startdata
KB170428$^{a}$ & KB170428 & $-4.30$ & 0.88 & \citet{logqlt-4} \\
OB171434       & KB170016 & $-4.24$ & 0.98 & \citet{ob171434} \\
OB170482       & KB170084 & $-3.87$ & 1.07 & \citet{ob170482} \\
KB170165       & KB170165 & $-3.87$ & 0.95 & \citet{kb170165} \\
OB170406       & KB170243 & $-3.16$ & 1.13 & \citet{ob170406} \\
OB171777$^{b}$ & KB170282 & $-2.86$ & 0.72 & This paper       \\
OB170640       & KB171726 & $-2.31$ & 0.61 & This paper       \\
OB171237       & KB170422 & $-2.10$ & 1.04 & This paper       \\
OB171275$^{c}$ & KB170314 & $-2.06$ & 1.09 & This paper       \\
OB171049       & KB170370 & $-2.02$ & 1.32 & \citet{ob171049} \\
OB171375$^{c}$ & KB170078 & $-1.88$ & 0.84 & \citet{kb162397} \\
OB171522       & KB170460 & $-1.80$ & 0.95 & \citet{ob171522} \\
\hline
\hline
OB170173$^{d}$  & KB171707 & $-4.61$ & 1.54 & \citet{ob170173} \\
OB170448$^{d}$  & KB170090 & $-4.30$ & 3.16 & Zhai et al., in prep \\
OB170373$^{d}$  & KB171529 & $-2.81$ & 1.38 & \citet{ob170373} \\
OB170543$^{e}$  & KB170140 & $-2.37$ & 1.52 & This paper       \\
OB171694$^{d,f}$ & KB172126 & $-1.64$ & 0.64 & This paper       \\
\enddata
\tablecomments{Event names are abbreviations for, e.g.,
KMT-2017-BLG-0165 and OGLE-2017-BLG-1434
a: Minor $s$ degeneracy.
b: Exceptionally complex model.
c: $s$ degeneracy. 
d: large $q$ degeneracy.
e: 1L2S/2L1S degeneracy.
f: planet/binary degeneracy.
}
\label{tab:2017events}
\end{deluxetable}